\documentclass[]{aa}

\usepackage{bm}
\usepackage{newtxtext}
\usepackage[varg]{newtxmath}
\usepackage{amsmath}
\usepackage{caption}
\usepackage{upgreek} 
\usepackage{booktabs}
\usepackage{multirow}
\usepackage{siunitx}
\usepackage{graphicx}
\usepackage{xcolor}
\usepackage{orcidlink}
\usepackage{natbib,twoopt}
\bibpunct{(}{)}{;}{a}{}{,} 

\usepackage{hyperref}
\hypersetup{
    colorlinks,
    linkcolor={red!50!black},
    citecolor={blue!50!black},
    urlcolor={blue!80!black},
    pdftitle={How much large dust could be present in hot exozodiacal dust
              systems?},
    pdfauthor={T. A. Stuber, F. Kirchschlager, T. D. Pearce, S. Ertel,
               S. Krivov, S. Wolf},
    pdfkeywords={
     debris disk, debris disc, hot dust, hot exozodiacal dust, hot exozodi,
     comets, planetesimals, exoplanets, polarimetry, near infrared, NIR, mid
     infrared, MIR, far infrared, FIR, submillimeter, millimeter, submm/mm,
     radio, observational simulations, interferometry, long baseline, nulling,
     VLT, VLTI, VINCI, MATISSE, NOTT, Hi-5, CHARA, FLUOR, Keck, KIN, ALMA,
     ngVLA, circumstellar matter, interplanetary medium, infrared: planetary
     systems, submillimeter: planetary systems, methods: numerical
    }
}

\newcommand{\sbs}[1]{\ensuremath{_\text{#1}}}
\newcommand{\DMS}{\texttt{DMS}}
\newcommand{\NratioMath}{N\sbs{L} / N\sbs{S}}
\newcommand{\RelMassLargeMath}{M\sbs{L} / M}
\newcommand{\Fratio}{\scalebox{0.9}{$F^\nu\sbs{L} / F^\nu\sbs{S}\ $}}
\newcommand{\FNIR}{\scalebox{0.9}{$F^\nu\sbs{NIR}$}}
\newcommand{\FMIR}{\scalebox{0.9}{$F^\nu\sbs{MIR}$}}
\newcommand{\FpNIR}{\scalebox{0.9}{$F^\nu\sbs{NIR} + \Delta F^\nu\sbs{NIR}$}}
\newcommand{\FmNIR}{\scalebox{0.9}{$F^\nu\sbs{NIR} - \Delta F^\nu\sbs{NIR}$}}
\newcommand{\FpMIR}{\scalebox{0.9}{$F^\nu\sbs{MIR} + \Delta F^\nu\sbs{MIR}$}}
\newcommand{\FmMIR}{\scalebox{0.9}{$F^\nu\sbs{MIR} - \Delta F^\nu\sbs{MIR}$}}
\newcommand{\FNIRunsca}{$F^\nu\sbs{NIR}$}
\newcommand{\FMIRunsca}{$F^\nu\sbs{MIR}$}
\newcommand{\FpNIRunsca}{$F^\nu\sbs{NIR} + \Delta F^\nu\sbs{NIR}$}
\newcommand{\FmNIRunsca}{$F^\nu\sbs{NIR} - \Delta F^\nu\sbs{NIR}$}
\newcommand{\FpMIRunsca}{$F^\nu\sbs{MIR} + \Delta F^\nu\sbs{MIR}$}
\newcommand{\FmMIRunsca}{$F^\nu\sbs{MIR} - \Delta F^\nu\sbs{MIR}$}

\begin{document}


\title{How much large dust could be present in hot exozodiacal dust systems?}
\author{T. A. Stuber\inst{\ref{inst_kiel}} \orcidlink{0000-0003-2185-0525}
        \and F. Kirchschlager\inst{\ref{inst_ghent}} \orcidlink{0000-0002-3036-0184}
        \and T. D. Pearce\inst{\ref{inst_jena}} \orcidlink{0000-0001-5653-5635}
        \and S. Ertel\inst{\ref{inst_arizona}, \ref{inst_LBTI}} \orcidlink{0000-0002-2314-7289}
        \and A. V. Krivov\inst{\ref{inst_jena}}
        \and S. Wolf\inst{\ref{inst_kiel}} \orcidlink{0000-0001-7841-3452}
}
\authorrunning{T. A. Stuber et al.}
\institute{Institut für Theoretische Physik und Astrophysik,
           Christian-Albrechts-Universität zu Kiel,
           Leibnizstr. 15, 24118 Kiel, Germany\\
           \email{tstuber@astrophysik.uni-kiel.de}
           \label{inst_kiel}
           \and Sterrenkundig Observatorium, Ghent University,
           Krijgslaan 281-S9, B-9000 Gent, Belgium\label{inst_ghent}
           \and Astrophysikalisches Institut und Universitätssternwarte,
           Friedrich-Schiller-Universität Jena, Schillergässchen 2–3,
           07745 Jena, Germany\label{inst_jena}
           \and Department of Astronomy and Steward Observatory, The University
           of Arizona, 933 North Cherry Ave, Tucson, AZ 85721, USA\label{inst_arizona}
           \and Large Binocular Telescope Observatory, The University of
           Arizona, 933 North Cherry Ave, Tucson, AZ 85721, USA\label{inst_LBTI}
          }

\date{Received 9 February 2023 / Accepted 19 July 2023}

\abstract
{
An infrared excess over the stellar photospheric emission of main-sequence
stars has been found in interferometric surveys, commonly attributed to the
presence of hot exozodiacal dust (HEZD). While submicrometer-sized grains in
close vicinity to their host star have been inferred to be responsible for the
found near-infrared excesses, the presence and amount of larger grains as part
of the dust distributions are weakly constrained.
}
{
We quantify how many larger grains (above-micrometer-sized) could be present in
addition to submicrometer-sized grains, while being consistent with
observational constraints. This is important in order to distinguish between various
scenarios for the origin of HEZD and to better estimate its
observational appearance when observed with future instruments.
}
{
We extended a model suitable to reproduce current observations
of HEZD to investigate a bimodal size distribution. By deriving the
characteristics of dust distributions whose observables are consistent with
observational limits from interferometric measurements in the $K$ and
$N$ bands we constrained the radii of sub- and above-micrometer-sized grains as
well as their mass, number, and flux density ratios.
}
{
In the most extreme cases of some of the investigated systems,
large grains $\gtrsim \SI{10}{\upmu\m}$ might dominate the mass budget of HEZD
while contributing up to \SI{25}{\percent} of the total flux density
originating from the dust at a wavelength of \SI{2.13}{\upmu\m} and up to
\SI{50}{\percent} at a wavelength of \SI{4.1}{\upmu\m}; at a wavelength of
\SI{11.1}{\upmu\m} their emission might clearly dominate over the emission of
small grains. While it is not possible to detect such hot-dust distributions
using ALMA, the ngVLA might allow us to detect HEZD at millimeter wavelengths.
}
{
Large dust grains (above-micrometer-sized) might have a more important impact on
the observational appearance of HEZD than previously assumed, especially at
longer wavelengths.
}


\maketitle


\section{Introduction}\label{sect_introduction}

Interferometric observations of main-sequence stars in the photometric bands $H$
to $L$ revealed an excess emission over the stellar photosphere. It was found
across spectral types G to A and over diverse stellar ages with an incidence rate of
around \SI{10}{\percent} \citep{ertel:2014} to \SI{30}{\percent} \citep{absil:2013}.
This excess emission has been attributed to the presence of hot circumstellar
dust distributions \citep[e.g.,][]{absil:2006, di_folco:2007, defrere:2011},
called hot exozodiacal dust or hot exozodis \citep[HEZD; for a review,
see][]{kral:2017}. In contrast, the corresponding excess measured in the
mid-infrared (MIR) $N$ band is weaker and often not even detected
\citep{millan_gabet:2011, mennesson:2014}.$\,$\footnote{The only significant
excess detection in the $N$ band has been observed in the system of HD~102647
($\beta$~Leo). However, this is in doubt due to recent observations (see
Sect.~\ref{subsect_discussion_mir_observations}).}

Further analyses show the need for small (submicrometer-sized) dust grains in
close (sub-au) vicinity of their host star to explain the measured near-infrared
(NIR) excess. While carbonaceous grains are consistent with the observations, a
dominant contribution of silicate grains can be excluded because the weak MIR
excess would be inconsistent with the otherwise expected strong $N$ band
emission feature \citep[e.g.,][]{absil:2006, akeson:2009, kirchschlager:2017}.

The physical processes behind the phenomenon of HEZD are not yet
well-constrained. While the classical dust replenishment by a steady-state
collisional cascade in an in situ debris belt is unlikely
\citep{wyatt:2007a, lebreton:2013},
submicrometer motes should be repelled by radiation pressure and rapidly
blown out of the system, in particular around luminous stars of spectral type A
\citep{burns:1979, backman:paresce:1993}. Multiple alternative solutions to this
problem have been proposed to date, such as dust migration and pile-up by the
combination of Poynting-Robertson (P--R) drag and sublimation \citep[see also
\citeauthor{poynting:1904} \citeyear{poynting:1904};
\citeauthor{robertson:1937} \citeyear{robertson:1937};
\citeauthor{wyatt:whipple:1950} \citeyear{wyatt:whipple:1950};
\citeauthor{burns:1979} \citeyear{burns:1979} for work on the P--R drag]{
belton:1966, mukai:1974, krivov:1998, kobayashi:2009, kobayashi:2010,
kobayashi:2011, van_lieshout:2014b, sezestre:2019}; in situ delivery by
exocomets \citep{bonsor:2012a, bonsor:2012b, bonsor:2014, raymond:bonsor:2014,
marboeuf:2016, faramaz:2017, sezestre:2019, pearce:2022b}; or trapping of dust
grains by gas \citep{lebreton:2013, pearce:2020}, magnetic fields
\citep{czechowski:mann:2010, su:2013, rieke:2016, stamm:2019, kimura:2020},
or the differential Doppler effect \citep{burns:1979, rusk:1987,
sezestre:2019}. However, none of these scenarios provides a comprehensive
explanation for the transport of the dust to the inner regions nor how it can
survive there or be replenished efficiently \citep[see overview
in][]{pearce:2022b}.

While it was possible to constrain the grain size and location of small dust
grains (i.e., $\lesssim \SI{1}{\upmu\m}$), the deducible constraints on dust
grains much larger than the NIR and MIR observing wavelengths are limited.
Paired with studies that tried to explain the MIR excess by
a second cooler dust distribution farther out \citep[e.g.,][]{lebreton:2013},
this led to the picture that the hot inner HEZD distributions consist solely of
small submicrometer-sized grains (or dust populations with an extremely steep
grain size distribution). Constraints on the prevalence or lack of larger
grains in HEZD distributions would have significant implications on the
theoretical models explaining the phenomenon. For example, if hot dust
originates farther out in the system and is deposited close to the star, its
size distribution upon arrival would depend on the transportation process;
the size distribution of grains released close to a star by star-grazing comets
\citep[e.g.,][]{sekanina:miller:1973, hanner:1984, harker:2002, blum:2017}
may differ considerably to that of grains originating in an asteroid belt,
migrating in under P--R drag and experiencing collisions on the way in
\citep[e.g.,][]{rigley:2020}. Hence the transportation mechanism would
set the ratio of large to small grains arriving at the hot-emission region.
Similarly, many hot-dust models posit that some mechanisms trap grains close to
the star; mechanisms such as gas trapping or magnetic trapping would
preferentially trap grains of certain sizes (generally trapping smaller grains
more efficiently), again affecting the ratio of large to small grains near the
star. Consequently, if there is an unjustified bias toward denying the presence
of larger grains, this bias transfers to unified theories seeking
to provide a thorough explanation of the enigmatic phenomenon of HEZD around
main-sequence stars. Furthermore, constraints on HEZD distributions are
important for exoplanetary studies because the radiation scattered and emitted
by the HEZD would potentially interfere with that of close-in exoplanets.

In this study we derived constraints on large ($\gtrsim \SI{1}{\upmu\m}$) dust
grains in HEZD distributions from existing observations. For this purpose we
revisited interferometric observations in both the NIR and MIR wavelength
ranges of the targets that were analyzed by \citet{kirchschlager:2017}. They
utilized a model with dust of a single grain size distributed in a thin ring
around a central star to constrain the HEZD characteristics; we expanded their
model to contain two dust grain sizes. We present the stellar sample and our
modeling approach in Sect.~\ref{sect_methods} and the derived constraints on
HEZD parameters in Sect.~\ref{sect_results}. In Sect.~\ref{sect_discussion} we
discuss our modeling approach and findings. Furthermore, we conduct a study of
feasibility to detect large grain distributions at submillimeter/millimeter
wavelengths. We close with a summary in Sect.~\ref{sect_summary}.


\section{Methods}\label{sect_methods}

A single population of submicrometer-sized grains was shown to reproduce the
inferred NIR $K$ band (hereafter NIR) excesses, while being consistent with the
inferred smaller MIR $N$ band (hereafter MIR) excesses \citep{kirchschlager:2017}.
The emissivity of such small grains drops steeply with increasing wavelength,
as derived from Mie theory \citep{mie:1908}, leading to a steep decrease in the
emission spectrum from the NIR to the MIR wavelength regime which is consistent
with the observations. In that wavelength range, larger grains (above-micrometer-sized)
show a less drastic decrease in emissivity and are mostly cooler at the same
stellocentric distance, and thus they produce an emission spectrum shallower than that
inferred from observations. Therefore, the possible amount of these larger grains
is limited. However, beyond this qualitative argumentation, it is unknown what
amount of larger grains can be added to the already inferred submicrometer-sized
grain population without violating the observational constraints. To derive these
upper limits on the amount of larger grains in HEZD distributions, we searched
for combinations of populations of smaller and larger grains whose radiation
reproduces the inferred flux density both in the NIR and MIR wavelength range.


\subsection{Stellar sample and observations}\label{subsect_stellar_sample}

We investigated nine hot-dust targets with interferometric measurements both in
the NIR and MIR wavelength range that are compiled in Group I in \citet[][see
there for more information about the sample selection criteria]{kirchschlager:2017}.
The stellar parameters and references are listed in Table~\ref{table_target_parameters}.
In the NIR all targets were part of a survey performed by \citet{absil:2013}
using the Fiber Linked Unit for Optical Recombination
\citep[FLUOR,][]{coude_du_foresto:1997, coude_du_foresto:1998, merand:2006} at
the Center for High Angular Resolution Astronomy (CHARA) Array
\citep{coude_du_foresto:2003, ten_brummelaar:2005} except HD~216956
(Fomal\-haut), which was investigated by \citet{absil:2009} using archival data
obtained with the Very Large Telescope Interferometer
\citep[VLTI,][]{schoeller:2003, schoeller:glindemann:2003, richichi:percheron:2005}
using the VLT Interferometer Commissioning Instrument \citep[VINCI,][]{kervella:2000,
kervella:2003}. For all NIR measurements, also for that of VLTI/VINCI, we assumed
an effective observing wavelength of the $K$ band to be $\lambda = \SI{2.13}{\upmu\m}$,
as given by \citet{absil:2013}. In the MIR, all targets were part of a survey by
\citet[][central bin wavelength of $\lambda = \SI{8.5}{\upmu\m}$]{mennesson:2014}
using the Keck Interferometer mid-infrared Nulling instrument
\citep[KIN,][]{serabyn:2012, colavita:2013}.

Due to the new Hipparcos reduction \citep{van_leeuwen:2007} and the Gaia Early
Data Release 3 \citep{gaia_collaboration:2021} the parallaxes of several targets
changed compared to those used in the literature, which slightly affects
the determined stellar properties. As these changes are usually small compared to
uncertainties originating from observational and modeling effects, we used the
stellar parameters that appear in the literature.


\subsection{Inferred values of flux density}\label{subsect_infer_flux_density}

\citet{absil:2009, absil:2013} and \citet{mennesson:2014} inferred the
radiation originating from the circumstellar environment relative to the
contribution of the central star alone. We transformed these relative measures
to values of the flux density originating from the circumstellar dust
distributions by multiplying by the stellar flux density $F^\nu_\star$ obtained
from the Planck function that corresponds to the effective temperature $T_\star$
of the respective host star. For the NIR measurements, the flux density of the
circumstellar emission is computed as
\begin{equation}\label{eq_flux_density_NIR}
 F^\nu\sbs{NIR} = f\sbs{NIR} F^\nu_{\star\textrm{, NIR}} \; ,
\end{equation}
with the disk-to-star flux ratio $f\sbs{NIR}$ taken from Table~2 of
\citet{absil:2009} and Table~4 of \citet[][denoted there as
$f\sbs{CSE}$]{absil:2013}. For the MIR measurements, we computed the flux density
of the circumstellar emission following \citet{kirchschlager:2017} as
\begin{equation}\label{eq_flux_density_MIR}
 F^\nu\sbs{MIR} = 2.5 E\sbs{MIR} F^\nu_{\star\textrm{, MIR}} \; ,
\end{equation}
with $E\sbs{MIR}$ being the measured excess leak in the \num{8} to \SI{9}{\upmu\m}
bin from Table~2 of \citet[][denoted there as $E\sbs{8--9}$]{mennesson:2014}
and the factor of \num{2.5} compensating for the radiation not transferred
through the KIN transmission pattern. To compute the corresponding uncertainties
of the derived flux densities, denoted by $\Delta F\sbs{NIR}$ and $\Delta F\sbs{MIR}$,
we took the mostly probabilistic uncertainties of the measured excesses
$\Delta f\sbs{NIR}$ \citep[denoted as $\sigma\sbs{f}$ in][]{absil:2013} and
$\Delta E\sbs{MIR}$ \citep[denoted as $\sigma\sbs{8--9}$ in][]{mennesson:2014}
into account; we neglected uncertainties of the stellar flux density
$F^\nu_{\star}$ in Eqs.~\ref{eq_flux_density_NIR} and \ref{eq_flux_density_MIR}.
The derived flux densities and corresponding quantities for all considered
sources are listed in Table~\ref{table_excesses}.
\begin{table*}
 \centering
 \caption{Stellar parameters of investigated targets
 \citep[group I from][]{kirchschlager:2017}.}
 \begin{tabular}{cccccccc}
  \toprule
  HD number & HIP number & Alternative name & $d$ /\SI{}{pc} & $T_\star$ /\SI{}{\K} & $L_\star$ /$L_\sun$ & $R_\star$ /$R_\sun$ & Spectral type \\
  \midrule
  \phantom{1}10700 & \phantom{11}8102 & $\phantom{1}\tau$~Cet     & \phantom{1}\num{3.7}$\,$\tablefootmark{a} & \phantom{1}\num{5290}$\,$\tablefootmark{b} & \phantom{1}\num{0.47}$\,$\tablefootmark{b} & \num{0.82}$\,$\tablefootmark{c}           & G8$\,$V\phantom{IV-}\tablefootmark{d} \\
  \phantom{1}22484 & \phantom{1}16852 & 10~Tau                    & \num{13.9}$\,$\tablefootmark{a}           & \phantom{1}\num{5998}$\,$\tablefootmark{b} & \phantom{1}\num{3.06}$\,$\tablefootmark{b} & \num{1.62}$\,$\tablefootmark{e}           & F9$\,$IV-V$\;$\tablefootmark{f} \\
  \phantom{1}56537 & \phantom{1}35350 & \phantom{0i}$\lambda$~Gem & \num{30.7}$\,$\tablefootmark{a}           & \phantom{1}\num{7932}$\,$\tablefootmark{b} & \num{27.39}$\,$\tablefootmark{b}           & \num{2.78}$\,$\tablefootmark{e}           & A4$\,$IV\phantom{V}$\;$\tablefootmark{d} \\
  102647           & \phantom{1}57632 & \phantom{1}$\beta$~Leo    & \num{11.0}$\,$\tablefootmark{g}           & \phantom{1}\num{8604}$\,$\tablefootmark{b} & \num{13.25}$\,$\tablefootmark{b}           & \num{1.66}$\,$\tablefootmark{h}           & A3$\,$V\phantom{IV}$\;$\tablefootmark{h} \\
  172167           & \phantom{1}91262 & \phantom{1}$\alpha$~Lyr   & \phantom{1}\num{7.7}$\,$\tablefootmark{g} & \phantom{1}\num{9620}$\,$\tablefootmark{i} & \num{37}\phantom{.00}$\,$\tablefootmark{j} & \num{2.19}$\,$\tablefootmark{\phantom{g}} & A0$\,$V\phantom{IV}$\;$\tablefootmark{k} \\
  177724           & \phantom{1}93747 & \phantom{1}$\zeta$~Aql    & \num{25.5}$\,$\tablefootmark{g}           & \phantom{1}\num{9078}$\,$\tablefootmark{b} & \num{36.56}$\,$\tablefootmark{b}           & \num{2.45}$\,$\tablefootmark{e}           & A0$\,$IV-V$\,$\tablefootmark{k} \\
  187642           & \phantom{1}97649 & \phantom{1}$\alpha$~Aql   & \phantom{1}\num{5.1}$\,$\tablefootmark{g} & $\phantom{1}$7680$\,$\tablefootmark{l}     & \num{10.09}$\,$\tablefootmark{\phantom{g}} & \num{1.79}$\,$\tablefootmark{l}           & A7$\,$V\phantom{IV}$\,\,$\tablefootmark{k} \\
  203280           & 105199           & \phantom{1}$\alpha$~Cep   & \num{15.0}$\,$\tablefootmark{g}           & \phantom{1}\num{7510}$\,$\tablefootmark{m} & \num{18.1}\phantom{0}$\,$\tablefootmark{m} & \num{2.51}$\,$\tablefootmark{\phantom{g}} & A8$\,$V\phantom{IV}$\,\,$\tablefootmark{k}\\
  216956           & 113368           & \phantom{1}$\alpha$~PsA   & \phantom{1}\num{7.7}$\,$\tablefootmark{g} & \phantom{1}\num{8590}$\,$\tablefootmark{n} & \num{16.63}$\,$\tablefootmark{n}           & \num{1.84}$\,$\tablefootmark{n}           & A4$\,$V\phantom{IV}$\;$\tablefootmark{o} \\
  \bottomrule
 \end{tabular}
 \tablefoot{Stellar parameters are distance $d$, effective temperature
 $T_\star$, luminosity $L_\star$, and radius $R_\star$. The stellar radius of
 HD~172167 and HD~203280, and luminosity of HD~187642 were computed from
 $L_\star = 4 \pi \sigma_{\textrm{sb}} R_\star^2 T_\star^4$
 ($\sigma_{\textrm{sb}}$ is the Stefan--Boltzmann constant). We used
 interferometric measurements of the stellar radii where possible and used
 determinations based on stellar luminosity in the other cases (see
 Sect.~\ref{subsubsect_discussion_stellar_model} for a discussion of the stellar
 model).}
 \tablebib{
           \tablefoottext{a}{\citet{gaia_collaboration:2021};}
           \tablefoottext{b}{\citet{boyajian:2013};}
           \tablefoottext{c}{\citet{di_folco:2004};}
           \tablefoottext{d}{\citet{keenan:1989};}
           \tablefoottext{e}{\citet{boyajian:2012a};}
           \tablefoottext{f}{\citet{gray:garrison:1989};}
           \tablefoottext{g}{\citet{van_leeuwen:2007};}
           \tablefoottext{h}{\citet{van_belle:von_braun:2009};}
           \tablefoottext{i}{\citet{habing:2001};}
           \tablefoottext{j}{\citet{aufdenberg:2006b, aufdenberg:2006a};}
           \tablefoottext{k}{\citet{gray:2003};}
           \tablefoottext{l}{\citet{van_belle:2001};}
           \tablefoottext{m}{\citet{zhao:2009};}
           \tablefoottext{n}{\citet{mamajek:2012};}
           \tablefoottext{o}{\citet{gray:2006}.}
           }
 \label{table_target_parameters}
\end{table*}
\begin{table*}
 \centering
 \caption{Measurements and inferred flux densities in the NIR and MIR wavelength
  range of investigated targets.}
 \begin{tabular}{ccccccccccccccc}
  \toprule
  HD number        & $f\sbs{NIR}$ /\SI{}{\percent} & $\Delta f\sbs{NIR}$ /\SI{}{\percent} & $E\sbs{MIR}$ /\SI{}{\percent} & $\Delta E\sbs{MIR}$ /\SI{}{\percent} & $F^\nu\sbs{NIR}$ /\SI{}{Jy} & $\Delta F^\nu\sbs{NIR}$ /\SI{}{Jy} & $F^\nu\sbs{MIR}$ /\SI{}{Jy} & $\Delta F^\nu\sbs{MIR}$ /\SI{}{Jy}\\
  \midrule
  \phantom{1}10700 & 0.98 & 0.18\phantom{$\,$\tablefootmark{a}} & -0.11           & 0.21 & \phantom{1}1.24 & 0.23 & -0.04           & 0.07\phantom{5} \\
  \phantom{1}22484 & 1.21 & 0.11\phantom{$\,$\tablefootmark{a}} & \phantom{-}0.76 & 0.41 & \phantom{1}0.52 & 0.05 & \phantom{-}0.08 & 0.04\phantom{5} \\
  \phantom{1}56537 & 0.74 & 0.17\phantom{$\,$\tablefootmark{a}} & -0.30           & 0.30 & \phantom{1}0.30 & 0.07 & -0.03           & 0.03\phantom{5} \\
  102647           & 0.94 & 0.26\phantom{$\,$\tablefootmark{a}} & \phantom{-}0.56 & 0.14 & \phantom{1}1.18 & 0.33 & \phantom{-}0.15 & 0.04\phantom{5} \\
  172167           & 1.26 & 0.27\phantom{$\,$\tablefootmark{a}} & \phantom{-}0.21 & 0.09 & \phantom{1}6.61 & 1.42 & \phantom{-}0.23 & 0.10\phantom{5} \\
  177724           & 1.69 & 0.31$\,$\tablefootmark{a}           & \phantom{-}0.36 & 0.44 & \phantom{1}0.93 & 0.17 & \phantom{-}0.04 & 0.05\phantom{5} \\
  187642           & 3.07 & 0.24\phantom{$\,$\tablefootmark{a}} & \phantom{-}0.21 & 0.14 &           17.48 & 1.37 & \phantom{-}0.27 & 0.18\phantom{5} \\
  203280           & 0.87 & 0.18\phantom{$\,$\tablefootmark{a}} & \phantom{-}0.03 & 0.20 & \phantom{1}1.09 & 0.23 & \phantom{-}0.01 & 0.06\phantom{5} \\
  216956           & 0.88 & 0.12\phantom{$\,$\tablefootmark{a}} & \phantom{-}0.15 & 0.14 & \phantom{1}2.77 & 0.38 & \phantom{-}0.10 & 0.095 \\
  \bottomrule
 \end{tabular}
 \tablefoot{
  Uncertainties of quantities are denoted by a preceding $\Delta$.
  Values of $f\sbs{NIR}$ and $\Delta f\sbs{NIR}$ were taken from Table~4 of
  \citet[][denoted there as $f\sbs{CSE}$ and $\sigma\sbs{f}$,
  respectively]{absil:2013}, except for HD~216956 in which case they were taken
  from Table 2 of \citet{absil:2009}. All values of $E\sbs{MIR}$ and
  $\Delta E\sbs{MIR}$ were taken from Table~2 of \citet[][denoted there as
  $E\sbs{8--9}$ and $\sigma\sbs{8--9}$, respectively]{mennesson:2014}.
  \tablefoottext{a}{Stated erroneously as 0.27 in Table~4 of
  \citet{{absil:2013}}; the correct value is given in their Sect. 4.2 and in
  \citet{absil:2008}.}
 }
 \label{table_excesses}
\end{table*}


\subsection{Model of hot exozodiacal dust}\label{subsect_HEZD_model}


\subsubsection{Motivation: underlying observational constraints}
\label{subsubsect_model_justification}

The available observational data on the investigated targets are strongly limited.
They typically consist of a significant detection in the NIR and an additional
photometric upper limit in the MIR plus the constraint that the radiation
has to originate from the region inside the respective working angles of the
instruments. Consequently, those data are not overly constraining regarding
characteristics of HEZD. To date there is no consistent
theory explaining the dynamics of hot dust nor how the dust is delivered and
trapped in the close vicinity of the central star. Therefore, due to the
limited constraints on possible models, the use of a simple model is indicated.

A thin ring with a flat surface density seen face-on that consists of dust with a
single grain size represents such a simple model. It has been proven to explain
the rich data set of observations of HEZD for the target HD~7788 ($\kappa$~Tuc),
including $L$ band interferometric measurements obtained with the Multi AperTure
mid-Infrared SpectroScopic Experiment \citep[MATISSE,][]{lopez:2014, lopez:2022}
at the VLTI \citep{kirchschlager:2020}. Compared to a broad ring, this model
allows dust temperatures to be more easily interpreted, especially in the context
of sublimation. Furthermore, it allows the results from this study and previous
results to be directly compared \citep{kirchschlager:2017,
kirchschlager:2018, kirchschlager:2020}. As motivated above, the observational
data hardly justify a more detailed model, nor would the observations allow us
to well constrain more parameters.

To be able to constrain the amount of above-micrometer-sized grains that
are in agreement with the two photometric data points, we expanded the single
grain size description used in \citet{kirchschlager:2017} to include a second
larger grain size. Due to the interplay of different dust transport and trapping
mechanisms that most efficiently deliver or trap grains of certain sizes, a
bimodal grain size distribution could be produced that differs from the
continuous power law shapes classically used in debris disk models. Consequently,
the use of a bimodal size distribution leads to conclusions on the possible
interplay of different dust delivery and trapping mechanisms. While there is no
known trapping mechanism that traps grains of different sizes at the same radial
location that would correspond to this model, there is also no mechanism fully
explaining the observations of HEZD, and hence the dust dynamics. In summary,
the application of a simple model is indicated by both the limited observational
constraints and the uncertainty with regard to the underlying physical
mechanisms that are responsible for the HEZD phenomenon.


\subsubsection{Model setup and simulated observations}
\label{subsubsect_HEZD_model_set-up}

Given the considerations presented in
Sect.~\ref{subsubsect_model_justification}, we chose the following
description of HEZD. As in \citet{kirchschlager:2017}, the dust is distributed
in a wedge-shaped ring with a half-opening angle of \SI{5}{\degree}, radially
restricted by an inner and outer radius $R\sbs{out} = 1.5 R\sbs{in}$; the number
density of the dust grains is assumed to decrease with increasing distance to
the star $R$ as $n\left( R \right) \propto R^{-1}$, resulting in a flat surface
density. The overall dust distribution is assumed to be optically thin and seen
in face-on orientation; it consists of two single dust populations
at the same location, but with different grain radii, denoted by $a\sbs{S}$
(S: smaller) and $a\sbs{L}$ (L: larger).

To compute the surface brightness maps of the dust rings, we used an
extended version of the numerical tool \textbf{D}ebris disks around
\textbf{M}ain-sequence \textbf{S}tars \citep[\DMS,][]{kim:2018},
considering both thermal dust emission and scattered stellar radiation.
To obtain dust absorption and scattering cross sections, and the scattering
phase function, we used the numerical tool \texttt{miex}
\citep{wolf:voshchinnikov:2004} based on Mie theory.
The single-mode fiber interferometers FLUOR and VINCI showed a sensitivity over
the field of view of a Gaussian shape with full width at half maximum (FWHM)
of \SI{0.8}{as} and \SI{1.6}{as}, respectively. We took this into
account when simulating the brightness of dust configurations in the NIR by
multiplying the maps of surface brightness with two-dimensional circular
Gaussian distributions centered on the host star with peak intensity of unity
and with the respective FWHM. This decreases the contribution of the outer disk
regions to the total flux density along with increasing stellocentric distance.
From these brightness distributions we computed values of total flux density by
summing the flux densities of the entire map.


\subsection{Constraining parameters of hot exozodiacal dust}
\label{subsect_approach_to_constrain_parameters}

In their approach to constraining parameters of HEZD, \citet{kirchschlager:2017}
considered the mostly insignificant MIR measurements as
upper limits. Single-sized grains with radii in the submicrometer range then
allowed them to reproduce the observed NIR excess. In our study we aimed to put
constraints on the amount and characteristics of larger
($\gtrsim \SI{1}{\upmu\m}$) grains as constituents of HEZD that are nonetheless
consistent with the present upper limits represented by the MIR measurements.
For this purpose, we assumed the MIR measurements to be the true values.

\begin{figure}[!ht]
 \resizebox{\hsize}{!}{
           \includegraphics{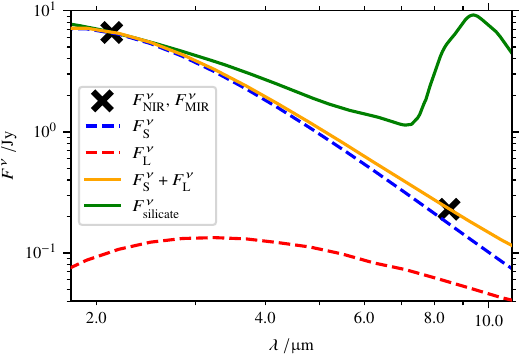}}
 \caption{
 Flux density $F^\nu$ originating from hot dust for the system HD~172167 (Vega).
 The black crosses denote the values inferred from the observations (see
 Table~\ref{table_excesses}). The orange line
 shows the radiation of a carbonaceous dust ring reproducing the observed
 flux densities with an inner disk radius of $R\sbs{in} = \SI{0.18}{au}$,
 a population of grains with radius $a\sbs{S} = \SI{0.2}{\upmu\m}$
 and a dust mass of $M\sbs{S} \approx \SI{6.2e-10}{M_\oplus}$,
 and a population of larger grains with radius $a\sbs{L} = \SI{10}{\upmu\m}$
 and a dust mass of $M\sbs{L} \approx \SI{8.9e-10}{M_\oplus}$; the dashed lines
 in blue and red denote the respective contributions of the two grain populations.
 For comparison, the solid green line depicts the radiation of a dust ring at the same
 location comprised of grains made of astronomical silicate with a grain radius
 of $a = \SI{0.2}{\upmu\m}$; the dust mass is scaled such that the flux density
 matches the value inferred from observations at $\lambda = \SI{2.13}{\upmu\m}$.
 This dust ring does not reproduce the observations due to the high predicted
 flux density in the MIR wavelength range.
 }
 \label{fig_illustrative_sed}
\end{figure}

To constrain the possible number of dust grains with certain radii, the two
dust populations with grain radii $a\sbs{S}$ and $a\sbs{L}$, respectively, are
assumed to produce the entire NIR and MIR radiation of the HEZD. With the
observed NIR to MIR flux density ratio
\begin{equation}
\label{eq_obs_flux_density_ratio}
 C^\textrm{NIR/MIR} = \frac{F\sbs{NIR}^\nu}{F\sbs{MIR}^\nu}
\end{equation}
and the ratios of the simulated flux densities
\scalebox{0.9}{$F^\nu_{\left\{\textrm{S, L}\right\}\textrm{, }\left\{\textrm{NIR, MIR}\right\}}$}
originating from the grain populations of the smaller and larger grains in the
NIR and MIR wavelength range, respectively,
\begin{equation}
\label{eq_sim_flux_density_ratio}
 C\sbs{S}^\textrm{NIR/MIR} = \frac{F\sbs{S, NIR}^\nu}{F\sbs{S, MIR}^\nu} \; ,\
 C\sbs{L}^\textrm{NIR/MIR} = \frac{F\sbs{L, NIR}^\nu}{F\sbs{L, MIR}^\nu} \; ,
\end{equation}
a dust configuration with a certain parameter combination of disk
radius and grain radii can reproduce the flux densities as inferred from the
observations if the condition
\begin{equation}
\label{eq_condition_ratio}
 \begin{aligned}
  C\sbs{S}^\textrm{NIR/MIR} &< C^\textrm{NIR/MIR} &< C\sbs{L}^\textrm{NIR/MIR}\\
  & \quad\quad\ \textrm{or} \\
  C\sbs{S}^\textrm{NIR/MIR} &> C^\textrm{NIR/MIR} &> C\sbs{L}^\textrm{NIR/MIR}
 \end{aligned}
\end{equation}
is fulfilled. In this case, the respective dust masses of the simulated
populations can be scaled such that the flux densities inferred from
observations are matched simultaneously at the NIR and MIR wavelengths. This is
achieved by solving the system of two linear equations
\begin{equation}
\label{eq_linear_eq_system}
 \begin{aligned}
  F^\nu\sbs{NIR} &= x\sbs{S} F^\nu\sbs{S, NIR}
                    + x\sbs{L} F^\nu\sbs{L, NIR} \; ,\\
  F^\nu\sbs{MIR} &= x\sbs{S} F^\nu\sbs{S, MIR}
                    + x\sbs{L} F^\nu\sbs{L, MIR}
 \end{aligned}
\end{equation}
for the scaling factors $x_{\left\{\textrm{S, L}\right\}}$.
With the obtained scaling factors we yielded the dust masses of the two grain
populations $M\sbs{S}$ and $M\sbs{L}$ (see
Fig.~\ref{fig_illustrative_sed}).

We investigated the impact of the interferometric measurement
uncertainties on the derived constraints on HEZD parameters. As the valid
parameter combinations are determined by the observed NIR-to-MIR flux density
ratio (Eq.~\ref{eq_condition_ratio}), we considered the cases of the
largest and smallest value thereof possible within the observational
uncertainties. This is achieved for the two cases (\FpNIR, \FmMIR) and vice
versa (\FmNIR, \FpMIR), respectively, for which we repeat the computation of
dust masses with Eq.~\ref{eq_linear_eq_system}.

Additionally to the discussed observations in the NIR and MIR wavelength range,
all investigated stellar systems have been observed at longer wavelengths
$\lambda \geq \SI{24}{\upmu\m}$. We collected those measurements from the
literature and used them as upper limits for valid dust configurations at the
respective wavelengths (see Table~\ref{table_obs_longer_wavelengths}). For
observations with the Atacama Large Millimeter/submillimeter Array
\citep[ALMA,][]{kurz:2002} at wavelengths of \SI{870}{\upmu\m} and
\SI{1300}{\upmu\m} (we assumed a wavelength of \SI{1300}{\upmu\m} for all
observations with ALMA band 6) and for the observation of HD~216956
with the \textit{Herschel} Space Observatory \citep{pilbratt:2010} by
\citet{acke:2012} we used the inferred flux density of the central unresolved
point source including stellar
photospheric emission. For all other observations we used the entire flux
density measured in the field of view, and thus the measurements possibly
include contributions of cold-dust emission originating from larger
stellocentric distances. Therefore, all of these measurements pose conservative
upper limits, and we neglected the corresponding measurement uncertainties in
this investigation.


\subsection{Parameter space}\label{subsect_parameter_space}

In the following we present the investigated parameter space of hot-dust
configurations.\\
\textit{Disk radii}:
We varied the disk inner and outer radii within the interval corresponding to the
angular on-sky region the instruments FLUOR, VINCI, and KIN were sensitive to.
In contrast to classical imaging instruments, additionally to an outer working
angle (OWA) set by the field of view, interferometric instruments also possess
an inner working angle (IWA) within which radiation is strongly attenuated.
Thus, constraints on dust distributions with an angular size smaller than the
IWA or larger than the OWA of any of the instruments cannot be derived using
our modeling approach. The KIN possessed an IWA of \SI{6}{mas}
\citep{mennesson:2014}, which is larger than the IWAs of CHARA/FLUOR and
VLTI/VINCI derived by \citet{kirchschlager:2017}. At the same time, the KIN OWA
was \SI{200}{mas,} which is smaller than those of CHARA/FLUOR and VLTI/VINCI.
Therefore, for each of the investigated systems, we can only constrain
dust configurations with
$R\sbs{in} \geq \SI{6}{mas}$ and
$R\sbs{out} \leq \SI{200}{mas}$. The
corresponding length scales depend on the stellar distance $d$ and are
individual for each target (see Table~\ref{table_target_parameters}).\\
\textit{Dust radii}:
We investigated dust grains with radii $a$ between \SI{1}{\nm} and \SI{1}{\mm},
logarithmically sampled with \num{103} values.$\,$\footnote{With \num{103}
values the grain radii $a = \SI{10}{\upmu\m}$, \SI{100}{\upmu\m}, and
\SI{1000}{\upmu\m} investigated in Sect.~\ref{subsect_results_large_grains}
are present in the grid exactly.} From this sample we built mutually exclusive
pairs of grain radii for the combinations of $a\sbs{S}$ and $a\sbs{L}$,
for a total of \num{5253} pairs.\\
\textit{Dust material:}
We investigated two different dust
materials using values of the wavelength-dependent complex refractive index
derived by \citet{draine:2003b}, astronomical silicate with a bulk density of
$\rho = \SI{3.8}{\g\per\cubic\cm}$ and graphite with a bulk density of
$\rho = \SI{2.2}{\g\per\cubic\cm}$. For the latter we used the catalog of
complex refractive indices that were computed for an assumed grain radius of
$a = \SI{0.1}{\upmu\m}$ and the 1/3--2/3 assumption to combine parallel- and
perpendicular-oriented grains \citep{draine:malhotra:1993}.

Such small dust grains in such close vicinity to their host star can
reach equilibrium temperatures high enough to sublimate. To decide whether a
dust grain of a specific material and size at a specific distance from
the host star is sublimated, we compute the equilibrium temperature at constant
distance while taking the wavelength-dependent absorption cross section into
account, and compare it to the sublimation temperatures $T\sbs{sub} =
\SI{1200}{\K}$ for astronomical silicate and $T\sbs{sub} = \SI{2000}{\K}$ for
graphite \citep{kobayashi:2009}. We considered every dust distribution with any
grains exceeding the sublimation temperature to be invalid to reproduce the
observations.


\section{Results}\label{sect_results}
\begin{figure}
 \resizebox{\hsize}{!}{
           \includegraphics{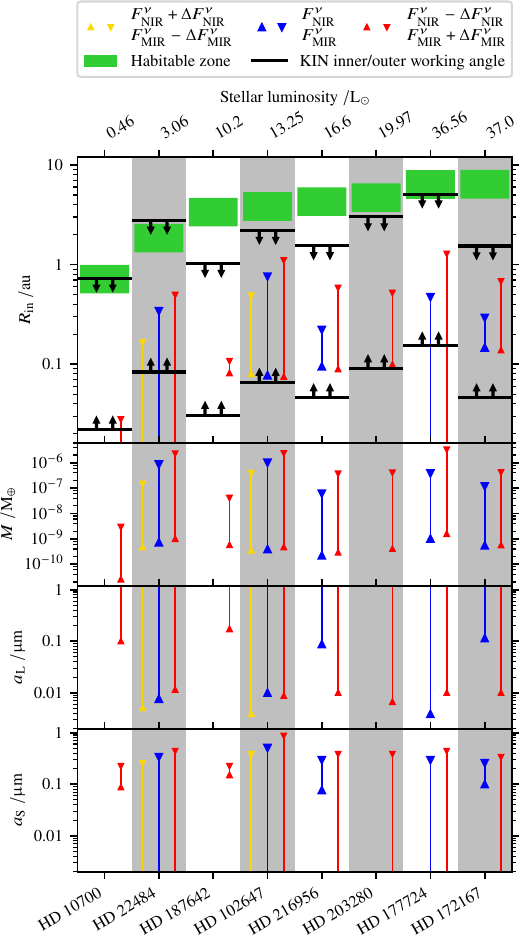}}
 \caption{
 Constraints on HEZD parameters of inner disk radius $R\sbs{in}$,
  total dust mass $M$, and
  grain radius of the populations of larger and smaller grains, $a\sbs{L}$ and
  $a\sbs{S}$. Three approaches were used to deal with observational
  uncertainties: no uncertainties (\protect\FNIR, \protect\FMIR) in blue; plus NIR,
  minus MIR uncertainty (\protect\FpNIR, \protect\FmMIR) in gold; and minus NIR, plus
  MIR uncertainty (\protect\FmNIR, \protect\FpMIR) in red. No markers are displayed
  when the respective quantity could not be constrained because the MIR
  excess leak was non-positive due to instrumental noise (see
  Sect.~\ref{subsect_results_no_analysis_possible}), no viable dust distribution
  was found in the investigated parameter space that is able to reproduce the
  observations (see Sect.~\ref{subsect_results_full_parameter_constraints}), or
  the boundary of the chosen simulated parameter space (see
  Sect.~\ref{subsect_parameter_space}) was reached (for $R\sbs{in}$, $a\sbs{L}$, or
  $a\sbs{S}$). In addition, the first panel shows the location of the habitable
  zones \citep[as defined in Sect.~2.5 of][]{kirchschlager:2017} and the inner and outer
  working angles of the KIN at the respective system distances that represent the
  borders of the parameter space of disk radii used in the simulations. The
  second panel shows the total dust mass $M$ with possible grain radii shown in
  the third and fourth panel. When no further restrictions were possible, the
  boundaries of the parameter space apply (i.e., is grain radii from \SI{1}{\nm}
  to \SI{1}{\mm}).
 }
 \label{fig_valid_param_ranges}
\end{figure}
\begin{figure*}
 \resizebox{\hsize}{!}{
           \includegraphics{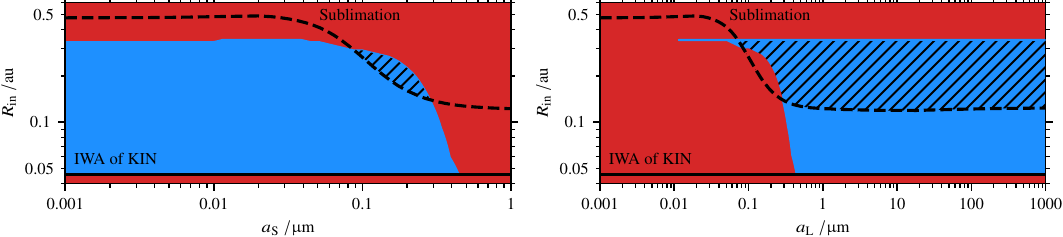}}
 \caption{
 Parameter combinations of inner disk radius $R\sbs{in}$ and the radii
 of the smaller and larger grains $a\sbs{S}$, $a\sbs{L}$ for the
 system HD~172167 (see the Fig.~\ref{fig_valid_Rin_and_a_all_stars}
 for all other investigated targets). The blue sections enclose all
 parameter combinations able to reproduce the flux densities inferred
 from observations in the NIR and MIR wavelength ranges. The red
 sections indicate parameter combinations that cannot reproduce the
 observations. The upper boundary of possible values of $R\sbs{in}$
 is given by the distance, where the radiation originating from the
 grain population of smaller grains no longer shows a NIR-to-MIR
 flux density ratio (Eq.~\ref{eq_sim_flux_density_ratio}) higher than the
 observed value (Eq.~\ref{eq_obs_flux_density_ratio}). The lower boundary is
 given by the physical scale corresponding to the IWA of the KIN, which
 is depicted by the solid black line.
 The dashed black line depicts the sublimation radii for different grain
 sizes. The intersecting set of parameter values for which, on the one hand,
 the dust configurations can reproduce the observations and, on the other hand,
 the dust grains of the respective grain sizes do not sublimate is indicated by
 the dashed area.\protect\\
 \textit{Right}:
 Parameter space of $R\sbs{in}$ and $a\sbs{L}$ for all possible values of
 $a\sbs{S}$.
 \textit{Left}:
 Parameter space of $R\sbs{in}$ and $a\sbs{S}$ with the value of $a\sbs{L}$
 fixed to $a\sbs{L}\in \left\{\SI{10}{\upmu\m}, \SI{100}{\upmu\m},
 \SI{1000}{\upmu\m} \right\}$; valid parameter combinations of $R\sbs{in}$ and
 $a\sbs{S}$ are the same for these three values.}
 \label{fig_valid_Rin_and_a_HD172167}
\end{figure*}
From the procedure described in Sect.~\ref{sect_methods} we obtained a set of
valid dust distributions, whose simulated observable quantities are consistent
with the observations.


\subsection{Systems where not all analyses were possible due to MIR
observations}
\label{subsect_results_no_analysis_possible}

Due to the measured excess leaks in the MIR ($E\sbs{MIR}$ in
Table~\ref{table_excesses}), which can be negative due to instrumental noise
\citep{mennesson:2014}, all combinations of inferred flux density and
corresponding uncertainty are non-positive for the system HD~56537 as
$E\sbs{MIR} = - \Delta E\sbs{MIR}$. Thus, this system could not be
constrained as our modeling approach requires a positive flux density. For the
same reason, not all combinations of inferred flux density and corresponding
uncertainties could be investigated for any system as $E\sbs{MIR} \leq \Delta
E\sbs{MIR}$; for the systems HD~177724 and HD~203280 we were only able to
investigate dust parameters for the cases of inferred flux densities of
\FMIR\ and \FpMIR; for the system HD~10700 only for the case of
\FpMIR.


\subsection{Dust composition}

For grains composed of pure astronomical silicate, there is no viable
combination of grain radii and disk radii in the entire investigated parameter
space \citep[the lack of a significant contribution of silicate grains was
already pointed out by, e.g.,][]{absil:2006, akeson:2009, lebreton:2013,
kirchschlager:2017}.
This is an expression of the fact that the characteristic $N$ band silicate
emission feature always results in a too high emission at $\lambda \approx
\SI{10}{\upmu\m}$ (see Fig.~\ref{fig_illustrative_sed});
none of the investigated sizes of the silicate grains at none of the
investigated disk radii fulfills the condition given in
Eq.~\ref{eq_condition_ratio}. Therefore, the following discussions will refer
solely to carbonaceous grains.


\subsection{No constraints from observations at wavelengths $\lambda \geq \SI{24}{\upmu\m}$}

Of those carbonaceous dust configurations able to reproduce the
flux densities inferred from observations in the NIR and MIR wavelength
range, none showed flux densities at longer wavelengths (i.e.,
$\lambda = \num{24}\ldots\SI{1300}{\upmu\m}$), which exceed any of the
photometric measurements listed in Table~\ref{table_obs_longer_wavelengths}.
Thus, the results are constrained solely by the interferometric measurements at
$\lambda = \SI{2.13}{\upmu\m}$ and $\lambda = \SI{8.5}{\upmu\m}$.


\subsection{Constraints on hot exozodiacal dust parameters}
\label{subsect_results_full_parameter_constraints}

For each stellar system we derived constraints on all parameters of hot dust
set by the smallest and largest value out of all valid parameter sets.
The results are displayed in Fig.~\ref{fig_valid_param_ranges}; the numerical
values are listed in Table~\ref{table_appdx_all_constraints} in the Appendix.

However, there are systems and combinations of observational uncertainties for
which there are no carbonaceous dust distributions in the investigated parameter
space that are able to reproduce the observations, either because the condition
given in Eq.~\ref{eq_condition_ratio} cannot be fulfilled or all the dust grains
exceed their sublimation temperature: HD~187642 in the case of inferred flux
densities of (\FpNIR, \FmMIR) and (\FNIR, \FMIR), HD~203280 in the case
of (\FNIR, \FMIR), and HD~172167 and HD~216956 in the case of (\FpNIR, \FmMIR).

The derived constraints on the inner disk radius and grain radius of the
small grain population are generally similar to those derived by
\citet{kirchschlager:2017}. This was expected because our modeling
approaches are similar (see Sect.~\ref{subsect_discussion_param_constraints} for
a discussion about the deviations and a comparison of the modeling approaches).
In the following, we present and discuss the results for all parameters
individually.


\subsubsection{Inner disk radius $R\sbs{in}$}

For the systems HD~10700, HD~22484, and HD~177724, the smallest possible inner
disk radius $R\sbs{in}$ (first panel of Fig.~\ref{fig_valid_param_ranges})
coincides with the smallest considered value, which is the
IWA of the KIN, and therefore remains unconstrained. However, relaxing the need
of MIR data and using solely data obtained with CHARA/FLUOR that provided higher
angular resolution, the innermost dust location can be further constrained
\citep{kirchschlager:2017}. For the systems with lower limits on the inner disk
radius $R\sbs{in}$ we found the same trend of values increasing with increasing
stellar luminosity. On the upper end all inner disk radii are constrained to
values smaller than $\sim \SI{1}{au}$.


\subsubsection{Dust mass $M$}

The constraints on the total dust mass in the systems
(second panel of Fig.~\ref{fig_valid_param_ranges})
encompass grains up to a grain radius of $a = \SI{1}{\mm}$. For all investigated
systems, except HD~10700, the possible dust masses are constrained to values of
$\sim \num{e-10}\ldots\SI{e-6}{M_\oplus}$ without any obvious trend with stellar
luminosity. For HD~10700, the dust masses are constrained to values of
$\approx \num{2.5e-11} \ldots \SI{3.0e-9}{M_\oplus}$ when taking observational
uncertainties into account. In each case the dust distributions with the highest
masses are those with the largest possible grain radius of the population of
large grains, that is $a = \SI{1}{\mm}$. The relative contribution of the larger
grains to the total mass is barely constrained; values of $\RelMassLargeMath
\sim \num{0.01}\ldots\SI{99.99}{\percent}$ are allowed.


\subsubsection{Larger grain radius $a\sbs{L}$}

The condition given in Eq.~\ref{eq_condition_ratio} sets lower limits on the
grain radius of the population of larger grains $a\sbs{L}$ (third
panel of Fig.~\ref{fig_valid_param_ranges}). It is constrained to values larger
than $\approx \num{0.004}\ldots\SI{0.17}{\upmu\m}$. On the upper end,
$a\sbs{L}$ is unconstrained because above a certain distance from the
central star arbitrarily large particles show a NIR-to-MIR flux density ratio
(Eq.~\ref{eq_sim_flux_density_ratio}) smaller than that derived from
observations (Eq.~\ref{eq_obs_flux_density_ratio}). For every system these
constraints allow grains with radii smaller and larger than the blowout grain
radius. An illustration can be seen of the
parameter space of valid combinations of the grain radius $a\sbs{L}$ and the
inner disk radius $R\sbs{in}$ for all possible values of $a\sbs{S}$ for the
example system HD~172167 in the right panel of
Fig.~\ref{fig_valid_Rin_and_a_HD172167} (see the
Fig.~\ref{fig_valid_Rin_and_a_all_stars} for all other investigated targets).


\subsubsection{Smaller grain radius $a\sbs{S}$}

For the grain radius of the population of smaller grains $a\sbs{S}$
(fourth panel of Fig.~\ref{fig_valid_param_ranges})
Eq.~\ref{eq_condition_ratio} sets upper limits; values are constrained to
be smaller than $\approx \num{0.2}\ldots\SI{0.9}{\upmu\m}$. For every
system the upper value is smaller than the respective blowout grain
radius, which shows the
need for submicrometer-sized, sub-blowout grains to reproduce the NIR excess.
We found lower limits being constrained to values of $\approx \num{0.08} \ldots
\SI{0.15}{\upmu\m}$ around the systems, HD~10700, HD~187642, HD~216956, and
HD~172167.


\subsection{Impact of observational uncertainties on derived constraints}
\label{subsect_results_observational_uncertainties}

The consideration of observational uncertainties does not alter the
qualitative nature of the resulting parameter constraints while the exact
values thereof change.
The lower constraint of the stellocentric distance is mostly invariant to
changes in the inferred MIR flux density as it is, if interferometrically
resolved, set by the sublimation radii in our model. As the largest possible
stellocentric distance is set by the distance where no grain of the population
of smaller grains can produce a NIR-to-MIR flux density ratio higher than the
one observed, this constraint increases with increasing true MIR excess and shrinks
for smaller true MIR excess. The possible dust masses increase with decreasing
NIR-to-MIR flux density ratio. At a fixed stellocentric distance, a smaller MIR
excess, and hence a higher NIR-to-MIR flux density ratio, would typically require
hotter, and thus smaller grains, while a larger MIR excess would require colder,
thus larger grains, effectively shifting the constraints on the grain radii $a\sbs{S}$
and $a\sbs{L}$ in the respective directions. However, exceptions from this general
trend with grain temperature can arise as the emission spectrum of a dust grain
depends on the complex interplay of the wavelength-dependent intensity of the
stellar light and wavelength, grain size, and material dependent
absorption (emission) efficiency of the respective dust grain. Such exceptions
occur for the systems HD~102647, HD~216956, and HD~172167 for which the case of
(\FmNIR, \FpMIR) permits smaller values of $a\sbs{L}$ than the case of
(\FNIR, \FMIR) (see third panel of Fig.~\ref{fig_valid_param_ranges}).


\subsection{Dust distributions with larger grains of the radii $a\sbs{L} = \SI{10}{\upmu\m}$, \SI{100}{\upmu\m}, \SI{1000}{\upmu\m}}\label{subsect_results_large_grains}

To explore the nature of valid dust distributions with large grain radii
$a\sbs{L} > \SI{1}{\upmu\m}$, we selected subsets of all valid parameter
combinations with the population of larger grains having a grain radius of $a\sbs{L} \in
\left\{ \SI{10}{\upmu\m}, \SI{100}{\upmu\m}, \SI{1000}{\upmu\m} \right\}$.
These grain radii are larger than the blowout radii for each system. The
constraints on the inner disk radius $R\sbs{in}$ are the same as for the full
parameter set, grains of all three sizes at locations as shown in
Fig.~\ref{fig_valid_param_ranges} are consistent with the observational constraints.
The same is true for the grain radius of the population of smaller grains $a\sbs{S}$
except for the system HD~177724 without observational uncertainties, there is
now a lower limit of $a\sbs{S} = \SI{0.004}{\upmu\m}$.
For the example system HD~172167 the combinations of $R\sbs{in}$ and
$a\sbs{S}$ of all dust distributions with $a\sbs{L}
\in \left\{ \SI{10}{\upmu\m}, \SI{100}{\upmu\m}, \SI{1000}{\upmu\m} \right\}$
that are able to reproduce the observations are shown for illustration in the
left panel of Fig.~\ref{fig_valid_Rin_and_a_HD172167} (see
Fig.~\ref{fig_valid_Rin_and_a_all_stars} for all other investigated targets).

\begin{figure}
 \resizebox{\hsize}{!}{
           \includegraphics{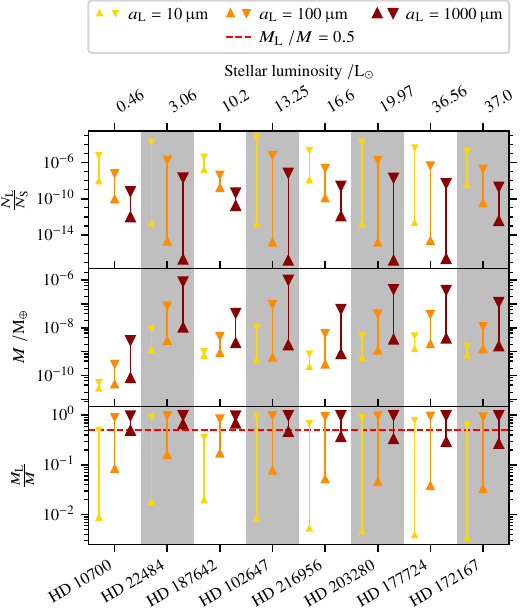}}
 \caption{
  Constraints for dust distributions with grain radius of the population of
  larger grains of $a\sbs{L} = \SI{10}{\upmu\m}$ (gold), $a\sbs{L} = \SI{100}{\upmu\m}$
  (orange), and $a\sbs{L} = \SI{1000}{\upmu\m}$ (dark red) on the HEZD parameters
  number ratio of larger to smaller grains $\NratioMath$, total dust mass $M$,
  and relative contribution of the population of larger grains to the total dust
  mass $\RelMassLargeMath$. The horizontal dashed red line denotes the case when
  the mass of the population of larger grains would make up half of the total
  dust mass (i.e., $\RelMassLargeMath = 0.5$); for higher values of
  $\RelMassLargeMath$ the large grains dominate the mass budget. Constraints are
  computed without taking observational uncertainties into account, except for
  the systems HD~10700, HD~187642, and HD~203280 for which no other constraints
  could be derived (see Sects.~\ref{subsect_results_no_analysis_possible} and
  \ref{subsect_results_full_parameter_constraints}).
  }
 \label{fig_valid_param_ranges_10_100_1000micron}
\end{figure}

From the mass and grain radius of each grain population we inferred the number of
grains making up each population, $N\sbs{S}$ and $N\sbs{L}$, and computed the
number ratio of larger to smaller grains $\NratioMath$. The constraints on this
quantity together with that on the total dust mass $M$ and the relative
contribution of the population of larger grains to the total dust mass
$\RelMassLargeMath$ are displayed in
Fig.~\ref{fig_valid_param_ranges_10_100_1000micron} (for all possible values of
$R\sbs{in}$ and $a\sbs{S}$). Constraints are shown
without taking observational uncertainties into account, except for the systems
HD~10700, HD~187642, and HD~203280 for which no other constraints could be
derived (see Sects.~\ref{subsect_results_no_analysis_possible} and
\ref{subsect_results_full_parameter_constraints}). Consistent across the
investigated stellar systems, along with increasing grain radius $a\sbs{L}$ the
values of $\NratioMath$ decrease: from a logarithmic mid value of the
parameter intervals of $\NratioMath \approx \num{3e-9}\ldots\num{8e-7}$ for
$a\sbs{L} = \SI{10}{\upmu\m}$, over
$\NratioMath \approx \num{3e-11}\ldots\num{8e-9}$ for $a\sbs{L} = \SI{100}{\upmu\m}$
to $\NratioMath \approx \num{3e-13}\ldots\num{9e-11}$ for
$a\sbs{L} = \SI{1000}{\upmu\m}$.

Along with an increasing larger grain radius $a\sbs{L}$, the total dust mass
increases as larger and more massive grains contribute disproportionately to the
total mass budget. The dust distributions with the
largest overall grain radius $a\sbs{L} = \SI{1000}{\upmu\m}$ are the most massive
ones in the overall set of valid dust distributions (cf. Fig.~\ref{fig_valid_param_ranges}).
Furthermore, the relative contribution of the population of larger grains to the
overall mass budget $\RelMassLargeMath$ increases: from
$\RelMassLargeMath \approx \SI{0.3}{\percent} \ldots \SI{96.2}{\percent}$
for $a\sbs{L} = \SI{10}{\upmu\m}$, over
$\RelMassLargeMath \approx \SI{3.3}{\percent} \ldots \SI{99.6}{\percent}$
for $a\sbs{L} = \SI{100}{\upmu\m}$ to
$\RelMassLargeMath \approx \SI{26.7}{\percent} \ldots \SI{99.96}{\percent}$
for $a\sbs{L} = \SI{1000}{\upmu\m}$. For each of the three investigated grain radii,
there are valid dust distributions for which the relative contribution to the total
disk mass of the population of larger grains exceeds that of the population of
smaller grains (red dashed line in bottom panel of
Fig.~\ref{fig_valid_param_ranges_10_100_1000micron}). The other HEZD parameters
of those dust distributions show the whole range of possible values. This finding
illustrates that the dust configurations for which the mass is dominated by
grains with radii $\geq \SI{10}{\upmu\m}$ are not located in isolated parts of
the parameter space.

\begin{figure}
 \resizebox{\hsize}{!}{
           \includegraphics{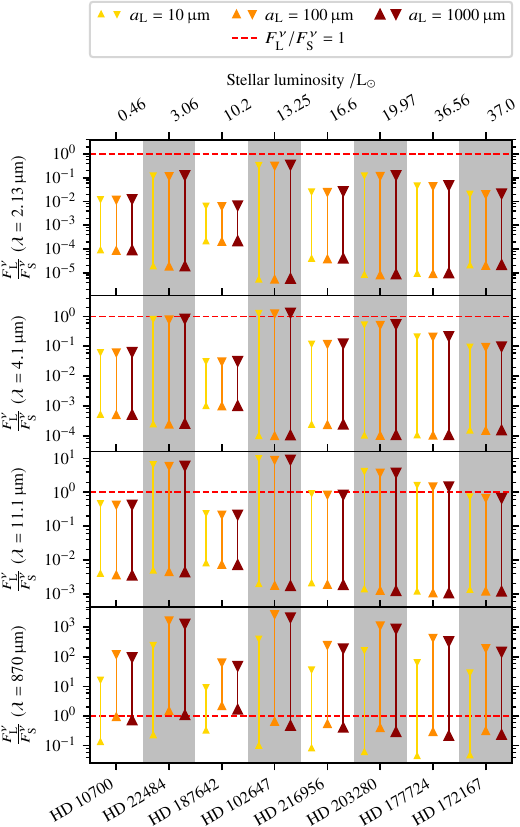}}
 \caption{
  Constraints on the ratio of flux density originating from the population
  of larger grains with the grain radius $a\sbs{L} = \SI{10}{\upmu\m}$ (gold),
  $a\sbs{L} = \SI{100}{\upmu\m}$ (orange), and $a\sbs{L} = \SI{1000}{\upmu\m}$
  (dark red) to that originating from the population of smaller grains with any
  possible grain radius $a\sbs{S}$ for the four observing wavelengths
  \SI{2.13}{\upmu\m}, \SI{4.1}{\upmu\m},
  \SI{11.1}{\upmu\m}, and \SI{870}{\upmu\m}. The horizontal dashed red lines denote
  equal contribution of the two grain populations to the total flux density (i.e.,
\protect\Fratio $= \num{1}$). Constraints are
  computed without taking observational uncertainties into account, except
  for the systems HD~10700, HD~187642, and HD~203280 for which no other
  constraints could be derived (see
  Sects.~\ref{subsect_results_no_analysis_possible} and
  \ref{subsect_results_full_parameter_constraints}).
  }
 \label{fig_flux_ratios_10_100_1000micron}
\end{figure}

To investigate the possible impact of the presence of larger grains in
HEZD systems on observable quantities, we computed the ratio of flux density
\Fratio originating from the larger grains of sizes $a\sbs{L} \in \left\{
\SI{10}{\upmu\m}, \SI{100}{\upmu\m}, \SI{1000}{\upmu\m} \right\}$ to that
originating from the population of smaller grains with any possible grain radius
$a\sbs{S}$ (see Fig.~\ref{fig_flux_ratios_10_100_1000micron}). For a
quantitative analysis we chose four example observing wavelengths. At first we
selected $\lambda = \SI{2.13}{\upmu\m}$ as the adopted wavelength for the CHARA/FLUOR
sample that is also accessible with GRAVITY \citep{gravity_collaboration:2017}
at the VLTI. Toward longer wavelengths, the impact of the larger grains
increases due to their lower temperatures and increasing emission efficiency.
For this wavelength range we chose $\lambda = \SI{4.1}{\upmu\m}$, which is the
central wavelength of the $L$\&$M$ band arm of VLTI/MATISSE and
$\lambda = \SI{11.1}{\upmu\m}$, which is the central wavelength
of the $N'$ filter of the Large Binocular Telescope Interferometer
\citep[LBTI,][] {hinz:2016, ertel:2020b}, that was used to perform the survey
Hunt for Observable Signatures of Terrestrial Systems
\citep[HOSTS,][]{ertel:2018a, ertel:2018b, ertel:2020a}. Finally, we selected
$\lambda = \SI{870}{\upmu\m}$ as the most promising ALMA band to observe
larger grains in HEZD (see Sect.~\ref{subsect_discussion_submm_mm_feasibility}
for a feasibility investigation regarding ALMA).

At the chosen wavelengths of \SI{2.13}{\upmu\m}, \SI{4.1}{\upmu\m}, and
\SI{11.1}{\upmu\m}, the ratio of flux
densities are constrained to similar values for all three grain radii. This can
be explained by the fact that the large grains are equally
inefficient emitters at the respective wavelengths $\lesssim a$. Consequently,
at the wavelength \SI{870}{\upmu\m} we see a split. There the grains with radius
$a\sbs{L} = \SI{100}{\upmu\m}$ are the most efficient emitters of the three
investigated sizes, followed by the grains with radius $a\sbs{L} =
\SI{1000}{\upmu\m}$. Across all stellar systems and all three grain radii, the
ratio of flux densities is constrained to values on the order of
\Fratio $\sim \num{e-5}$ to at most $\approx \num{0.35}$ at a wavelength of
$\lambda = \SI{2.13}{\upmu\m}$,
to values of \Fratio $\sim \num{e-4}\ldots\num{1}$ at a wavelength of
$\lambda = \SI{4.1}{\upmu\m}$, and to values of \Fratio $\sim \num{e-3}\ldots\num{10}$
at $\lambda = \SI{11.1}{\upmu\m}$.
At $\lambda = \SI{870}{\upmu\m}$, we found
\Fratio $\approx \num{4e-2}\ldots\num{4e2}$ for $a\sbs{L} = \SI{10}{\upmu\m}$,
\Fratio $\approx \num{3e-1}\ldots\num{3e3}$ for $a\sbs{L} = \SI{100}{\upmu\m}$,
and intermediate values of \Fratio $\approx \num{2e-1}\ldots\num{2e3}$ for
$a\sbs{L} = \SI{1000}{\upmu\m}$.
Across all four wavelengths, we generally found the ratios of flux density to be
more tightly constrained around closer systems, but without a clear monotonic
trend with stellar distance.
Even at a wavelength of \SI{4.1}{\upmu\m}, there are dust configurations
consistent with the observations, for which the population of larger grains with
a grain radius of $a\sbs{L} \geq \SI{10}{\upmu\m}$ contributes as much to the
total dust emission as the population of smaller grains or even dominates it.


\section{Discussion}\label{sect_discussion}


\subsection{Employed model of hot exozodiacal dust}
\label{subsect_discussion_disk_model}

Above-micrometer-sized grains as constituents of HEZD have
already been inferred by \citet{lebreton:2013} to model interferometric
observations of the system HD~216956. They used a different dust
description and assumed a two-ring model. This approach is
based on the work of \citet{mennesson:2013}; using only one dust
ring in the modeling they were not able to reasonably fit their whole data set
of MIR observations obtained with the KIN. However, they modeled the grain size
distribution with a single continuous power law. To reproduce the high NIR
excess, the minimum grain size was found to lie in the submicrometer range and
the slope of the power law was found to be extremely steep; however, this model
was unable to reproduce the rising slope of the MIR excess. They solved this
issue by invoking a second ring with a second
grain size distribution; this ring was found to have a larger radius and
to contain a dust distribution with above-micrometer-sized grains, although again
with an extremely steep slope of the grain size distribution. Using this model
together with a new treatment of grain sublimation
\citet{lebreton:2013} then confirmed these general results (see
Sect.~\ref{subsect_discussion_param_constraints} for a comparison of their
findings to ours). However, the need for two independent (i.e., in the two
rings) power-law grain size distributions, each with an extremely steep slope,
may also be interpreted as an indication of a bimodal grain size distribution
as applied in this work. As the distance from the central star is degenerate
with the grain size in its effect on the grain temperature and hence emission
spectrum of the dust, the two-ring description in the modeling approach of
\citet{mennesson:2013} and \citet{lebreton:2013} has a similar impact on the
resulting spectral distribution of flux density as a bimodal or at least broken
grain size distribution.

The use of different dust descriptions in our investigation would
not cause qualitative changes of the
derived parameter constraints, in particular of the main result: large
amounts of above-micrometer-sized grains are consistent with current
observations of HEZD and can dominate the dust mass budget.
However, quantitative changes would arise, and in the following we discuss
the individual model assumptions of this study in detail.


\subsubsection{Thin dust ring seen face-on}

We assumed the HEZD to be distributed in a thin ring with $R\sbs{in} =
1.5 R\sbs{out}$. A broader ring would allow for grains at
larger stellocentric distances, and hence lower equilibrium temperatures. On the
one hand, the presence of lower dust temperatures would result in a
lower NIR-to-MIR flux density ratio of the emission of both grain populations,
decreasing the possible relative number of large grains. On the other hand, it
would result in a generally lower emission power of the dust grains, leading to
higher dust masses to compensate for the lower emissivity. The relative impact
of these effects on the overall emission spectrum depends on the radial
dependency of the density distribution.

We considered the dust to be seen in face-on orientation. For an
optically thin system of spherical particles, different spatial dust
configurations and orientations with respect to the line of sight change the
amount of stellar light scattered toward the observer, while the amount of
observed thermal emission remains unchanged (neglecting minor effects of dust
coverage by the host star). However, if the total dust radiation (thermal
emission and scattered stellar light) is inferred from interferometric
observations, most of the forward-scattered light typical of Mie scattering
is strongly attenuated due to the IWAs of the instruments. Consequently, for a
dust configuration comprising solely grains of a specific grain size, a dust
ring seen edge-on and a spherical dust shell do not result in clearly different
constraints on the dust characteristics compared to a ring seen face-on
\citep{kirchschlager:2017}. Furthermore, for the
A-type star HD~56537 the relative contribution of scattered stellar light to the
total brightness peaks in the NIR wavelength regime for grains with radii
in the range \num{0.2}\ldots\SI{1}{\upmu\m} \citep[][see their
Fig.~4]{kirchschlager:2017}. Thus, for our dust
configurations additionally comprising grains up to a radius of \SI{1}{\mm},
the impact of scattered stellar light and hence the impact of the spatial dust
configuration and orientation is negligible in the NIR wavelength regime.
\citet{lebreton:2013} also note that in their modeling of the inner debris disk of
HD~216956 the contribution of scattered stellar light is always negligible in
the NIR--MIR wavelength range.
At an observing wavelength of $\SI{870}{\upmu\m,}$ large grains of the
size $\sim \SI{100}{\upmu\m}$ can be efficient in scattering stellar light, while
the dust thermal emission is weaker because of lower equilibrium temperatures of
the dust grains. However, at such long wavelengths the stellar emission is also
weaker, and the contribution of scattered stellar emission to the total dust
radiation remains negligible.


\subsubsection{Bimodal grain size distribution}

We modeled the grain size distribution with two distinct
grain sizes. The use of grains with single radii $a$ can cause interference in
the spectral distribution of radiation due to the shape of the wavelength-dependent
cross sections derived from Mie theory; the effect can be mitigated by the use
of a narrow grain size distribution centered around $a$ so the interference
cancel each other out \citep[e.g.,][]{kirchschlager:2019}.
The effect of this interference on our model results is small and we refrained
from using narrow grain size distributions around the distinct grain sizes for
easier comparison with the results of \citet{kirchschlager:2017}.

When replacing our two grain sizes
in the model by a continuous distribution ranging from the smaller to the
larger grain size, the constraints on the stellocentric distance of the HEZD as
derived would be weakly affected; the inner boundary is determined by the
modeled dust sublimation and the outer boundary is determined by the distance
beyond which no grain of any size produces radiation with a NIR-to-MIR
flux density ratio higher than the one observed. The constraints on the smaller
grain radius would transform to constraints on the minimum grain radius of the
continuous distribution and remain qualitatively unaltered as they are set
predominantly by the large observed NIR-to-MIR flux density ratio. This is
supported by the findings of \citet{kirchschlager:2017}, who used only one grain
radius in the modeling of various HEZD systems, and by \citet{lebreton:2013}, who
modeled the HEZD around HD~216956 using a continuous grain size
distribution; both showed constraints on the size of smallest grains similar
to ours (see Sect~\ref{subsect_discussion_param_constraints}). The
amount of the largest grains would be reduced as the grain size regime between the
minimum and maximum grain size would then be populated by dust motes
contributing to the total radiation that is constrained by the observations.
This effect would be strongest in the case of the most separated smaller and
larger grain sizes. Nonetheless, while a continuous, single power-law size
distribution is a conventional
choice to model cold debris disks, in the case of HEZD it is doubtful as
possibly different dust transport and trapping mechanisms might shape the size
distribution (see Sects.~\ref{sect_introduction} and
\ref{subsubsect_model_justification}).


\subsubsection{Dust composition}

We used graphite as a proxy for a carbonaceous material; other possible
choices include amorphous carbon \citep[e.g.,][]{rouleau:martin:1991,
jaeger:1998}. While those different materials do not produce a strongly
different spectral distribution of flux density (see Fig.~6 in
\citeauthor{kirchschlager:2017} \citeyear{kirchschlager:2017} or the similar
fit results in \citeauthor{kirchschlager:2020} \citeyear{kirchschlager:2020}
for different carbonaceous materials) their bulk densities differ from our
choice, which would result in a linear re-scaling of our presented absolute mass
estimates.

With grains composed of graphite we can reproduce the NIR and MIR measurements,
while with grains composed of astronomical silicate we cannot as such grains
generate MIR emission that is far to high, due to their $\sim \SI{10}{\upmu\m}$
spectral feature. This was also found by \citet{kirchschlager:2017} and
indicated by previous observational analyses \citep[e.g.,][]{absil:2006,
akeson:2009, lebreton:2013}. Across all grain radii and investigated targets
the graphite grains of our valid dust distributions show temperatures ranging from
$\sim \SI{450}{\K}$ up to the assumed sublimation temperature of \SI{2000}{\K}.
Based on \citet{kobayashi:2009}, who indicate a sublimation temperature of silicates
of \SI{1200}{\K}, this temperature range allows silicates to survive.
Therefore, a dust configuration with carbonaceous grains that incorporate at
least some amount of silicates is possible \citep[also indicated
by][]{lebreton:2013}. Furthermore, with decreasing stellocentric distance to
the central star the silicate components could sublimate, leaving potentially
porous carbonaceous grains (see below).

We assumed compact and hence non-porous dust grains. Analyses of the
aftermath of the collision created during the Deep Impact mission
\citep{a'hearn:2005a, a'hearn:2005b, meech:2005} showed that the upper layer
of comets (which are discussed as suppliers of HEZD grains) can be highly
porous \citep{kobayashi:2013}, which increases the blowout grain size
\citep{arnold:2019}. Furthermore, from investigations of porous silicate grains
it was concluded that porous dust grains show a lower equilibrium
temperature at a fixed stellocentric distance \citep{kirchschlager:wolf:2013,
brunngraeber:2017}. With the caveat of a non-silicate material, in our modeling
approach lower grain temperatures would result in lower
NIR-to-MIR flux density ratios and reduce the
sublimation radii, allowing grains to survive closer to the central star. While
the exact equilibrium temperature and emissivity of porous grains depends on
the individual combination of stellar spectrum, grain size and porosity,
compared to our results we expect an increase in the grain number density with a
roughly conserved absolute dust mass for low porosity values. Highly porous
grains are generally disfavored due to the large blowout sizes, and hence the
difficulty to define reasonable supply and trapping mechanisms to explain the
permanent nature of the phenomenon of HEZD. \citet{lebreton:2013} suggested
that the submicrometer-sized population around HD~216956 is composed of solid
residuals from the disruption of larger particles. Following this idea, a dust
distribution with two phases is also possible, for example a population of
highly porous large grains supplied by comet layers that get disrupted and
create a second population of solid submicrometer-sized grains. In the
case of a complete disintegration of a porous aggregate into its constituents
\citep{blum:muench:1993} this would cause a discontinuous grain size
distribution similar to that used in our model. Nonetheless, as the
available scarce observational data result in a highly degenerate set of
possible dust parameters, more observational constraints are needed to conclude
on the porosity of HEZD.


\subsubsection{Stellar model}\label{subsubsect_discussion_stellar_model}

We modeled the central stars as spherical blackbody emitters with a specific
effective temperature, luminosity, and radius. This assumption neglects the
rapid rotator nature of some A-type stars (see \citeauthor{royer:2007}
\citeyear{royer:2007} for a compilation of rotational velocities of A-type
stars). The rapid spinning can
produce a rather oblate shape of the star and a varying surface brightness
across the stellar surface, which has been shown for the following stars from
our sample: HD~172167 \citep[e.g.,][]{gulliver:1994, peterson:2006b,
aufdenberg:2006b, aufdenberg:2006a, monnier:2012}, HD~177724 \citep{zhao:2009},
HD~187642 \citep[e.g.,][]{van_belle:2001, peterson:2006a}, and
HD~203280 \citep{van_belle:2006}.
This deviation from a spherical blackbody emitter can change our results
depending on the dust morphology and its orientation with respect to
the stellar rotation axis as a different surface brightness of the stellar
surface changes the equilibrium temperatures of the dust grains, and hence their
sublimation radii and emitted radiation, and the characteristics of the
scattered stellar emission as well. As the true dust morphology is not constrained
for any of the known HEZD systems, this issue persists until stronger
observational constraints are available. However, we do not expect a major
influence on the results by this as the general characteristics of our findings
are consistent across the investigated targets, which include A-type stars with
different orientations of their rotational axis as well as a G- and an F-type
star.


\subsubsection{Dust sublimation}

The derived constraints on the minima of stellocentric distance and dust grain radii
depend on the assumed fixed sublimation temperature of \SI{2000}{\K} for
graphite \citep{kobayashi:2009}. The assumption of a fixed temperature for
the phase change, which is in fact a continuous process, is justified as the
process becomes rapid once a certain temperature is reached \citep{lamy:1974b},
and we neglected grain dynamics in our modeling \citep[for a more thorough
treatment of dust grain sublimation, see, e.g.,][]{lamy:1974b, kobayashi:2008,
kobayashi:2009, lebreton:2013, van_lieshout:2014a, van_lieshout:2014b, pearce:2022b}.
\citet{lebreton:2013} found that the additional consideration of the processes
collisions, PR-drag, and blowout by radiation pressure (but no trapping
mechanism) alters the sublimation distances around HD~216956 only for grains
with radii smaller than \SI{10}{\upmu\m}. Thus, our results for large grains
$\geq \SI{10}{\upmu\m}$ would be not affected. Furthermore, the constraints on
the grain radii are mainly determined by the observed ratio of NIR-to-MIR flux
density.

However, the lower constraint on the inner disk radius $R\sbs{in}$ is determined
by the particular choice of the sublimation temperature as it defines the
stellocentric distance inside which no grains of a certain size can survive.
Consequently, a higher value of the assumed sublimation temperature would lead
to smaller possible inner disk radii and vice versa.


\subsection{Derived constraints on parameters of hot exozodiacal dust}
\label{subsect_discussion_param_constraints}

We constrained the stellocentric distance of the hot dust to
$\lesssim \SI{1}{au}$ and the grain radius of one dust grain population to
$\lesssim \SI{1}{\upmu\m}$, similarly to what was found by
\citet{kirchschlager:2017}. Deviations arise due to a different treatment of
observational uncertainties and due to the way to include the MIR observations
in the modeling; while our aim was to reproduce the inferred MIR flux densities
as measured and explore the uncertainties separately, \citet{kirchschlager:2017}
used the significance limits (except for HD~102647 for which the excess leak
was measured significantly) and considered these MIR measurements solely as
upper limits. Furthermore, we modeled only
dust distributions with an angular scale larger than the IWA of the KIN which
is larger than the IWAs of CHARA and VINCI, and thus did not consider dust at
smaller stellocentric distances as \citet{kirchschlager:2017} did. For the
systems with lower limits on the inner disk radius $R\sbs{in}$ we found
the same rough trend of increasing values along with stellar luminosity as
\citet{kirchschlager:2017}; using the stellar ages given in their Table~1, we
also found no clear trend for the dust mass. We found dust masses of
$\sim \num{e-11}\ldots\SI{e-6}{M_\oplus}$, similar to the results of
\citet[][their Fig. 7]{kirchschlager:2017}.

\citet{lebreton:2013} found the inner dust distribution around HD~216956 to be
located inside $\num{0.1}\ldots\SI{0.3}{au,}$ which closely matches our
constraints on the inner disk radius of $R\sbs{in} =
\num{0.1}\ldots\SI{0.22}{au}$. They determined the dust mass up to grain radii of
\SI{1}{\mm} to be $\SI{2.5e-10}{M_\oplus}$ ($\SI{2.6e-10}{M_\oplus}$ with a
different modeling approach). This is consistent with the low mass end of our results
$\num{2.2e-10}\ldots\SI{6.1e-8}{M_\oplus}$, which is defined by dust
distributions where large grains do not dominate the mass budget. The grain size
distribution of the inner ring appears extremely steep with a minimum grain radius of
$\num{0.01}\ldots\SI{0.5}{\um}$, similar to our constraints on the grain radius
of the population of smaller grains $a\sbs{S} = \num{0.08}\ldots\SI{0.3}{\um}$.
In addition to the population of
submicrometer-sized carbonaceous grains, their modeling approach also required 
above-micrometer-sized grains ($\gtrapprox \SI{2.3}{\upmu\m}$ or
\SI{3.5}{\upmu\m}, depending on the exact modeling approach); the composition was
assumed to be half silicate and half carbonaceous. This second grain population
resides by design in a second ring that was constrained to be located at a
stellocentric distance of $\sim \SI{2}{au}$ with
a mass of
$\sim \SI{e-6}{M_\oplus}$ up to grain radii of \SI{1}{\mm}. Both distance and
dust mass are greater than the
upper limits we found on the inner disk radius, \SI{0.22}{au}, and dust mass,
\SI{6.1e-8}{M_\oplus} (\SI{0.58}{au} and \SI{3.8e-7}{M_\oplus} in the case of
\FmNIR, \FpMIR). However, in our modeling approach the population of large
grains shares the same location with the small grains.

In principle, the presence of arbitrarily large grains $a \gg \SI{1}{\upmu\m}$ in
HEZD systems is consistent with the present observational constraints because
arbitrarily large particles at a sufficient distance from the central star
produce a NIR-to-MIR flux density ratio lower than that derived from
observations. The relative mass contribution of the large grain population
($a\sbs{L} \leq \SI{1}{\mm}$) is in the range
$\sim \num{0.01}\ldots\SI{99.99}{\percent}$. Thus, above-micrometer-sized
grains might indeed dominate the HEZD mass budget. Furthermore, these larger
grains might contribute up to $\sim \SI{25}{\percent}$ to the total radiation
originating from the HEZD at a wavelength of \SI{2.13}{\upmu\m} (e.g.,
CHARA/FLUOR, VLTI/GRAVITY) and up to $\sim \SI{50}{\percent}$ at a wavelength of
\SI{4.1}{\upmu\m} (e.g., VLTI/MATISSE); at a wavelength of \SI{11.1}{\upmu\m}
(e.g., LBTI) they might even clearly dominate the total radiation with a
contribution ten times larger than that of smaller submicrometer-sized grains
(see Sect.~\ref{subsect_results_large_grains}).
However, a possible large relative contribution of these large grains to the
total dust mass and radiation budgets has only been identified within a rather
limited fraction of the investigated parameter space in the case of about half of the
investigated systems. Nevertheless, in future attempts to explain the HEZD
phenomenon it must be considered that large grains (i.e., above-micrometer-sized),
might have a significant impact on the mass budget and observational appearance
of HEZD distributions.


\subsection{Significance of the MIR observations and use in the modeling approach}
\label{subsect_discussion_mir_observations}
The stellar set compiled in group I of \citet{kirchschlager:2017} that we
investigated further in this study represents the complete ensemble of systems
with detected NIR excess and additional interferometric MIR measurements with a
sufficiently small IWA (excluding the variable HEZD system HD~7788~A).
We investigated what amount of above-micrometer-sized grains is still
consistent with the mostly insignificant flux densities inferred from the
MIR observations. For this purpose, we assumed the measured excess leaks in the
MIR at the wavelengths bin of \num{8} to \SI{9}{\upmu\m} \citep{mennesson:2014}
to be true values.

Out of the set of systems, only HD~102647 shows a significant
MIR excess with a measured excess significance (ratio of the measured
excess to its uncertainty) higher than three, with the caveat that this excess
could mainly originate from the outer regions of the KIN field of view (see
below). However, \citet{mennesson:2014} note that from statistical
considerations not all measurements with an excess significance between \num{1}
and \num{2} can be produced by instrumental noise, hinting at true dust
emission. This is supported by the fact that the excess measurements
presented by \citet{mennesson:2014} are averaged values from multiple
observations. The intensity of the radiation originating from the circumstellar
environment that is transferred through the nulling transmission pattern depends
on the source morphology and its orientation with respect to the transmission
pattern during observation, which also changes with time as the source traverses
the sky. Therefore, for an asymmetric dust configuration, for example a disk
seen edge-on, there is a statistical error of the excess values resulting from
averaging the measurements of observations performed at different hour angles.
Nonetheless, as the HEZD is not safely detected by the KIN, the
results of this study as presented provide upper limits of the presence and
impact of above-micrometer-sized grains in HEZD distributions.

From the measured excess leaks we inferred values of flux density originating
from circumstellar dust. The applied factor of \num{2.5} to rectify for the
circumstellar radiation not transmitted through the KIN transmission pattern is
an approximation suitable for an average system, while the true factor again
depends on the source morphology and its orientation with respect to the
transmission pattern (see above). Therefore, the true MIR radiation from any of
the observed systems can deviate in both directions from the inferred values.
This would alter the observed NIR-to-MIR flux density ratio, causing the
same effects on the derived HEZD parameter constraints as when taking
observational uncertainties into account (see
Sect.~\ref{subsect_results_observational_uncertainties}).

For the special case of HD~102647, with recent LBTI
measurements \citep{ertel:2020a} a similar excess to that found with the KIN
was inferred. As the LBTI possesses an IWA of $\sim \SI{70}{mas}$
\citep{ertel:2018a}, and thus larger than that of the KIN (\SI{6}{mas}),
this indicates that most of the excess inferred with the KIN originated from the
outer parts of its field of view $\gtrsim \SI{70}{mas}$, and hence less
radiation originates from the on-sky region in the range \num{6}\ldots
\SI{70}{mas}. Thus, the total flux density inferred from KIN
measurements (listed in Table~\ref{table_excesses}) cannot be attributed to hot
dust at locations corresponding to the angular scale of \num{6}\ldots
\SI{70}{mas} around HD~102647 \citep{pearce:2022b}. This would result
in a higher NIR-to-MIR flux density ratio derived from the observations. However,
to maintain uniformity of the analyzed data set, we decided not to use these
recent LBTI measurements.

Altogether, our results emphasize the need for a larger number of high
sensitivity MIR measurements of systems, with already significantly detected
HEZD distributions in the NIR wavelength regime, to either verify the MIR
detections or to derive robust upper limits. As the LBTI has an IWA of
$\sim \SI{70}{mas}$, too large
to investigate the innermost circumstellar regions, to date, VLTI/MATISSE is
the only instrument worldwide providing the required instrument capabilities
\citep{lopez:2014, lopez:2022, kirchschlager:2018, kirchschlager:2020}.
Possible future capabilities
to observe HEZD may be offered by the Nulling Observations of Exoplanets and
Dust \citep[NOTT, formerly Hi-5,][]{defrere:2018b, defrere:2018a, defrere:2022,
laugier:2023} which is proposed to work as part of the Asgard suite
\citep{martinod:2022, martinod:2023} at the VLTI.


\subsection{Feasibility to observe large grains at submillimeter and millimeter
wavelengths}
\label{subsect_discussion_submm_mm_feasibility}

To date, HEZD has only been convincingly detected with interferometers operating
in the $H$, $K$, and $L$ bands.
However, in this study we showed that a non-negligible contribution to the
total mass of HEZD distributions by above-micrometer-sized grains is
consistent with the existing observations. While the confirmation of the
presence of such large grains appears infeasible in the NIR and MIR wavelength
range, observations with comparable angular resolution in the far-infrared
(FIR) and submillimeter/millimeter wavelength regimes are more appropriate to
measure the emission of such large grains. Currently, there are no
observatories providing such a high angular resolution in the FIR; however, in
the submillimeter/millimeter wavelength regime there is ALMA.

\begin{figure}
 \resizebox{\hsize}{!}{
           \includegraphics{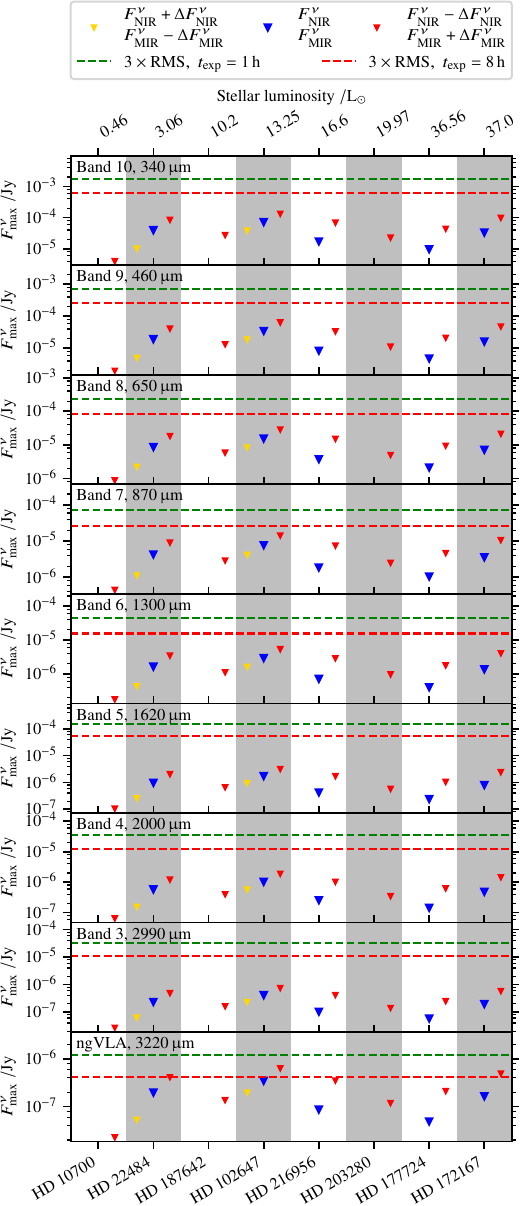}}
 \caption{
  Maximum flux density of all valid dust distributions at observing wavelengths
  corresponding to ALMA bands 10 to 3 and the ngVLA at an observing frequency
  of \SI{93}{\GHz} ($\lambda \approx \SI{3220}{\upmu\m}$). Three approaches were used
  to deal with the observational uncertainties: no uncertainties
  (\protect\FNIR, \protect\FMIR) in blue; plus NIR, minus MIR uncertainty
  (\protect\FpNIR, \protect\FmMIR) in gold; and minus NIR, plus MIR uncertainty
  (\protect\FmNIR, \protect\FpMIR) in red. Dashed horizontal lines indicate
  instrument sensitivity as three times the RMS for exposure times of
  $t\sbs{exp} = \SI{1}{\hour}$ (green) and \SI{8}{\hour} (red).
  }
 \label{fig_submm_mm_feasibility}
\end{figure}

To explore the general feasibility of observing HEZD that partly consists of
larger grains using
ALMA, we assumed an unresolved observation and compared the maximum flux densities
produced by dust thermal emission and scattered stellar light (in the face-on
case) of all valid dust distributions to the sensitivity of ALMA for a source at zenith
in all bands (see Fig.~\ref{fig_submm_mm_feasibility}). We estimated the instrument
sensitivity as three times the root mean square (RMS) for exposure times of
$t\sbs{exp} = \SI{1}{\hour}$ and \SI{8}{\hour}; these values were obtained
from the ALMA sensitivity
calculator.$\,$\footnote{\href{https://almascience.eso.org/proposing/sensitivity-calculator}{https://almascience.eso.org/proposing/sensitivity-calculator}}

The system with the highest possible flux density in any ALMA band is HD~102647,
followed by HD~22484 and HD~172167. Before band 6 and 8, the most promising
one is band 7 at an observing wavelength of \SI{870}{\upmu\m}. For HD~102647
and taking no observational uncertainties into account, the maximum flux density
of all valid dust distributions is $\approx \num{3}$ times smaller than the sensitivity
for an exposure time of $t\sbs{exp} = \SI{8}{\hour}$; when taking observational
uncertainties into account it is $\approx \num{2}$ times smaller. As this is the
most favorable case in our investigation, detecting HEZD with ALMA seems to be
infeasible. Across all other systems, the dust mass and hence the flux density
would have to be one to two orders of magnitude higher to enable detections with
ALMA.

A future instrument with sufficient resolving power is the ngVLA
\citep{selina:2018}. Performance estimates from December
2021$\,$\footnote{\href{https://ngvla.nrao.edu/page/performance}{https://ngvla.nrao.edu/page/performance}}
are given with a continuum RMS of \SI{0.40}{\upmu Jy/\textrm{beam}} at an
observing frequency of \SI{93}{\giga\Hz} (corresponding to a wavelength of
$\lambda \approx \SI{3.22}{\mm}$) for an exposure time of $t\sbs{exp} = \SI{1}{\hour}$.
Assuming all HEZD emission is collected in one beam, the possible presence of
some of the brightest dust distributions containing larger grains could be
tested with the ngVLA (see Fig.~\ref{fig_submm_mm_feasibility}, bottom
panel).


\subsection{Implications on HEZD theories}
\label{subsect_discussion_hezd_theory}
Better constraints on the dust populations responsible for the excess emission
over the stellar photosphere are essential if we are to understand this
phenomenon. Until now such populations were generally considered to comprise
only small grains, and one of the major technical difficulties encountered by
theoretical models is explaining how such sub-blowout grains can exist in large
quantities close to stars. If a significant fraction of the dust mass could
actually be hidden in larger (non-blowout) grains, as our work suggests, then
this would allow for a broader range of dynamical possibilities that could
make it easier to find a viable mechanism. Similarly, some existing models
(e.g., cometary supply) predict that grains would be present at all
sizes, while other models (e.g., gas trapping) suggest that the size
distribution is dominated by dust at a single size. Our work has shown that it
is possible (though by no means guaranteed) that large grains are present,
which unfortunately means that we cannot exclude different models from our
current large-grain constraints at the moment. However, we showed that future
observations, for example with the ngVLA, could directly measure the large-grain
component (see Sect.~\ref{subsect_discussion_submm_mm_feasibility}), which
would allow us to exclude various dynamical models.


\subsection{Implications on exoplanetary science}
\label{subsect_discussion_exoplanets}
We were able to constrain dust configurations with the largest extent matching
the OWA of the KIN. As shown in Fig.~\ref{fig_valid_param_ranges}, for the
system HD~22484 this completely covers the habitable zone \citep[as defined in
Sect.~2.5 of][]{kirchschlager:2017}, while a partial coverage results
for HD~10700 and HD~177724. The habitable zones of the other systems
did not fall in the sensitivity window of the KIN. However, the radii were
constrained to be smaller than the OWA. Thus, in the case that any additional dust
reservoirs are present outside the OWA and inside the habitable zone, these are
not connected to the HEZD populations in view of the sensitivities of the KIN.
However, \citet{ahlers:2022} showed that the habitable zones around rapid
rotators (see Sect.~\ref{subsubsect_discussion_stellar_model}) can reside closer to
the host star than for an equivalent non-rotating star, which we assumed in our
modeling. Furthermore, as we used a simple approximation of a blackbody in
thermal equilibrium for a potential celestial body to compute the location of
the habitable zones, a more thorough treatment of this topic including effects
such as light scattering in a potential atmosphere or off the surface, the
greenhouse effect, or inner heating sources might result in deviating locations
of the habitable zones.

In the case of exoplanet transit observations any additional source of emission,
such as hot circumstellar dust, causes an offset of the monitored light curve,
both before and during the transit, which translates into the planetary radius
being determined too small. In the $H$ and $K$ bands, the dust-to-star flux
ratio of HEZD distributions can be as high as several percent
\citep{absil:2013, ertel:2014} and in the $L$ band as high as $\num{5}\ldots
\SI{7}{\percent}$ \citep{kirchschlager:2020}, comparable to achievable
precision in the determination of stellar radii \citep[e.g.,][]{baines:2009,
torres:2010, white:2013, ligi:2016}.

In addition to HEZD as contributor to the net flux of extrasolar planetary systems,
we also have to take into account that the radiation disseminating from an
exoplanet is partially polarized. This is the result of different
light scattering processes, such as stellar light scattering off the planetary
atmosphere \citep[e.g.,][]{kattawar:1971, seager:2000, stam:2004, stam:2008,
buenzli:schmid:2009, fluri:berdyugina:2010, zugger:2010, zugger:2011a,
zugger:2011b, karalidi:2011, karalidi:2012, karalidi:2013, karalidi:stam:2012,
lietzow:2022}, an ocean surface \citep[e.g.,][]{stam:2008, williams:gaidos:2008,
zugger:2010}, or circumplanetary rings \citep[e.g.,][]{lietzow:2021,
lietzow:2023}, and by the scattering of planetary thermal emission in the
planetary atmosphere \citep[e.g.,][]{de_kok:2011,
stolker:2017}, with periodically changing magnitudes of the total scattered
flux and polarization degree thereof as the planet orbits its host star and
rotates around its own axis. This effect has been used to search for exoplanets
and the first exoplanet detected in this way, HD~189733b, was reported by
\citet[][]{berdyugina:2008}. However, the large amplitude of the periodic
polarization signal, also reported by \citet{berdyugina:2011}, was not
confirmed by \citet{wiktorowicz:2015} and \citet{bott:2016}, and according to
\citet{bailey:2021} could also be caused by the magnetic activity of the host
star. In addition to magnetic
activity of the host star, a co-existing HEZD distribution could also produce a
polarization signal by scattering the stellar light. The signal can be variable
in time in the case of an asymmetric HEZD distribution, which could be produced
for example by dust clumps formed by interaction with the close-by exoplanet or
a stochastic dust delivery process such as the disintegration of star grazing
comets. As the true dust morphology of any of the known HEZD systems is
unknown so far, HEZD must be considered as an additional radiation source in
the efforts to detect and characterize exoplanets via polarimetric
observations.


\section{Summary}\label{sect_summary}
Using a bimodal grain size distribution we investigated how many large
(above-micrometer-sized) dust grains in addition to the already inferred
submicrometer-sized
dust grains could be present in HEZD systems, while still being
consistent with interferometric observations in the $K$ and $N$ bands.
In confirmation of previous findings, we found that while dust
grains consisting of pure silicates are ruled out due to their high $N$ band
emission, carbonaceous grains are consistent with the observational
constraints. We constrained the dust location to be $\lesssim \SI{1}{au}$,
consistent with previous findings and, given our bimodal grain size model, also
constrained the number ratios of larger to smaller grains. In the most extreme
systems and dust configurations, large grains
$\gtrsim \SI{10}{\upmu\m}$ might dominate the mass budget of the considered
HEZD systems while contributing up to \SI{25}{\percent} to the total flux
density originating from the dust at a wavelength of \SI{2.13}{\upmu\m}, up to
\SI{50}{\percent} at a wavelength of \SI{4.1}{\upmu\m}, and clearly dominating
at a wavelength of \SI{11.1}{\upmu\m}. While it is not possible to detect
the radiation stemming from such hot-dust distributions with ALMA, the ngVLA
could potentially allow us to detect HEZD at millimeter wavelengths.


\section*{ORCID iDs}
T. A. Stuber \orcidlink{0000-0003-2185-0525}
\href{https://orcid.org/0000-0003-2185-0525}
     {https://orcid.org/0000-0003-2185-0525}\\
F. Kirchschlager \orcidlink{0000-0002-3036-0184}
\href{https://orcid.org/0000-0002-3036-0184}
     {https://orcid.org/0000-0002-3036-0184}\\
T. D. Pearce \orcidlink{0000-0001-5653-5635}
\href{https://orcid.org/0000-0001-5653-5635}
     {https://orcid.org/0000-0001-5653-5635}\\
S. Ertel \orcidlink{0000-0002-2314-7289}
\href{https://orcid.org/0000-0002-2314-7289}
     {https://orcid.org/0000-0002-2314-7289}\\
S. Wolf \orcidlink{0000-0001-7841-3452}
\href{https://orcid.org/0000-0001-7841-3452}
     {https://orcid.org/0000-0001-7841-3452}


\begin{acknowledgements}
 We thank the anonymous referee for various suggestions that helped to improve the
 presentation and discussion of our results, particularly in the context of
 previous studies. The author TS thanks all members of the Astrophysics
 Department Kiel for discussions and comments about this work in general.
 This research has made use of NASA's Astrophysics Data System Bibliographic
 Services, a modified A\&A bibliography style file for the preprint version of
 the article (\href{https://github.com/yangcht/AA-bibstyle-with-hyperlink}
 {https://github.com/yangcht/AA-bibstyle-with-hyperlink}), Ipython
 \citep{perez:granger:2007}, Jupyter notebooks \citep{jupyter},
 \texttt{Astropy} (\href{https://www.astropy.org}{https://www.astropy.org}), a
 community-developed core Python package for Astronomy
 \citep{astropy_collaboration:2013, astropy_collaboration:2018}, Matplotlib
 \citep{hunter:2007}, and Numpy \citep{harris:2020}.
 This work was supported by the Research Unit FOR 2285 “Debris Disks in
 Planetary Systems” of the Deutsche Forschungsgemeinschaft (DFG). The authors
 TS and SW acknowledge the DFG for financial support under grant WO 857/15-2,
 TDP and AVK under grants KR 2164/14-2 and KR 2164/15-2.
 FK has received funding from the European Research Council (ERC) under the
 EU’s Horizon 2020 research and innovation programme \mbox{DustOrigin}
 (ERC-2019-StG-851622). SE is supported by the National Aeronautics and Space
 Administration through the Exoplanet Research Program (Grant No.
 80NSSC21K0394).
\end{acknowledgements}


\bibliographystyle{aa_url}
\bibliography{bibliography}

\begin{thebibliography}{171}
\expandafter\ifx\csname natexlab\endcsname\relax\def\natexlab#1{#1}\fi

\bibitem[{{Absil} {et~al.}(2013){Absil}, {Defr{\`e}re}, {Coud{\'e} du Foresto},
  {Di Folco}, {M{\'e}rand}, {Augereau}, {Ertel}, {Hanot}, {Kervella},
  {Mollier}, {Scott}, {Che}, {Monnier}, {Thureau}, {Tuthill}, {ten Brummelaar},
  {McAlister}, {Sturmann}, {Sturmann}, \& {Turner}}]{absil:2013}
{Absil}, O., {Defr{\`e}re}, D., {Coud{\'e} du Foresto}, V., {et~al.} 2013,
  \href{http://dx.doi.org/10.1051/0004-6361/201321673}{\color{magenta}\aap},
  \href{https://ui.adsabs.harvard.edu/abs/2013A&A...555A.104A}{555, A104}

\bibitem[{{Absil} {et~al.}(2006){Absil}, {di Folco}, {M{\'e}rand}, {Augereau},
  {Coud{\'e} du Foresto}, {Aufdenberg}, {Kervella}, {Ridgway}, {Berger}, {ten
  Brummelaar}, {Sturmann}, {Sturmann}, {Turner}, \& {McAlister}}]{absil:2006}
{Absil}, O., {di Folco}, E., {M{\'e}rand}, A., {et~al.} 2006,
  \href{http://dx.doi.org/10.1051/0004-6361:20054522}{\color{magenta}\aap},
  \href{https://ui.adsabs.harvard.edu/abs/2006A&A...452..237A}{452, 237}

\bibitem[{{Absil} {et~al.}(2008){Absil}, {di Folco}, {M{\'e}rand}, {Augereau},
  {Coud{\'e} du Foresto}, {Defr{\`e}re}, {Kervella}, {Aufdenberg}, {Desort},
  {Ehrenreich}, {Lagrange}, {Montagnier}, {Olofsson}, {ten Brummelaar},
  {McAlister}, {Sturmann}, {Sturmann}, \& {Turner}}]{absil:2008}
{Absil}, O., {di Folco}, E., {M{\'e}rand}, A., {et~al.} 2008,
  \href{http://dx.doi.org/10.1051/0004-6361:200810008}{\color{magenta}\aap},
  \href{https://ui.adsabs.harvard.edu/abs/2008A&A...487.1041A}{487, 1041}

\bibitem[{{Absil} {et~al.}(2009){Absil}, {Mennesson}, {Le Bouquin}, {Di Folco},
  {Kervella}, \& {Augereau}}]{absil:2009}
{Absil}, O., {Mennesson}, B., {Le Bouquin}, J.-B., {et~al.} 2009,
  \href{http://dx.doi.org/10.1088/0004-637X/704/1/150}{\color{magenta}\apj},
  \href{https://ui.adsabs.harvard.edu/abs/2009ApJ...704..150A}{704, 150}

\bibitem[{{Acke} {et~al.}(2012){Acke}, {Min}, {Dominik}, {Vandenbussche},
  {Sibthorpe}, {Waelkens}, {Olofsson}, {Degroote}, {Smolders}, {Pantin},
  {Barlow}, {Blommaert}, {Brandeker}, {De Meester}, {Dent}, {Exter}, {Di
  Francesco}, {Fridlund}, {Gear}, {Glauser}, {Greaves}, {Harvey}, {Henning},
  {Hogerheijde}, {Holland}, {Huygen}, {Ivison}, {Jean}, {Liseau}, {Naylor},
  {Pilbratt}, {Polehampton}, {Regibo}, {Royer}, {Sicilia-Aguilar}, \&
  {Swinyard}}]{acke:2012}
{Acke}, B., {Min}, M., {Dominik}, C., {et~al.} 2012,
  \href{http://dx.doi.org/10.1051/0004-6361/201118581}{\color{magenta}\aap},
  \href{https://ui.adsabs.harvard.edu/abs/2012A&A...540A.125A}{540, A125}

\bibitem[{{A'Hearn} {et~al.}(2005{\natexlab{a}}){A'Hearn}, {Belton},
  {Delamere}, \& {Blume}}]{a'hearn:2005a}
{A'Hearn}, M.~F., {Belton}, M. J.~S., {Delamere}, A., \& {Blume}, W.~H.
  2005{\natexlab{a}},
  \href{http://dx.doi.org/10.1007/s11214-005-3387-3}{\color{magenta}\ssr},
  \href{https://ui.adsabs.harvard.edu/abs/2005SSRv..117....1A}{117, 1}

\bibitem[{{A'Hearn} {et~al.}(2005{\natexlab{b}}){A'Hearn}, {Belton},
  {Delamere}, {Kissel}, {Klaasen}, {McFadden}, {Meech}, {Melosh}, {Schultz},
  {Sunshine}, {Thomas}, {Veverka}, {Yeomans}, {Baca}, {Busko}, {Crockett},
  {Collins}, {Desnoyer}, {Eberhardy}, {Ernst}, {Farnham}, {Feaga}, {Groussin},
  {Hampton}, {Ipatov}, {Li}, {Lindler}, {Lisse}, {Mastrodemos}, {Owen},
  {Richardson}, {Wellnitz}, \& {White}}]{a'hearn:2005b}
{A'Hearn}, M.~F., {Belton}, M.~J.~S., {Delamere}, W.~A., {et~al.}
  2005{\natexlab{b}},
  \href{http://dx.doi.org/10.1126/science.1118923}{\color{magenta}Science},
  \href{https://ui.adsabs.harvard.edu/abs/2005Sci...310..258A}{310, 258}

\bibitem[{{Ahlers} {et~al.}(2022){Ahlers}, {Fromont}, {Kopparappu}, {Cauley},
  \& {Haqq-Misra}}]{ahlers:2022}
{Ahlers}, J.~P., {Fromont}, E.~F., {Kopparappu}, R., {Cauley}, P.~W., \&
  {Haqq-Misra}, J. 2022,
  \href{http://dx.doi.org/10.3847/1538-4357/ac5596}{\color{magenta}\apj},
  \href{https://ui.adsabs.harvard.edu/abs/2022ApJ...928...35A}{928, 35}

\bibitem[{{Akeson} {et~al.}(2009){Akeson}, {Ciardi}, {Millan-Gabet}, {Merand},
  {di Folco}, {Monnier}, {Beichman}, {Absil}, {Aufdenberg}, {McAlister}, {ten
  Brummelaar}, {Sturmann}, {Sturmann}, \& {Turner}}]{akeson:2009}
{Akeson}, R.~L., {Ciardi}, D.~R., {Millan-Gabet}, R., {et~al.} 2009,
  \href{http://dx.doi.org/10.1088/0004-637X/691/2/1896}{\color{magenta}\apj},
  \href{https://ui.adsabs.harvard.edu/abs/2009ApJ...691.1896A}{691, 1896}

\bibitem[{{Arnold} {et~al.}(2019){Arnold}, {Weinberger}, {Videen}, \&
  {Zubko}}]{arnold:2019}
{Arnold}, J.~A., {Weinberger}, A.~J., {Videen}, G., \& {Zubko}, E.~S. 2019,
  \href{http://dx.doi.org/10.3847/1538-3881/ab095e}{\color{magenta}\aj},
  \href{https://ui.adsabs.harvard.edu/abs/2019AJ....157..157A}{157, 157}

\bibitem[{{Astropy Collaboration} {et~al.}(2018){Astropy Collaboration},
  {Price-Whelan}, {Sip{\H{o}}cz}, {G{\"u}nther}, {Lim}, {Crawford}, {Conseil},
  {Shupe}, {Craig}, {Dencheva}, {Ginsburg}, {VanderPlas}, {Bradley},
  {P{\'e}rez-Su{\'a}rez}, {de Val-Borro}, {Aldcroft}, {Cruz}, {Robitaille},
  {Tollerud}, {Ardelean}, {Babej}, {Bach}, {Bachetti}, {Bakanov}, {Bamford},
  {Barentsen}, {Barmby}, {Baumbach}, {Berry}, {Biscani}, {Boquien}, {Bostroem},
  {Bouma}, {Brammer}, {Bray}, {Breytenbach}, {Buddelmeijer}, {Burke},
  {Calderone}, {Cano Rodr{\'\i}guez}, {Cara}, {Cardoso}, {Cheedella}, {Copin},
  {Corrales}, {Crichton}, {D'Avella}, {Deil}, {Depagne}, {Dietrich}, {Donath},
  {Droettboom}, {Earl}, {Erben}, {Fabbro}, {Ferreira}, {Finethy}, {Fox},
  {Garrison}, {Gibbons}, {Goldstein}, {Gommers}, {Greco}, {Greenfield},
  {Groener}, {Grollier}, {Hagen}, {Hirst}, {Homeier}, {Horton}, {Hosseinzadeh},
  {Hu}, {Hunkeler}, {Ivezi{\'c}}, {Jain}, {Jenness}, {Kanarek}, {Kendrew},
  {Kern}, {Kerzendorf}, {Khvalko}, {King}, {Kirkby}, {Kulkarni}, {Kumar},
  {Lee}, {Lenz}, {Littlefair}, {Ma}, {Macleod}, {Mastropietro}, {McCully},
  {Montagnac}, {Morris}, {Mueller}, {Mumford}, {Muna}, {Murphy}, {Nelson},
  {Nguyen}, {Ninan}, {N{\"o}the}, {Ogaz}, {Oh}, {Parejko}, {Parley}, {Pascual},
  {Patil}, {Patil}, {Plunkett}, {Prochaska}, {Rastogi}, {Reddy Janga},
  {Sabater}, {Sakurikar}, {Seifert}, {Sherbert}, {Sherwood-Taylor}, {Shih},
  {Sick}, {Silbiger}, {Singanamalla}, {Singer}, {Sladen}, {Sooley},
  {Sornarajah}, {Streicher}, {Teuben}, {Thomas}, {Tremblay}, {Turner},
  {Terr{\'o}n}, {van Kerkwijk}, {de la Vega}, {Watkins}, {Weaver}, {Whitmore},
  {Woillez}, {Zabalza}, \& {Astropy Contributors}}]{astropy_collaboration:2018}
{Astropy Collaboration}, {Price-Whelan}, A.~M., {Sip{\H{o}}cz}, B.~M., {et~al.}
  2018, \href{http://dx.doi.org/10.3847/1538-3881/aabc4f}{\color{magenta}\aj},
  \href{https://ui.adsabs.harvard.edu/abs/2018AJ....156..123A}{156, 123}

\bibitem[{{Astropy Collaboration} {et~al.}(2013){Astropy Collaboration},
  {Robitaille}, {Tollerud}, {Greenfield}, {Droettboom}, {Bray}, {Aldcroft},
  {Davis}, {Ginsburg}, {Price-Whelan}, {Kerzendorf}, {Conley}, {Crighton},
  {Barbary}, {Muna}, {Ferguson}, {Grollier}, {Parikh}, {Nair}, {Unther},
  {Deil}, {Woillez}, {Conseil}, {Kramer}, {Turner}, {Singer}, {Fox}, {Weaver},
  {Zabalza}, {Edwards}, {Azalee Bostroem}, {Burke}, {Casey}, {Crawford},
  {Dencheva}, {Ely}, {Jenness}, {Labrie}, {Lim}, {Pierfederici}, {Pontzen},
  {Ptak}, {Refsdal}, {Servillat}, \& {Streicher}}]{astropy_collaboration:2013}
{Astropy Collaboration}, {Robitaille}, T.~P., {Tollerud}, E.~J., {et~al.} 2013,
  \href{http://dx.doi.org/10.1051/0004-6361/201322068}{\color{magenta}\aap},
  \href{https://ui.adsabs.harvard.edu/abs/2013A&A...558A..33A}{558, A33}

\bibitem[{{Aufdenberg} {et~al.}(2006{\natexlab{a}}){Aufdenberg}, {M{\'e}rand},
  {Coud{\'e} du Foresto}, {Absil}, {Di Folco}, {Kervella}, {Ridgway}, {Berger},
  {ten Brummelaar}, {McAlister}, {Sturmann}, {Sturmann}, \&
  {Turner}}]{aufdenberg:2006b}
{Aufdenberg}, J.~P., {M{\'e}rand}, A., {Coud{\'e} du Foresto}, V., {et~al.}
  2006{\natexlab{a}},
  \href{http://dx.doi.org/10.1086/507484}{\color{magenta}\apj},
  \href{https://ui.adsabs.harvard.edu/abs/2006ApJ...651..617A}{651, 617}

\bibitem[{{Aufdenberg} {et~al.}(2006{\natexlab{b}}){Aufdenberg}, {M{\'e}rand},
  {Coud{\'e} du Foresto}, {Absil}, {Di Folco}, {Kervella}, {Ridgway}, {Berger},
  {ten Brummelaar}, {McAlister}, {Sturmann}, {Sturmann}, \&
  {Turner}}]{aufdenberg:2006a}
{Aufdenberg}, J.~P., {M{\'e}rand}, A., {Coud{\'e} du Foresto}, V., {et~al.}
  2006{\natexlab{b}},
  \href{http://dx.doi.org/10.1086/504149}{\color{magenta}\apj},
  \href{https://ui.adsabs.harvard.edu/abs/2006ApJ...645..664A}{645, 664}

\bibitem[{{Backman} \& {Paresce}(1993)}]{backman:paresce:1993}
{Backman}, D.~E. \& {Paresce}, F. 1993, in Protostars and Planets III, ed.
  E.~H. {Levy} \& J.~I. {Lunine},
  \href{https://ui.adsabs.harvard.edu/abs/1993prpl.conf.1253B}{1253}

\bibitem[{{Bailey} {et~al.}(2021){Bailey}, {Bott}, {Cotton},
  {Kedziora-Chudczer}, {Zhao}, {Evensberget}, {Marshall}, {Wright}, \&
  {Lucas}}]{bailey:2021}
{Bailey}, J., {Bott}, K., {Cotton}, D.~V., {et~al.} 2021,
  \href{http://dx.doi.org/10.1093/mnras/stab172}{\color{magenta}\mnras},
  \href{https://ui.adsabs.harvard.edu/abs/2021MNRAS.502.2331B}{502, 2331}

\bibitem[{{Baines} {et~al.}(2009){Baines}, {McAlister}, {ten Brummelaar},
  {Sturmann}, {Sturmann}, {Turner}, \& {Ridgway}}]{baines:2009}
{Baines}, E.~K., {McAlister}, H.~A., {ten Brummelaar}, T.~A., {et~al.} 2009,
  \href{http://dx.doi.org/10.1088/0004-637X/701/1/154}{\color{magenta}\apj},
  \href{https://ui.adsabs.harvard.edu/abs/2009ApJ...701..154B}{701, 154}

\bibitem[{{Belton}(1966)}]{belton:1966}
{Belton}, M. J.~S. 1966,
  \href{http://dx.doi.org/10.1126/science.151.3706.35}{\color{magenta}Science},
  \href{https://ui.adsabs.harvard.edu/abs/1966Sci...151...35B}{151, 35}

\bibitem[{{Berdyugina} {et~al.}(2008){Berdyugina}, {Berdyugin}, {Fluri}, \&
  {Piirola}}]{berdyugina:2008}
{Berdyugina}, S.~V., {Berdyugin}, A.~V., {Fluri}, D.~M., \& {Piirola}, V. 2008,
  \href{http://dx.doi.org/10.1086/527320}{\color{magenta}\apjl},
  \href{https://ui.adsabs.harvard.edu/abs/2008ApJ...673L..83B}{673, L83}

\bibitem[{{Berdyugina} {et~al.}(2011){Berdyugina}, {Berdyugin}, {Fluri}, \&
  {Piirola}}]{berdyugina:2011}
{Berdyugina}, S.~V., {Berdyugin}, A.~V., {Fluri}, D.~M., \& {Piirola}, V. 2011,
  \href{http://dx.doi.org/10.1088/2041-8205/728/1/L6}{\color{magenta}\apjl},
  \href{https://ui.adsabs.harvard.edu/abs/2011ApJ...728L...6B}{728, L6}

\bibitem[{{Blum} {et~al.}(2017){Blum}, {Gundlach}, {Krause}, {Fulle},
  {Johansen}, {Agarwal}, {von Borstel}, {Shi}, {Hu}, {Bentley}, {Capaccioni},
  {Colangeli}, {Della Corte}, {Fougere}, {Green}, {Ivanovski}, {Mannel},
  {Merouane}, {Migliorini}, {Rotundi}, {Schmied}, \& {Snodgrass}}]{blum:2017}
{Blum}, J., {Gundlach}, B., {Krause}, M., {et~al.} 2017,
  \href{http://dx.doi.org/10.1093/mnras/stx2741}{\color{magenta}\mnras},
  \href{https://ui.adsabs.harvard.edu/abs/2017MNRAS.469S.755B}{469, S755}

\bibitem[{{Blum} \& {M{\"u}nch}(1993)}]{blum:muench:1993}
{Blum}, J. \& {M{\"u}nch}, M. 1993,
  \href{http://dx.doi.org/10.1006/icar.1993.1163}{\color{magenta}\icarus},
  \href{https://ui.adsabs.harvard.edu/abs/1993Icar..106..151B}{106, 151}

\bibitem[{{Bonsor} {et~al.}(2012){Bonsor}, {Augereau}, \&
  {Th{\'e}bault}}]{bonsor:2012b}
{Bonsor}, A., {Augereau}, J.~C., \& {Th{\'e}bault}, P. 2012,
  \href{http://dx.doi.org/10.1051/0004-6361/201220005}{\color{magenta}\aap},
  \href{https://ui.adsabs.harvard.edu/abs/2012A&A...548A.104B}{548, A104}

\bibitem[{{Bonsor} {et~al.}(2014){Bonsor}, {Raymond}, {Augereau}, \&
  {Ormel}}]{bonsor:2014}
{Bonsor}, A., {Raymond}, S.~N., {Augereau}, J.-C., \& {Ormel}, C.~W. 2014,
  \href{http://dx.doi.org/10.1093/mnras/stu721}{\color{magenta}\mnras},
  \href{https://ui.adsabs.harvard.edu/abs/2014MNRAS.441.2380B}{441, 2380}

\bibitem[{{Bonsor} \& {Wyatt}(2012)}]{bonsor:2012a}
{Bonsor}, A. \& {Wyatt}, M.~C. 2012,
  \href{http://dx.doi.org/10.1111/j.1365-2966.2011.20156.x}{\color{magenta}\mnras},
  \href{https://ui.adsabs.harvard.edu/abs/2012MNRAS.420.2990B}{420, 2990}

\bibitem[{{Bott} {et~al.}(2016){Bott}, {Bailey}, {Kedziora-Chudczer}, {Cotton},
  {Lucas}, {Marshall}, \& {Hough}}]{bott:2016}
{Bott}, K., {Bailey}, J., {Kedziora-Chudczer}, L., {et~al.} 2016,
  \href{http://dx.doi.org/10.1093/mnrasl/slw046}{\color{magenta}\mnras},
  \href{https://ui.adsabs.harvard.edu/abs/2016MNRAS.459L.109B}{459, L109}

\bibitem[{{Boyajian} {et~al.}(2012){Boyajian}, {McAlister}, {van Belle},
  {Gies}, {ten Brummelaar}, {von Braun}, {Farrington}, {Goldfinger}, {O'Brien},
  {Parks}, {Richardson}, {Ridgway}, {Schaefer}, {Sturmann}, {Sturmann},
  {Touhami}, {Turner}, \& {White}}]{boyajian:2012a}
{Boyajian}, T.~S., {McAlister}, H.~A., {van Belle}, G., {et~al.} 2012,
  \href{http://dx.doi.org/10.1088/0004-637X/746/1/101}{\color{magenta}\apj},
  \href{https://ui.adsabs.harvard.edu/abs/2012ApJ...746..101B}{746, 101}

\bibitem[{{Boyajian} {et~al.}(2013){Boyajian}, {von Braun}, {van Belle},
  {Farrington}, {Schaefer}, {Jones}, {White}, {McAlister}, {ten Brummelaar},
  {Ridgway}, {Gies}, {Sturmann}, {Sturmann}, {Turner}, {Goldfinger}, \&
  {Vargas}}]{boyajian:2013}
{Boyajian}, T.~S., {von Braun}, K., {van Belle}, G., {et~al.} 2013,
  \href{http://dx.doi.org/10.1088/0004-637X/771/1/40}{\color{magenta}\apj},
  \href{https://ui.adsabs.harvard.edu/abs/2013ApJ...771...40B}{771, 40}

\bibitem[{{Brunngr{\"a}ber} {et~al.}(2017){Brunngr{\"a}ber}, {Wolf},
  {Kirchschlager}, \& {Ertel}}]{brunngraeber:2017}
{Brunngr{\"a}ber}, R., {Wolf}, S., {Kirchschlager}, F., \& {Ertel}, S. 2017,
  \href{http://dx.doi.org/10.1093/mnras/stw2675}{\color{magenta}\mnras},
  \href{https://ui.adsabs.harvard.edu/abs/2017MNRAS.464.4383B}{464, 4383}

\bibitem[{{Buenzli} \& {Schmid}(2009)}]{buenzli:schmid:2009}
{Buenzli}, E. \& {Schmid}, H.~M. 2009,
  \href{http://dx.doi.org/10.1051/0004-6361/200911760}{\color{magenta}\aap},
  \href{https://ui.adsabs.harvard.edu/abs/2009A&A...504..259B}{504, 259}

\bibitem[{{Burns} {et~al.}(1979){Burns}, {Lamy}, \& {Soter}}]{burns:1979}
{Burns}, J.~A., {Lamy}, P.~L., \& {Soter}, S. 1979,
  \href{http://dx.doi.org/10.1016/0019-1035(79)90050-2}{\color{magenta}\icarus},
  \href{https://ui.adsabs.harvard.edu/abs/1979Icar...40....1B}{40, 1}

\bibitem[{{Churcher} {et~al.}(2011){Churcher}, {Wyatt}, {Duch{\^e}ne},
  {Sibthorpe}, {Kennedy}, {Matthews}, {Kalas}, {Greaves}, {Su}, \&
  {Rieke}}]{churcher:2011}
{Churcher}, L.~J., {Wyatt}, M.~C., {Duch{\^e}ne}, G., {et~al.} 2011,
  \href{http://dx.doi.org/10.1111/j.1365-2966.2011.19341.x}{\color{magenta}\mnras},
  \href{https://ui.adsabs.harvard.edu/abs/2011MNRAS.417.1715C}{417, 1715}

\bibitem[{{Colavita} {et~al.}(2013){Colavita}, {Wizinowich}, {Akeson},
  {Ragland}, {Woillez}, {Millan-Gabet}, {Serabyn}, {Abajian}, {Acton},
  {Appleby}, {Beletic}, {Beichman}, {Bell}, {Berkey}, {Berlin}, {Boden},
  {Booth}, {Boutell}, {Chaffee}, {Chan}, {Chin}, {Chock}, {Cohen}, {Cooper},
  {Crawford}, {Creech-Eakman}, {Dahl}, {Eychaner}, {Fanson}, {Felizardo},
  {Garcia-Gathright}, {Gathright}, {Hardy}, {Henderson}, {Herstein}, {Hess},
  {Hovland}, {Hrynevych}, {Johansson}, {Johnson}, {Kelley}, {Kendrick},
  {Koresko}, {Kurpis}, {Le Mignant}, {Lewis}, {Ligon}, {Lupton}, {McBride},
  {Medeiros}, {Mennesson}, {Moore}, {Morrison}, {Nance}, {Neyman}, {Niessner},
  {Paine}, {Palmer}, {Panteleeva}, {Papin}, {Parvin}, {Reder}, {Rudeen},
  {Saloga}, {Sargent}, {Shao}, {Smith}, {Smythe}, {Stomski}, {Summers},
  {Swain}, {Swanson}, {Thompson}, {Tsubota}, {Tumminello}, {Tyau}, {van Belle},
  {Vasisht}, {Vause}, {Vescelus}, {Walker}, {Wallace}, {Wehmeier}, \&
  {Wetherell}}]{colavita:2013}
{Colavita}, M.~M., {Wizinowich}, P.~L., {Akeson}, R.~L., {et~al.} 2013,
  \href{http://dx.doi.org/10.1086/673475}{\color{magenta}\pasp},
  \href{https://ui.adsabs.harvard.edu/abs/2013PASP..125.1226C}{125, 1226}

\bibitem[{{Coud{\'e} du Foresto} {et~al.}(2003){Coud{\'e} du Foresto}, {Borde},
  {Merand}, {Baudouin}, {Remond}, {Perrin}, {Ridgway}, {ten Brummelaar}, \&
  {McAlister}}]{coude_du_foresto:2003}
{Coud{\'e} du Foresto}, V., {Borde}, P.~J., {Merand}, A., {et~al.} 2003, in
  Society of Photo-Optical Instrumentation Engineers (SPIE) Conference Series,
  Vol. 4838, Interferometry for Optical Astronomy II, ed. W.~A. {Traub},
  \href{https://ui.adsabs.harvard.edu/abs/2003SPIE.4838..280C}{280--285}

\bibitem[{{Coud{\'e} du Foresto} {et~al.}(1998){Coud{\'e} du Foresto},
  {Perrin}, {Ruilier}, {Mennesson}, {Traub}, \&
  {Lacasse}}]{coude_du_foresto:1998}
{Coud{\'e} du Foresto}, V., {Perrin}, G., {Ruilier}, C., {et~al.} 1998, in
  Society of Photo-Optical Instrumentation Engineers (SPIE) Conference Series,
  Vol. 3350, Astronomical Interferometry, ed. R.~D. {Reasenberg},
  \href{https://ui.adsabs.harvard.edu/abs/1998SPIE.3350..856C}{856--863}

\bibitem[{{Coud{\'e} du Foresto} {et~al.}(1997){Coud{\'e} du Foresto},
  {Ridgway}, \& {Mariotti}}]{coude_du_foresto:1997}
{Coud{\'e} du Foresto}, V., {Ridgway}, S., \& {Mariotti}, J.~M. 1997,
  \href{http://dx.doi.org/10.1051/aas:1997290}{\color{magenta}\aaps},
  \href{https://ui.adsabs.harvard.edu/abs/1997A&AS..121..379C}{121, 379}

\bibitem[{{Czechowski} \& {Mann}(2010)}]{czechowski:mann:2010}
{Czechowski}, A. \& {Mann}, I. 2010,
  \href{http://dx.doi.org/10.1088/0004-637X/714/1/89}{\color{magenta}\apj},
  \href{https://ui.adsabs.harvard.edu/abs/2010ApJ...714...89C}{714, 89}

\bibitem[{{de Kok} {et~al.}(2011){de Kok}, {Stam}, \& {Karalidi}}]{de_kok:2011}
{de Kok}, R.~J., {Stam}, D.~M., \& {Karalidi}, T. 2011,
  \href{http://dx.doi.org/10.1088/0004-637X/741/1/59}{\color{magenta}\apj},
  \href{https://ui.adsabs.harvard.edu/abs/2011ApJ...741...59D}{741, 59}

\bibitem[{{Defr{\`e}re} {et~al.}(2011){Defr{\`e}re}, {Absil}, {Augereau}, {di
  Folco}, {Berger}, {Coud{\'e} du Foresto}, {Kervella}, {Le Bouquin},
  {Lebreton}, {Millan-Gabet}, {Monnier}, {Olofsson}, \& {Traub}}]{defrere:2011}
{Defr{\`e}re}, D., {Absil}, O., {Augereau}, J.~C., {et~al.} 2011,
  \href{http://dx.doi.org/10.1051/0004-6361/201117017}{\color{magenta}\aap},
  \href{https://ui.adsabs.harvard.edu/abs/2011A&A...534A...5D}{534, A5}

\bibitem[{{Defr{\`e}re} {et~al.}(2018{\natexlab{a}}){Defr{\`e}re}, {Absil},
  {Berger}, {Boulet}, {Danchi}, {Ertel}, {Gallenne}, {H{\'e}nault}, {Hinz},
  {Huby}, {Ireland}, {Kraus}, {Labadie}, {Le Bouquin}, {Martin}, {Matter},
  {M{\'e}rand}, {Mennesson}, {Minardi}, {Monnier}, {Norris}, {Orban de Xivry},
  {Pedretti}, {Pott}, {Reggiani}, {Serabyn}, {Surdej}, {Tristram}, \&
  {Woillez}}]{defrere:2018b}
{Defr{\`e}re}, D., {Absil}, O., {Berger}, J.~P., {et~al.} 2018{\natexlab{a}},
  \href{http://dx.doi.org/10.1007/s10686-018-9593-2}{\color{magenta}Experimental
  Astronomy}, \href{https://ui.adsabs.harvard.edu/abs/2018ExA....46..475D}{46,
  475}

\bibitem[{{Defr{\`e}re} {et~al.}(2022){Defr{\`e}re}, {Bigioli}, {Dandumont},
  {Garreau}, {Laugier}, {Martinod}, {Absil}, {Berger}, {Bouzerand},
  {Courtney-Barrer}, {Emsenhuber}, {Ertel}, {Gagne}, {Glauser}, {Gross},
  {Ireland}, {Kenchington}, {Kluska}, {Kraus}, {Labadie}, {Laborde},
  {L{\'e}ger}, {Leisenring}, {Loicq}, {Martin}, {Morren}, {Matter}, {Mazzoli},
  {Missiaen}, {Muhammad}, {Ollivier}, {Raskin}, {Rousseau}, {Sanny},
  {Verlinden}, {Vandenbussche}, \& {Woillez}}]{defrere:2022}
{Defr{\`e}re}, D., {Bigioli}, A., {Dandumont}, C., {et~al.} 2022, in Society of
  Photo-Optical Instrumentation Engineers (SPIE) Conference Series, Vol. 12183,
  Optical and Infrared Interferometry and Imaging VIII, ed. A.~{M{\'e}rand},
  S.~{Sallum}, \& J.~{Sanchez-Bermudez},
  \href{https://ui.adsabs.harvard.edu/abs/2022SPIE12183E..0HD}{121830H}

\bibitem[{{Defr{\`e}re} {et~al.}(2018{\natexlab{b}}){Defr{\`e}re}, {Ireland},
  {Absil}, {Berger}, {Danchi}, {Ertel}, {Gallenne}, {H{\'e}nault}, {Hinz},
  {Huby}, {Kraus}, {Labadie}, {Le Bouquin}, {Martin}, {Matter}, {Mennesson},
  {M{\'e}rand}, {Minardi}, {Monnier}, {Norris}, {Orban de Xivry}, {Pedretti},
  {Pott}, {Reggiani}, {Serabyn}, {Surdej}, {Tristram}, \&
  {Woillez}}]{defrere:2018a}
{Defr{\`e}re}, D., {Ireland}, M., {Absil}, O., {et~al.} 2018{\natexlab{b}}, in
  Society of Photo-Optical Instrumentation Engineers (SPIE) Conference Series,
  Vol. 10701, Optical and Infrared Interferometry and Imaging VI, ed. M.~J.
  {Creech-Eakman}, P.~G. {Tuthill}, \& A.~{M{\'e}rand},
  \href{https://ui.adsabs.harvard.edu/abs/2018SPIE10701E..0UD}{107010U}

\bibitem[{{di Folco} {et~al.}(2007){di Folco}, {Absil}, {Augereau},
  {M{\'e}rand}, {Coud{\'e} du Foresto}, {Th{\'e}venin}, {Defr{\`e}re},
  {Kervella}, {ten Brummelaar}, {McAlister}, {Ridgway}, {Sturmann}, {Sturmann},
  \& {Turner}}]{di_folco:2007}
{di Folco}, E., {Absil}, O., {Augereau}, J.~C., {et~al.} 2007,
  \href{http://dx.doi.org/10.1051/0004-6361:20077625}{\color{magenta}\aap},
  \href{https://ui.adsabs.harvard.edu/abs/2007A&A...475..243D}{475, 243}

\bibitem[{{Di Folco} {et~al.}(2004){Di Folco}, {Th{\'e}venin}, {Kervella},
  {Domiciano de Souza}, {Coud{\'e} du Foresto}, {S{\'e}gransan}, \&
  {Morel}}]{di_folco:2004}
{Di Folco}, E., {Th{\'e}venin}, F., {Kervella}, P., {et~al.} 2004,
  \href{http://dx.doi.org/10.1051/0004-6361:20047189}{\color{magenta}\aap},
  \href{https://ui.adsabs.harvard.edu/abs/2004A&A...426..601D}{426, 601}

\bibitem[{{Draine}(2003)}]{draine:2003b}
{Draine}, B.~T. 2003,
  \href{http://dx.doi.org/10.1086/379123}{\color{magenta}\apj},
  \href{https://ui.adsabs.harvard.edu/abs/2003ApJ...598.1026D}{598, 1026}

\bibitem[{{Draine} \& {Malhotra}(1993)}]{draine:malhotra:1993}
{Draine}, B.~T. \& {Malhotra}, S. 1993,
  \href{http://dx.doi.org/10.1086/173109}{\color{magenta}\apj},
  \href{https://ui.adsabs.harvard.edu/abs/1993ApJ...414..632D}{414, 632}

\bibitem[{{Ertel} {et~al.}(2014){Ertel}, {Absil}, {Defr{\`e}re}, {Le Bouquin},
  {Augereau}, {Marion}, {Blind}, {Bonsor}, {Bryden}, {Lebreton}, \&
  {Milli}}]{ertel:2014}
{Ertel}, S., {Absil}, O., {Defr{\`e}re}, D., {et~al.} 2014,
  \href{http://dx.doi.org/10.1051/0004-6361/201424438}{\color{magenta}\aap},
  \href{https://ui.adsabs.harvard.edu/abs/2014A&A...570A.128E}{570, A128}

\bibitem[{{Ertel} {et~al.}(2020{\natexlab{a}}){Ertel}, {Defr{\`e}re}, {Hinz},
  {Mennesson}, {Kennedy}, {Danchi}, {Gelino}, {Hill}, {Hoffmann}, {Mazoyer},
  {Rieke}, {Shannon}, {Stapelfeldt}, {Spalding}, {Stone}, {Vaz}, {Weinberger},
  {Willems}, {Absil}, {Arbo}, {Bailey}, {Beichman}, {Bryden}, {Downey},
  {Durney}, {Esposito}, {Gaspar}, {Grenz}, {Haniff}, {Leisenring}, {Marion},
  {McMahon}, {Millan-Gabet}, {Montoya}, {Morzinski}, {Perera}, {Pinna}, {Pott},
  {Power}, {Puglisi}, {Roberge}, {Serabyn}, {Skemer}, {Su}, {Vaitheeswaran}, \&
  {Wyatt}}]{ertel:2020a}
{Ertel}, S., {Defr{\`e}re}, D., {Hinz}, P., {et~al.} 2020{\natexlab{a}},
  \href{http://dx.doi.org/10.3847/1538-3881/ab7817}{\color{magenta}\aj},
  \href{https://ui.adsabs.harvard.edu/abs/2020AJ....159..177E}{159, 177}

\bibitem[{{Ertel} {et~al.}(2018{\natexlab{a}}){Ertel}, {Defr{\`e}re}, {Hinz},
  {Mennesson}, {Kennedy}, {Danchi}, {Gelino}, {Hill}, {Hoffmann}, {Rieke},
  {Shannon}, {Spalding}, {Stone}, {Vaz}, {Weinberger}, {Willems}, {Absil},
  {Arbo}, {Bailey}, {Beichman}, {Bryden}, {Downey}, {Durney}, {Esposito},
  {Gaspar}, {Grenz}, {Haniff}, {Leisenring}, {Marion}, {McMahon},
  {Millan-Gabet}, {Montoya}, {Morzinski}, {Pinna}, {Power}, {Puglisi},
  {Roberge}, {Serabyn}, {Skemer}, {Stapelfeldt}, {Su}, {Vaitheeswaran}, \&
  {Wyatt}}]{ertel:2018a}
{Ertel}, S., {Defr{\`e}re}, D., {Hinz}, P., {et~al.} 2018{\natexlab{a}},
  \href{http://dx.doi.org/10.3847/1538-3881/aab717}{\color{magenta}\aj},
  \href{https://ui.adsabs.harvard.edu/abs/2018AJ....155..194E}{155, 194}

\bibitem[{{Ertel} {et~al.}(2020{\natexlab{b}}){Ertel}, {Hinz}, {Stone}, {Vaz},
  {Montoya}, {West}, {Durney}, {Grenz}, {Spalding}, {Leisenring}, {Wagner},
  {Anugu}, {Power}, {Maier}, {Defr{\`e}re}, {Hoffmann}, {Perera}, {Brown},
  {Skemer}, {Mennesson}, {Kennedy}, {Downey}, {Hill}, {Pinna}, {Puglisi}, \&
  {Rossi}}]{ertel:2020b}
{Ertel}, S., {Hinz}, P.~M., {Stone}, J.~M., {et~al.} 2020{\natexlab{b}}, in
  Society of Photo-Optical Instrumentation Engineers (SPIE) Conference Series,
  Vol. 11446, Society of Photo-Optical Instrumentation Engineers (SPIE)
  Conference Series,
  \href{https://ui.adsabs.harvard.edu/abs/2020SPIE11446E..07E}{1144607}

\bibitem[{{Ertel} {et~al.}(2018{\natexlab{b}}){Ertel}, {Kennedy},
  {Defr{\`e}re}, {Hinz}, {Shannon}, {Mennesson}, {Danchi}, {Gelino}, {Hill},
  {Hoffman}, {Rieke}, {Spalding}, {Stone}, {Vaz}, {Weinberger}, {Willems},
  {Absil}, {Arbo}, {Bailey}, {Beichman}, {Bryden}, {Downey}, {Durney},
  {Esposito}, {Gaspar}, {Grenz}, {Haniff}, {Leisenring}, {Marion}, {McMahon},
  {Millan-Gabet}, {Montoya}, {Morzinski}, {Pinna}, {Power}, {Puglisi},
  {Roberge}, {Serabyn}, {Skemer}, {Stapelfeldt}, {Su}, {Vaitheeswaran}, \&
  {Wyatt}}]{ertel:2018b}
{Ertel}, S., {Kennedy}, G.~M., {Defr{\`e}re}, D., {et~al.} 2018{\natexlab{b}},
  in Society of Photo-Optical Instrumentation Engineers (SPIE) Conference
  Series, Vol. 10698, Space Telescopes and Instrumentation 2018: Optical,
  Infrared, and Millimeter Wave, ed. M.~{Lystrup}, H.~A. {MacEwen}, G.~G.
  {Fazio}, N.~{Batalha}, N.~{Siegler}, \& E.~C. {Tong},
  \href{https://ui.adsabs.harvard.edu/abs/2018SPIE10698E..1VE}{106981V}

\bibitem[{{Faramaz} {et~al.}(2017){Faramaz}, {Ertel}, {Booth}, {Cuadra}, \&
  {Simmonds}}]{faramaz:2017}
{Faramaz}, V., {Ertel}, S., {Booth}, M., {Cuadra}, J., \& {Simmonds}, C. 2017,
  \href{http://dx.doi.org/10.1093/mnras/stw2846}{\color{magenta}\mnras},
  \href{https://ui.adsabs.harvard.edu/abs/2017MNRAS.465.2352F}{465, 2352}

\bibitem[{{Fluri} \& {Berdyugina}(2010)}]{fluri:berdyugina:2010}
{Fluri}, D.~M. \& {Berdyugina}, S.~V. 2010,
  \href{http://dx.doi.org/10.1051/0004-6361/200809970}{\color{magenta}\aap},
  \href{https://ui.adsabs.harvard.edu/abs/2010A&A...512A..59F}{512, A59}

\bibitem[{{Gaia Collaboration} {et~al.}(2021){Gaia Collaboration}, {Brown},
  {Vallenari}, {Prusti}, {de Bruijne}, {Babusiaux}, {Biermann}, {Creevey},
  {Evans}, {Eyer}, {Hutton}, {Jansen}, {Jordi}, {Klioner}, {Lammers},
  {Lindegren}, {Luri}, {Mignard}, {Panem}, {Pourbaix}, {Randich}, {Sartoretti},
  {Soubiran}, {Walton}, {Arenou}, {Bailer-Jones}, {Bastian}, {Cropper},
  {Drimmel}, {Katz}, {Lattanzi}, {van Leeuwen}, {Bakker}, {Cacciari},
  {Casta{\~n}eda}, {De Angeli}, {Ducourant}, {Fabricius}, {Fouesneau},
  {Fr{\'e}mat}, {Guerra}, {Guerrier}, {Guiraud}, {Jean-Antoine Piccolo},
  {Masana}, {Messineo}, {Mowlavi}, {Nicolas}, {Nienartowicz}, {Pailler},
  {Panuzzo}, {Riclet}, {Roux}, {Seabroke}, {Sordo}, {Tanga}, {Th{\'e}venin},
  {Gracia-Abril}, {Portell}, {Teyssier}, {Altmann}, {Andrae}, {Bellas-Velidis},
  {Benson}, {Berthier}, {Blomme}, {Brugaletta}, {Burgess}, {Busso}, {Carry},
  {Cellino}, {Cheek}, {Clementini}, {Damerdji}, {Davidson}, {Delchambre},
  {Dell'Oro}, {Fern{\'a}ndez-Hern{\'a}ndez}, {Galluccio}, {Garc{\'\i}a-Lario},
  {Garcia-Reinaldos}, {Gonz{\'a}lez-N{\'u}{\~n}ez}, {Gosset}, {Haigron},
  {Halbwachs}, {Hambly}, {Harrison}, {Hatzidimitriou}, {Heiter},
  {Hern{\'a}ndez}, {Hestroffer}, {Hodgkin}, {Holl}, {Jan{\ss}en}, {Jevardat de
  Fombelle}, {Jordan}, {Krone-Martins}, {Lanzafame}, {L{\"o}ffler}, {Lorca},
  {Manteiga}, {Marchal}, {Marrese}, {Moitinho}, {Mora}, {Muinonen}, {Osborne},
  {Pancino}, {Pauwels}, {Petit}, {Recio-Blanco}, {Richards}, {Riello},
  {Rimoldini}, {Robin}, {Roegiers}, {Rybizki}, {Sarro}, {Siopis}, {Smith},
  {Sozzetti}, {Ulla}, {Utrilla}, {van Leeuwen}, {van Reeven}, {Abbas}, {Abreu
  Aramburu}, {Accart}, {Aerts}, {Aguado}, {Ajaj}, {Altavilla}, {{\'A}lvarez},
  {{\'A}lvarez Cid-Fuentes}, {Alves}, {Anderson}, {Anglada Varela}, {Antoja},
  {Audard}, {Baines}, {Baker}, {Balaguer-N{\'u}{\~n}ez}, {Balbinot}, {Balog},
  {Barache}, {Barbato}, {Barros}, {Barstow}, {Bartolom{\'e}}, {Bassilana},
  {Bauchet}, {Baudesson-Stella}, {Becciani}, {Bellazzini}, {Bernet}, {Bertone},
  {Bianchi}, {Blanco-Cuaresma}, {Boch}, {Bombrun}, {Bossini}, {Bouquillon},
  {Bragaglia}, {Bramante}, {Breedt}, {Bressan}, {Brouillet}, {Bucciarelli},
  {Burlacu}, {Busonero}, {Butkevich}, {Buzzi}, {Caffau}, {Cancelliere},
  {C{\'a}novas}, {Cantat-Gaudin}, {Carballo}, {Carlucci}, {Carnerero},
  {Carrasco}, {Casamiquela}, {Castellani}, {Castro-Ginard}, {Castro Sampol},
  {Chaoul}, {Charlot}, {Chemin}, {Chiavassa}, {Cioni}, {Comoretto}, {Cooper},
  {Cornez}, {Cowell}, {Crifo}, {Crosta}, {Crowley}, {Dafonte}, {Dapergolas},
  {David}, {David}, {de Laverny}, {De Luise}, {De March}, {De Ridder}, {de
  Souza}, {de Teodoro}, {de Torres}, {del Peloso}, {del Pozo}, {Delbo},
  {Delgado}, {Delgado}, {Delisle}, {Di Matteo}, {Diakite}, {Diener},
  {Distefano}, {Dolding}, {Eappachen}, {Edvardsson}, {Enke}, {Esquej}, {Fabre},
  {Fabrizio}, {Faigler}, {Fedorets}, {Fernique}, {Fienga}, {Figueras},
  {Fouron}, {Fragkoudi}, {Fraile}, {Franke}, {Gai}, {Garabato},
  {Garcia-Gutierrez}, {Garc{\'\i}a-Torres}, {Garofalo}, {Gavras}, {Gerlach},
  {Geyer}, {Giacobbe}, {Gilmore}, {Girona}, {Giuffrida}, {Gomel}, {Gomez},
  {Gonzalez-Santamaria}, {Gonz{\'a}lez-Vidal}, {Granvik},
  {Guti{\'e}rrez-S{\'a}nchez}, {Guy}, {Hauser}, {Haywood}, {Helmi}, {Hidalgo},
  {Hilger}, {H{\l}adczuk}, {Hobbs}, {Holland}, {Huckle}, {Jasniewicz},
  {Jonker}, {Juaristi Campillo}, {Julbe}, {Karbevska}, {Kervella}, {Khanna},
  {Kochoska}, {Kontizas}, {Kordopatis}, {Korn}, {Kostrzewa-Rutkowska},
  {Kruszy{\'n}ska}, {Lambert}, {Lanza}, {Lasne}, {Le Campion}, {Le Fustec},
  {Lebreton}, {Lebzelter}, {Leccia}, {Leclerc}, {Lecoeur-Taibi}, {Liao},
  {Licata}, {Lindstr{\o}m}, {Lister}, {Livanou}, {Lobel}, {Madrero Pardo},
  {Managau}, {Mann}, {Marchant}, {Marconi}, {Marcos Santos}, {Marinoni},
  {Marocco}, {Marshall}, {Martin Polo}, {Mart{\'\i}n-Fleitas}, {Masip},
  {Massari}, {Mastrobuono-Battisti}, {Mazeh}, {McMillan}, {Messina},
  {Michalik}, {Millar}, {Mints}, {Molina}, {Molinaro}, {Moln{\'a}r},
  {Montegriffo}, {Mor}, {Morbidelli}, {Morel}, {Morris}, {Mulone}, {Munoz},
  {Muraveva}, {Murphy}, {Musella}, {Noval}, {Ord{\'e}novic}, {Orr{\`u}},
  {Osinde}, {Pagani}, {Pagano}, {Palaversa}, {Palicio}, {Panahi}, {Pawlak},
  {Pe{\~n}alosa Esteller}, {Penttil{\"a}}, {Piersimoni}, {Pineau}, {Plachy},
  {Plum}, {Poggio}, {Poretti}, {Poujoulet}, {Pr{\v{s}}a}, {Pulone}, {Racero},
  {Ragaini}, {Rainer}, {Raiteri}, {Rambaux}, {Ramos}, {Ramos-Lerate}, {Re
  Fiorentin}, {Regibo}, {Reyl{\'e}}, {Ripepi}, {Riva}, {Rixon}, {Robichon},
  {Robin}, {Roelens}, {Rohrbasser}, {Romero-G{\'o}mez}, {Rowell}, {Royer},
  {Rybicki}, {Sadowski}, {Sagrist{\`a} Sell{\'e}s}, {Sahlmann}, {Salgado},
  {Salguero}, {Samaras}, {Sanchez Gimenez}, {Sanna}, {Santove{\~n}a},
  {Sarasso}, {Schultheis}, {Sciacca}, {Segol}, {Segovia}, {S{\'e}gransan},
  {Semeux}, {Shahaf}, {Siddiqui}, {Siebert}, {Siltala}, {Slezak}, {Smart},
  {Solano}, {Solitro}, {Souami}, {Souchay}, {Spagna}, {Spoto}, {Steele},
  {Steidelm{\"u}ller}, {Stephenson}, {S{\"u}veges}, {Szabados}, {Szegedi-Elek},
  {Taris}, {Tauran}, {Taylor}, {Teixeira}, {Thuillot}, {Tonello}, {Torra},
  {Torra}, {Turon}, {Unger}, {Vaillant}, {van Dillen}, {Vanel}, {Vecchiato},
  {Viala}, {Vicente}, {Voutsinas}, {Weiler}, {Wevers}, {Wyrzykowski}, {Yoldas},
  {Yvard}, {Zhao}, {Zorec}, {Zucker}, {Zurbach}, \&
  {Zwitter}}]{gaia_collaboration:2021}
{Gaia Collaboration}, {Brown}, A.~G.~A., {Vallenari}, A., {et~al.} 2021,
  \href{http://dx.doi.org/10.1051/0004-6361/202039657}{\color{magenta}\aap},
  \href{https://ui.adsabs.harvard.edu/abs/2021A&A...649A...1G}{649, A1}

\bibitem[{{G{\'a}sp{\'a}r} {et~al.}(2013){G{\'a}sp{\'a}r}, {Rieke}, \&
  {Balog}}]{gaspar:2013}
{G{\'a}sp{\'a}r}, A., {Rieke}, G.~H., \& {Balog}, Z. 2013,
  \href{http://dx.doi.org/10.1088/0004-637X/768/1/25}{\color{magenta}\apj},
  \href{https://ui.adsabs.harvard.edu/abs/2013ApJ...768...25G}{768, 25}

\bibitem[{{GRAVITY Collaboration} {et~al.}(2017){GRAVITY Collaboration},
  {Abuter}, {Accardo}, {Amorim}, {Anugu}, {{\'A}vila}, {Azouaoui}, {Benisty},
  {Berger}, {Blind}, {Bonnet}, {Bourget}, {Brandner}, {Brast}, {Buron},
  {Burtscher}, {Cassaing}, {Chapron}, {Choquet}, {Cl{\'e}net}, {Collin},
  {Coud{\'e} Du Foresto}, {de Wit}, {de Zeeuw}, {Deen},
  {Delplancke-Str{\"o}bele}, {Dembet}, {Derie}, {Dexter}, {Duvert}, {Ebert},
  {Eckart}, {Eisenhauer}, {Esselborn}, {F{\'e}dou}, {Finger}, {Garcia}, {Garcia
  Dabo}, {Garcia Lopez}, {Gendron}, {Genzel}, {Gillessen}, {Gonte}, {Gordo},
  {Grould}, {Gr{\"o}zinger}, {Guieu}, {Haguenauer}, {Hans}, {Haubois}, {Haug},
  {Haussmann}, {Henning}, {Hippler}, {Horrobin}, {Huber}, {Hubert}, {Hubin},
  {Hummel}, {Jakob}, {Janssen}, {Jochum}, {Jocou}, {Kaufer}, {Kellner},
  {Kendrew}, {Kern}, {Kervella}, {Kiekebusch}, {Klein}, {Kok}, {Kolb}, {Kulas},
  {Lacour}, {Lapeyr{\`e}re}, {Lazareff}, {Le Bouquin}, {L{\`e}na}, {Lenzen},
  {L{\'e}v{\^e}que}, {Lippa}, {Magnard}, {Mehrgan}, {Mellein}, {M{\'e}rand},
  {Moreno-Ventas}, {Moulin}, {M{\"u}ller}, {M{\"u}ller}, {Neumann}, {Oberti},
  {Ott}, {Pallanca}, {Panduro}, {Pasquini}, {Paumard}, {Percheron}, {Perraut},
  {Perrin}, {Pfl{\"u}ger}, {Pfuhl}, {Phan Duc}, {Plewa}, {Popovic}, {Rabien},
  {Ram{\'\i}rez}, {Ramos}, {Rau}, {Riquelme}, {Rohloff}, {Rousset},
  {Sanchez-Bermudez}, {Scheithauer}, {Sch{\"o}ller}, {Schuhler}, {Spyromilio},
  {Straubmeier}, {Sturm}, {Suarez}, {Tristram}, {Ventura}, {Vincent},
  {Waisberg}, {Wank}, {Weber}, {Wieprecht}, {Wiest}, {Wiezorrek}, {Wittkowski},
  {Woillez}, {Wolff}, {Yazici}, {Ziegler}, \&
  {Zins}}]{gravity_collaboration:2017}
{GRAVITY Collaboration}, {Abuter}, R., {Accardo}, M., {et~al.} 2017,
  \href{http://dx.doi.org/10.1051/0004-6361/201730838}{\color{magenta}\aap},
  \href{https://ui.adsabs.harvard.edu/abs/2017A&A...602A..94G}{602, A94}

\bibitem[{{Gray} {et~al.}(2006){Gray}, {Corbally}, {Garrison}, {McFadden},
  {Bubar}, {McGahee}, {O'Donoghue}, \& {Knox}}]{gray:2006}
{Gray}, R.~O., {Corbally}, C.~J., {Garrison}, R.~F., {et~al.} 2006,
  \href{http://dx.doi.org/10.1086/504637}{\color{magenta}\aj},
  \href{https://ui.adsabs.harvard.edu/abs/2006AJ....132..161G}{132, 161}

\bibitem[{{Gray} {et~al.}(2003){Gray}, {Corbally}, {Garrison}, {McFadden}, \&
  {Robinson}}]{gray:2003}
{Gray}, R.~O., {Corbally}, C.~J., {Garrison}, R.~F., {McFadden}, M.~T., \&
  {Robinson}, P.~E. 2003,
  \href{http://dx.doi.org/10.1086/378365}{\color{magenta}\aj},
  \href{https://ui.adsabs.harvard.edu/abs/2003AJ....126.2048G}{126, 2048}

\bibitem[{{Gray} \& {Garrison}(1989)}]{gray:garrison:1989}
{Gray}, R.~O. \& {Garrison}, R.~F. 1989,
  \href{http://dx.doi.org/10.1086/191349}{\color{magenta}\apjs},
  \href{https://ui.adsabs.harvard.edu/abs/1989ApJS...70..623G}{70, 623}

\bibitem[{{Greaves} {et~al.}(2004){Greaves}, {Wyatt}, {Holland}, \&
  {Dent}}]{greaves:2004b}
{Greaves}, J.~S., {Wyatt}, M.~C., {Holland}, W.~S., \& {Dent}, W.~R.~F. 2004,
  \href{http://dx.doi.org/10.1111/j.1365-2966.2004.07957.x}{\color{magenta}\mnras},
  \href{https://ui.adsabs.harvard.edu/abs/2004MNRAS.351L..54G}{351, L54}

\bibitem[{{Gulliver} {et~al.}(1994){Gulliver}, {Hill}, \&
  {Adelman}}]{gulliver:1994}
{Gulliver}, A.~F., {Hill}, G., \& {Adelman}, S.~J. 1994,
  \href{http://dx.doi.org/10.1086/187418}{\color{magenta}\apjl},
  \href{https://ui.adsabs.harvard.edu/abs/1994ApJ...429L..81G}{429, L81}

\bibitem[{{Habing} {et~al.}(2001){Habing}, {Dominik}, {Jourdain de Muizon},
  {Laureijs}, {Kessler}, {Leech}, {Metcalfe}, {Salama}, {Siebenmorgen},
  {Trams}, \& {Bouchet}}]{habing:2001}
{Habing}, H.~J., {Dominik}, C., {Jourdain de Muizon}, M., {et~al.} 2001,
  \href{http://dx.doi.org/10.1051/0004-6361:20000075}{\color{magenta}\aap},
  \href{https://ui.adsabs.harvard.edu/abs/2001A&A...365..545H}{365, 545}

\bibitem[{{Hanner}(1984)}]{hanner:1984}
{Hanner}, M.~S. 1984,
  \href{http://dx.doi.org/10.1016/0273-1177(84)90025-5}{\color{magenta}Advances
  in Space Research},
  \href{https://ui.adsabs.harvard.edu/abs/1984AdSpR...4i.189H}{4, 189}

\bibitem[{{Harker} {et~al.}(2002){Harker}, {Wooden}, {Woodward}, \&
  {Lisse}}]{harker:2002}
{Harker}, D.~E., {Wooden}, D.~H., {Woodward}, C.~E., \& {Lisse}, C.~M. 2002,
  \href{http://dx.doi.org/10.1086/343091}{\color{magenta}\apj},
  \href{https://ui.adsabs.harvard.edu/abs/2002ApJ...580..579H}{580, 579}

\bibitem[{{Harris} {et~al.}(2020){Harris}, {Millman}, {van der Walt},
  {Gommers}, {Virtanen}, {Cournapeau}, {Wieser}, {Taylor}, {Berg}, {Smith},
  {Kern}, {Picus}, {Hoyer}, {van Kerkwijk}, {Brett}, {Haldane}, {del R{\'\i}o},
  {Wiebe}, {Peterson}, {G{\'e}rard-Marchant}, {Sheppard}, {Reddy}, {Weckesser},
  {Abbasi}, {Gohlke}, \& {Oliphant}}]{harris:2020}
{Harris}, C.~R., {Millman}, K.~J., {van der Walt}, S.~J., {et~al.} 2020,
  \href{http://dx.doi.org/10.1038/s41586-020-2649-2}{\color{magenta}\nat},
  \href{https://ui.adsabs.harvard.edu/abs/2020Natur.585..357H}{585, 357}

\bibitem[{{Hindsley} \& {Harrington}(1994)}]{hindsley:harrington:1994}
{Hindsley}, R.~B. \& {Harrington}, R.~S. 1994,
  \href{http://dx.doi.org/10.1086/116852}{\color{magenta}\aj},
  \href{https://ui.adsabs.harvard.edu/abs/1994AJ....107..280H}{107, 280}

\bibitem[{{Hinz} {et~al.}(2016){Hinz}, {Defr{\`e}re}, {Skemer}, {Bailey},
  {Stone}, {Spalding}, {Vaz}, {Pinna}, {Puglisi}, {Esposito}, {Montoya},
  {Downey}, {Leisenring}, {Durney}, {Hoffmann}, {Hill}, {Millan-Gabet},
  {Mennesson}, {Danchi}, {Morzinski}, {Grenz}, {Skrutskie}, \&
  {Ertel}}]{hinz:2016}
{Hinz}, P.~M., {Defr{\`e}re}, D., {Skemer}, A., {et~al.} 2016, in Society of
  Photo-Optical Instrumentation Engineers (SPIE) Conference Series, Vol. 9907,
  Optical and Infrared Interferometry and Imaging V, ed. F.~{Malbet}, M.~J.
  {Creech-Eakman}, \& P.~G. {Tuthill},
  \href{https://ui.adsabs.harvard.edu/abs/2016SPIE.9907E..04H}{990704}

\bibitem[{{Holland} {et~al.}(2017){Holland}, {Matthews}, {Kennedy}, {Greaves},
  {Wyatt}, {Booth}, {Bastien}, {Bryden}, {Butner}, {Chen}, {Chrysostomou},
  {Davies}, {Dent}, {Di Francesco}, {Duch{\^e}ne}, {Gibb}, {Friberg}, {Ivison},
  {Jenness}, {Kavelaars}, {Lawler}, {Lestrade}, {Marshall}, {Moro-Martin},
  {Pani{\'c}}, {Phillips}, {Serjeant}, {Schieven}, {Sibthorpe}, {Vican},
  {Ward-Thompson}, {van der Werf}, {White}, {Wilner}, \&
  {Zuckerman}}]{holland:2017}
{Holland}, W.~S., {Matthews}, B.~C., {Kennedy}, G.~M., {et~al.} 2017,
  \href{http://dx.doi.org/10.1093/mnras/stx1378}{\color{magenta}\mnras},
  \href{https://ui.adsabs.harvard.edu/abs/2017MNRAS.470.3606H}{470, 3606}

\bibitem[{{Hunter}(2007)}]{hunter:2007}
{Hunter}, J.~D. 2007,
  \href{http://dx.doi.org/10.1109/MCSE.2007.55}{\color{magenta}Computing in
  Science and Engineering},
  \href{https://ui.adsabs.harvard.edu/abs/2007CSE.....9...90H}{9, 90}

\bibitem[{{J{\"a}ger} {et~al.}(1998){J{\"a}ger}, {Mutschke}, \&
  {Henning}}]{jaeger:1998}
{J{\"a}ger}, C., {Mutschke}, H., \& {Henning}, T. 1998, \aap,
  \href{https://ui.adsabs.harvard.edu/abs/1998A&A...332..291J}{332, 291}

\bibitem[{{Karalidi} \& {Stam}(2012)}]{karalidi:stam:2012}
{Karalidi}, T. \& {Stam}, D.~M. 2012,
  \href{http://dx.doi.org/10.1051/0004-6361/201219297}{\color{magenta}\aap},
  \href{https://ui.adsabs.harvard.edu/abs/2012A&A...546A..56K}{546, A56}

\bibitem[{{Karalidi} {et~al.}(2013){Karalidi}, {Stam}, \&
  {Guirado}}]{karalidi:2013}
{Karalidi}, T., {Stam}, D.~M., \& {Guirado}, D. 2013,
  \href{http://dx.doi.org/10.1051/0004-6361/201321492}{\color{magenta}\aap},
  \href{https://ui.adsabs.harvard.edu/abs/2013A&A...555A.127K}{555, A127}

\bibitem[{{Karalidi} {et~al.}(2011){Karalidi}, {Stam}, \&
  {Hovenier}}]{karalidi:2011}
{Karalidi}, T., {Stam}, D.~M., \& {Hovenier}, J.~W. 2011,
  \href{http://dx.doi.org/10.1051/0004-6361/201116449}{\color{magenta}\aap},
  \href{https://ui.adsabs.harvard.edu/abs/2011A&A...530A..69K}{530, A69}

\bibitem[{{Karalidi} {et~al.}(2012){Karalidi}, {Stam}, \&
  {Hovenier}}]{karalidi:2012}
{Karalidi}, T., {Stam}, D.~M., \& {Hovenier}, J.~W. 2012,
  \href{http://dx.doi.org/10.1051/0004-6361/201220245}{\color{magenta}\aap},
  \href{https://ui.adsabs.harvard.edu/abs/2012A&A...548A..90K}{548, A90}

\bibitem[{{Kattawar} \& {Adams}(1971)}]{kattawar:1971}
{Kattawar}, G.~W. \& {Adams}, C.~N. 1971,
  \href{http://dx.doi.org/10.1086/151017}{\color{magenta}\apj},
  \href{https://ui.adsabs.harvard.edu/abs/1971ApJ...167..183K}{167, 183}

\bibitem[{{Keenan} \& {McNeil}(1989)}]{keenan:1989}
{Keenan}, P.~C. \& {McNeil}, R.~C. 1989,
  \href{http://dx.doi.org/10.1086/191373}{\color{magenta}\apjs},
  \href{https://ui.adsabs.harvard.edu/abs/1989ApJS...71..245K}{71, 245}

\bibitem[{{Kervella} {et~al.}(2000){Kervella}, {Coud{\'e} du Foresto},
  {Glindemann}, \& {Hofmann}}]{kervella:2000}
{Kervella}, P., {Coud{\'e} du Foresto}, V., {Glindemann}, A., \& {Hofmann}, R.
  2000, in Society of Photo-Optical Instrumentation Engineers (SPIE) Conference
  Series, Vol. 4006, Interferometry in Optical Astronomy, ed. P.~{L{\'e}na} \&
  A.~{Quirrenbach},
  \href{https://ui.adsabs.harvard.edu/abs/2000SPIE.4006...31K}{31--42}

\bibitem[{{Kervella} {et~al.}(2003){Kervella}, {Gitton}, {Segransan}, {di
  Folco}, {Kern}, {Kiekebusch}, {Duc}, {Longinotti}, {Coud{\'e} du Foresto},
  {Ballester}, {Sabet}, {Cotton}, {Schoeller}, \& {Wilhelm}}]{kervella:2003}
{Kervella}, P., {Gitton}, P.~B., {Segransan}, D., {et~al.} 2003, in Society of
  Photo-Optical Instrumentation Engineers (SPIE) Conference Series, Vol. 4838,
  Interferometry for Optical Astronomy II, ed. W.~A. {Traub},
  \href{https://ui.adsabs.harvard.edu/abs/2003SPIE.4838..858K}{858--869}

\bibitem[{{Kim} {et~al.}(2018){Kim}, {Wolf}, {L{\"o}hne}, {Kirchschlager}, \&
  {Krivov}}]{kim:2018}
{Kim}, M., {Wolf}, S., {L{\"o}hne}, T., {Kirchschlager}, F., \& {Krivov}, A.~V.
  2018,
  \href{http://dx.doi.org/10.1051/0004-6361/201833061}{\color{magenta}\aap},
  \href{https://ui.adsabs.harvard.edu/abs/2018A&A...618A..38K}{618, A38}

\bibitem[{{Kimura} {et~al.}(2020){Kimura}, {Kunitomo}, {Suzuki}, {Robrade},
  {Thebault}, \& {Mitsuishi}}]{kimura:2020}
{Kimura}, H., {Kunitomo}, M., {Suzuki}, T.~K., {et~al.} 2020,
  \href{http://dx.doi.org/10.1016/j.pss.2018.07.010}{\color{magenta}\planss},
  \href{https://ui.adsabs.harvard.edu/abs/2020P&SS..18304581K}{183, 104581}

\bibitem[{{Kirchschlager} {et~al.}(2019){Kirchschlager}, {Bertrang}, \&
  {Flock}}]{kirchschlager:2019}
{Kirchschlager}, F., {Bertrang}, G. H.~M., \& {Flock}, M. 2019,
  \href{http://dx.doi.org/10.1093/mnras/stz1763}{\color{magenta}\mnras},
  \href{https://ui.adsabs.harvard.edu/abs/2019MNRAS.488.1211K}{488, 1211}

\bibitem[{{Kirchschlager} {et~al.}(2020){Kirchschlager}, {Ertel}, {Wolf},
  {Matter}, \& {Krivov}}]{kirchschlager:2020}
{Kirchschlager}, F., {Ertel}, S., {Wolf}, S., {Matter}, A., \& {Krivov}, A.~V.
  2020, \href{http://dx.doi.org/10.1093/mnrasl/slaa156}{\color{magenta}\mnras},
  \href{https://ui.adsabs.harvard.edu/abs/2020MNRAS.499L..47K}{499, L47}

\bibitem[{{Kirchschlager} \& {Wolf}(2013)}]{kirchschlager:wolf:2013}
{Kirchschlager}, F. \& {Wolf}, S. 2013,
  \href{http://dx.doi.org/10.1051/0004-6361/201220486}{\color{magenta}\aap},
  \href{https://ui.adsabs.harvard.edu/abs/2013A&A...552A..54K}{552, A54}

\bibitem[{{Kirchschlager} {et~al.}(2018){Kirchschlager}, {Wolf},
  {Brunngr{\"a}ber}, {Matter}, {Krivov}, \& {Labdon}}]{kirchschlager:2018}
{Kirchschlager}, F., {Wolf}, S., {Brunngr{\"a}ber}, R., {et~al.} 2018,
  \href{http://dx.doi.org/10.1093/mnras/stx2515}{\color{magenta}\mnras},
  \href{https://ui.adsabs.harvard.edu/abs/2018MNRAS.473.2633K}{473, 2633}

\bibitem[{{Kirchschlager} {et~al.}(2017){Kirchschlager}, {Wolf}, {Krivov},
  {Mutschke}, \& {Brunngr{\"a}ber}}]{kirchschlager:2017}
{Kirchschlager}, F., {Wolf}, S., {Krivov}, A.~V., {Mutschke}, H., \&
  {Brunngr{\"a}ber}, R. 2017,
  \href{http://dx.doi.org/10.1093/mnras/stx202}{\color{magenta}\mnras},
  \href{https://ui.adsabs.harvard.edu/abs/2017MNRAS.467.1614K}{467, 1614}

\bibitem[{Kluyver {et~al.}(2016)Kluyver, Ragan-Kelley, P{\'e}rez, Granger,
  Bussonnier, Frederic, Kelley, Hamrick, Grout, Corlay, Ivanov, Avila, Abdalla,
  Willing, \& development team}]{jupyter}
Kluyver, T., Ragan-Kelley, B., P{\'e}rez, F., {et~al.} 2016, Jupyter Notebooks
  - a publishing format for reproducible computational workflows, ed.
  F.~{Loizides} \& B.~{Schmidt} (Netherlands: IOS Press), 87--90

\bibitem[{{Kobayashi} {et~al.}(2011){Kobayashi}, {Kimura}, {Watanabe},
  {Yamamoto}, \& {M{\"u}ller}}]{kobayashi:2011}
{Kobayashi}, H., {Kimura}, H., {Watanabe}, S.~i., {Yamamoto}, T., \&
  {M{\"u}ller}, S. 2011,
  \href{http://dx.doi.org/10.5047/eps.2011.03.012}{\color{magenta}Earth,
  Planets and Space},
  \href{https://ui.adsabs.harvard.edu/abs/2011EP&S...63.1067K}{63, 1067}

\bibitem[{{Kobayashi} {et~al.}(2013){Kobayashi}, {Kimura}, \&
  {Yamamoto}}]{kobayashi:2013}
{Kobayashi}, H., {Kimura}, H., \& {Yamamoto}, S. 2013,
  \href{http://dx.doi.org/10.1051/0004-6361/201220464}{\color{magenta}\aap},
  \href{https://ui.adsabs.harvard.edu/abs/2013A&A...550A..72K}{550, A72}

\bibitem[{{Kobayashi} {et~al.}(2010){Kobayashi}, {Kimura}, {Yamamoto},
  {Watanabe}, \& {Yamamoto}}]{kobayashi:2010}
{Kobayashi}, H., {Kimura}, H., {Yamamoto}, S., {Watanabe}, S.~I., \&
  {Yamamoto}, T. 2010,
  \href{http://dx.doi.org/10.5047/eps.2009.03.001}{\color{magenta}Earth,
  Planets and Space},
  \href{https://ui.adsabs.harvard.edu/abs/2010EP&S...62...57K}{62, 57}

\bibitem[{{Kobayashi} {et~al.}(2008){Kobayashi}, {Watanabe}, {Kimura}, \&
  {Yamamoto}}]{kobayashi:2008}
{Kobayashi}, H., {Watanabe}, S.-i., {Kimura}, H., \& {Yamamoto}, T. 2008,
  \href{http://dx.doi.org/10.1016/j.icarus.2008.02.005}{\color{magenta}\icarus},
  \href{https://ui.adsabs.harvard.edu/abs/2008Icar..195..871K}{195, 871}

\bibitem[{{Kobayashi} {et~al.}(2009){Kobayashi}, {Watanabe}, {Kimura}, \&
  {Yamamoto}}]{kobayashi:2009}
{Kobayashi}, H., {Watanabe}, S.-i., {Kimura}, H., \& {Yamamoto}, T. 2009,
  \href{http://dx.doi.org/10.1016/j.icarus.2009.01.002}{\color{magenta}\icarus},
  \href{https://ui.adsabs.harvard.edu/abs/2009Icar..201..395K}{201, 395}

\bibitem[{{Kral} {et~al.}(2017){Kral}, {Krivov}, {Defr{\`e}re}, {van Lieshout},
  {Bonsor}, {Augereau}, {Th{\'e}bault}, {Ertel}, {Lebreton}, \&
  {Absil}}]{kral:2017}
{Kral}, Q., {Krivov}, A.~V., {Defr{\`e}re}, D., {et~al.} 2017,
  \href{http://dx.doi.org/10.1080/21672857.2017.1353202}{\color{magenta}The
  Astronomical Review},
  \href{https://ui.adsabs.harvard.edu/abs/2017AstRv..13...69K}{13, 69}

\bibitem[{{Krivov} {et~al.}(1998){Krivov}, {Kimura}, \& {Mann}}]{krivov:1998}
{Krivov}, A., {Kimura}, H., \& {Mann}, I. 1998,
  \href{http://dx.doi.org/10.1006/icar.1998.5949}{\color{magenta}\icarus},
  \href{https://ui.adsabs.harvard.edu/abs/1998Icar..134..311K}{134, 311}

\bibitem[{{Kurz} {et~al.}(2002){Kurz}, {Guilloteau}, \& {Shaver}}]{kurz:2002}
{Kurz}, R., {Guilloteau}, S., \& {Shaver}, P. 2002, The Messenger,
  \href{https://ui.adsabs.harvard.edu/abs/2002Msngr.107....7K}{107, 7}

\bibitem[{{Lamy}(1974)}]{lamy:1974b}
{Lamy}, P.~L. 1974, \aap,
  \href{https://ui.adsabs.harvard.edu/abs/1974A&A....35..197L}{35, 197}

\bibitem[{{Laugier} {et~al.}(2023){Laugier}, {Defr{\`e}re}, {Woillez},
  {Courtney-Barrer}, {Dannert}, {Matter}, {Dandumont}, {Gross}, {Absil},
  {Bigioli}, {Garreau}, {Labadie}, {Loicq}, {Martinod}, {Mazzoli}, {Raskin}, \&
  {Sanny}}]{laugier:2023}
{Laugier}, R., {Defr{\`e}re}, D., {Woillez}, J., {et~al.} 2023,
  \href{http://dx.doi.org/10.1051/0004-6361/202244351}{\color{magenta}\aap},
  \href{https://ui.adsabs.harvard.edu/abs/2023A&A...671A.110L}{671, A110}

\bibitem[{{Laureijs} {et~al.}(2002){Laureijs}, {Jourdain de Muizon}, {Leech},
  {Siebenmorgen}, {Dominik}, {Habing}, {Trams}, \& {Kessler}}]{laureijs:2002}
{Laureijs}, R.~J., {Jourdain de Muizon}, M., {Leech}, K., {et~al.} 2002,
  \href{http://dx.doi.org/10.1051/0004-6361:20020366}{\color{magenta}\aap},
  \href{https://ui.adsabs.harvard.edu/abs/2002A&A...387..285L}{387, 285}

\bibitem[{{Lawler} {et~al.}(2014){Lawler}, {Di Francesco}, {Kennedy},
  {Sibthorpe}, {Booth}, {Vandenbussche}, {Matthews}, {Holland}, {Greaves},
  {Wilner}, {Tuomi}, {Blommaert}, {de Vries}, {Dominik}, {Fridlund}, {Gear},
  {Heras}, {Ivison}, \& {Olofsson}}]{lawler:2014}
{Lawler}, S.~M., {Di Francesco}, J., {Kennedy}, G.~M., {et~al.} 2014,
  \href{http://dx.doi.org/10.1093/mnras/stu1641}{\color{magenta}\mnras},
  \href{https://ui.adsabs.harvard.edu/abs/2014MNRAS.444.2665L}{444, 2665}

\bibitem[{{Lebreton} {et~al.}(2013){Lebreton}, {van Lieshout}, {Augereau},
  {Absil}, {Mennesson}, {Kama}, {Dominik}, {Bonsor}, {Vandeportal}, {Beust},
  {Defr{\`e}re}, {Ertel}, {Faramaz}, {Hinz}, {Kral}, {Lagrange}, {Liu}, \&
  {Th{\'e}bault}}]{lebreton:2013}
{Lebreton}, J., {van Lieshout}, R., {Augereau}, J.~C., {et~al.} 2013,
  \href{http://dx.doi.org/10.1051/0004-6361/201321415}{\color{magenta}\aap},
  \href{https://ui.adsabs.harvard.edu/abs/2013A&A...555A.146L}{555, A146}

\bibitem[{{Lietzow} \& {Wolf}(2022)}]{lietzow:2022}
{Lietzow}, M. \& {Wolf}, S. 2022,
  \href{http://dx.doi.org/10.1051/0004-6361/202142521}{\color{magenta}\aap},
  \href{https://ui.adsabs.harvard.edu/abs/2022A&A...663A..55L}{663, A55}

\bibitem[{{Lietzow} \& {Wolf}(2023)}]{lietzow:2023}
{Lietzow}, M. \& {Wolf}, S. 2023,
  \href{http://dx.doi.org/10.1051/0004-6361/202245474}{\color{magenta}\aap},
  \href{https://ui.adsabs.harvard.edu/abs/2023A&A...671A.113L}{671, A113}

\bibitem[{{Lietzow} {et~al.}(2021){Lietzow}, {Wolf}, \&
  {Brunngr{\"a}ber}}]{lietzow:2021}
{Lietzow}, M., {Wolf}, S., \& {Brunngr{\"a}ber}, R. 2021,
  \href{http://dx.doi.org/10.1051/0004-6361/202038932}{\color{magenta}\aap},
  \href{https://ui.adsabs.harvard.edu/abs/2021A&A...645A.146L}{645, A146}

\bibitem[{{Ligi} {et~al.}(2016){Ligi}, {Creevey}, {Mourard}, {Crida},
  {Lagrange}, {Nardetto}, {Perraut}, {Schultheis}, {Tallon-Bosc}, \& {ten
  Brummelaar}}]{ligi:2016}
{Ligi}, R., {Creevey}, O., {Mourard}, D., {et~al.} 2016,
  \href{http://dx.doi.org/10.1051/0004-6361/201527054}{\color{magenta}\aap},
  \href{https://ui.adsabs.harvard.edu/abs/2016A&A...586A..94L}{586, A94}

\bibitem[{{Lopez} {et~al.}(2014){Lopez}, {Lagarde}, {Jaffe}, {Petrov},
  {Sch{\"o}ller}, {Antonelli}, {Beckmann}, {Berio}, {Bettonvil}, {Glindemann},
  {Gonzalez}, {Graser}, {Hofmann}, {Millour}, {Robbe-Dubois}, {Venema}, {Wolf},
  {Henning}, {Lanz}, {Weigelt}, {Agocs}, {Bailet}, {Bresson}, {Bristow},
  {Dugu{\'e}}, {Heininger}, {Kroes}, {Laun}, {Lehmitz}, {Neumann}, {Augereau},
  {Avila}, {Behrend}, {van Belle}, {Berger}, {van Boekel}, {Bonhomme},
  {Bourget}, {Brast}, {Clausse}, {Connot}, {Conzelmann}, {Cruzal{\`e}bes},
  {Csepany}, {Danchi}, {Delbo}, {Delplancke}, {Dominik}, {van Duin}, {Elswijk},
  {Fantei}, {Finger}, {Gabasch}, {Gay}, {Girard}, {Girault}, {Gitton},
  {Glazenborg}, {Gont{\'e}}, {Guitton}, {Guniat}, {De Haan}, {Haguenauer},
  {Hanenburg}, {Hogerheijde}, {ter Horst}, {Hron}, {Hugues}, {Hummel},
  {Idserda}, {Ives}, {Jakob}, {Jasko}, {Jolley}, {Kiraly}, {K{\"o}hler},
  {Kragt}, {Kroener}, {Kuindersma}, {Labadie}, {Leinert}, {Le Poole}, {Lizon},
  {Lucuix}, {Marcotto}, {Martinache}, {Martinot-Lagarde}, {Mathar}, {Matter},
  {Mauclert}, {Mehrgan}, {Meilland}, {Meisenheimer}, {Meisner}, {Mellein},
  {Menardi}, {Menut}, {Merand}, {Morel}, {Mosoni}, {Navarro}, {Nussbaum},
  {Ottogalli}, {Palsa}, {Panduro}, {Pantin}, {Parra}, {Percheron}, {Duc},
  {Pott}, {Pozna}, {Przygodda}, {Rabbia}, {Richichi}, {Rigal}, {Roelfsema},
  {Rupprecht}, {Schertl}, {Schmidt}, {Schuhler}, {Schuil}, {Spang},
  {Stegmeier}, {Thiam}, {Tromp}, {Vakili}, {Vannier}, {Wagner}, \&
  {Woillez}}]{lopez:2014}
{Lopez}, B., {Lagarde}, S., {Jaffe}, W., {et~al.} 2014, The Messenger,
  \href{https://ui.adsabs.harvard.edu/abs/2014Msngr.157....5L}{157, 5}

\bibitem[{{Lopez} {et~al.}(2022){Lopez}, {Lagarde}, {Petrov}, {Jaffe},
  {Antonelli}, {Allouche}, {Berio}, {Matter}, {Meilland}, {Millour},
  {Robbe-Dubois}, {Henning}, {Weigelt}, {Glindemann}, {Agocs}, {Bailet},
  {Beckmann}, {Bettonvil}, {van Boekel}, {Bourget}, {Bresson}, {Bristow},
  {Cruzal{\`e}bes}, {Eldswijk}, {Fante{\"\i} Caujolle}, {Gonz{\'a}lez Herrera},
  {Graser}, {Guajardo}, {Heininger}, {Hofmann}, {Kroes}, {Laun}, {Lehmitz},
  {Leinert}, {Meisenheimer}, {Morel}, {Neumann}, {Paladini}, {Percheron},
  {Riquelme}, {Schoeller}, {Stee}, {Venema}, {Woillez}, {Zins},
  {{\'A}brah{\'a}m}, {Abadie}, {Abuter}, {Accardo}, {Adler}, {Alonso},
  {Augereau}, {B{\"o}hm}, {Bazin}, {Beltran}, {Bensberg}, {Boland}, {Brast},
  {Burtscher}, {Castillo}, {Chelli}, {Cid}, {Clausse}, {Connot}, {Conzelmann},
  {Danchi}, {Delbo}, {Drevon}, {Dominik}, {van Duin}, {Ebert}, {Eisenhauer},
  {Flament}, {Frahm}, {G{\'a}mez Rosas}, {Gabasch}, {Gallenne}, {Garces},
  {Girard}, {Glazenborg}, {Gont{\'e}}, {Guitton}, {de Haan}, {Hanenburg},
  {Haubois}, {Hocd{\'e}}, {Hogerheijde}, {ter Horst}, {Hron}, {Hummel},
  {Hubin}, {Huerta}, {Idserda}, {Isbell}, {Ives}, {Jakob}, {Jask{\'o}},
  {Jochum}, {Klarmann}, {Klein}, {Kragt}, {Kuindersma}, {Kokoulina}, {Labadie},
  {Lacour}, {Leftley}, {Le Poole}, {Lizon}, {Lopez}, {Lykou}, {M{\'e}rand},
  {Marcotto}, {Mauclert}, {Maurer}, {Mehrgan}, {Meisner}, {Meixner}, {Mellein},
  {Menut}, {Mohr}, {Mosoni}, {Navarro}, {Nu{\ss}baum}, {Pallanca}, {Pantin},
  {Pasquini}, {Phan Duc}, {Pott}, {Pozna}, {Richichi}, {Ridinger}, {Rigal},
  {Rivinius}, {Roelfsema}, {Rohloff}, {Rousseau}, {Salabert}, {Schertl},
  {Schuhler}, {Schuil}, {Shabun}, {Soulain}, {Stephan}, {Toledo}, {Tristram},
  {Tromp}, {Vakili}, {Varga}, {Vinther}, {Waters}, {Wittkowski}, {Wolf},
  {Wrhel}, \& {Yoffe}}]{lopez:2022}
{Lopez}, B., {Lagarde}, S., {Petrov}, R.~G., {et~al.} 2022,
  \href{http://dx.doi.org/10.1051/0004-6361/202141785}{\color{magenta}\aap},
  \href{https://ui.adsabs.harvard.edu/abs/2022A&A...659A.192L}{659, A192}

\bibitem[{{MacGregor} {et~al.}(2016){MacGregor}, {Lawler}, {Wilner},
  {Matthews}, {Kennedy}, {Booth}, \& {Di Francesco}}]{macgregor:2016}
{MacGregor}, M.~A., {Lawler}, S.~M., {Wilner}, D.~J., {et~al.} 2016,
  \href{http://dx.doi.org/10.3847/0004-637X/828/2/113}{\color{magenta}\apj},
  \href{https://ui.adsabs.harvard.edu/abs/2016ApJ...828..113M}{828, 113}

\bibitem[{{Mamajek}(2012)}]{mamajek:2012}
{Mamajek}, E.~E. 2012,
  \href{http://dx.doi.org/10.1088/2041-8205/754/2/L20}{\color{magenta}\apjl},
  \href{https://ui.adsabs.harvard.edu/abs/2012ApJ...754L..20M}{754, L20}

\bibitem[{{Marboeuf} {et~al.}(2016){Marboeuf}, {Bonsor}, \&
  {Augereau}}]{marboeuf:2016}
{Marboeuf}, U., {Bonsor}, A., \& {Augereau}, J.~C. 2016,
  \href{http://dx.doi.org/10.1016/j.pss.2016.03.014}{\color{magenta}\planss},
  \href{https://ui.adsabs.harvard.edu/abs/2016P&SS..133...47M}{133, 47}

\bibitem[{{Martinod} {et~al.}(2023){Martinod}, {Defr{\`e}re}, {Ireland},
  {Kraus}, {Martinache}, {Tuthill}, {Bigioli}, {Bouzerand}, {Bryant},
  {Chhabra}, {Courtney-Barrer}, {Crous}, {Cvetojevic}, {Dandumont}, {Ertel},
  {Gardner}, {Garreau}, {Glauser}, {Labadie}, {Lagadec}, {Laugier}, {Mazzoli},
  {Mortimer}, {Norris}, {Raskin}, {Robertson}, {Sanny}, \&
  {Taras}}]{martinod:2023}
{Martinod}, M.-A., {Defr{\`e}re}, D., {Ireland}, M., {et~al.} 2023,
  \href{http://dx.doi.org/10.1117/1.JATIS.9.2.025007}{\color{magenta}Journal of
  Astronomical Telescopes, Instruments, and Systems},
  \href{https://ui.adsabs.harvard.edu/abs/2023JATIS...9b5007M}{9, 025007}

\bibitem[{{Martinod} {et~al.}(2022){Martinod}, {Defr{\`e}re}, {Ireland},
  {Kraus}, {Martinache}, {Tuthill}, {Bigioli}, {Bryant}, {Chhabra},
  {Courtney-Barrer}, {Crous}, {Cvetojevic}, {Dandumont}, {Garreau}, {Lagadec},
  {Laugier}, {Mortimer}, {Norris}, {Robertson}, \& {Taras}}]{martinod:2022}
{Martinod}, M.-A., {Defr{\`e}re}, D., {Ireland}, M.~J., {et~al.} 2022, in
  Society of Photo-Optical Instrumentation Engineers (SPIE) Conference Series,
  Vol. 12183, Optical and Infrared Interferometry and Imaging VIII, ed.
  A.~{M{\'e}rand}, S.~{Sallum}, \& J.~{Sanchez-Bermudez},
  \href{https://ui.adsabs.harvard.edu/abs/2022SPIE12183E..10M}{1218310}

\bibitem[{{Matr{\`a}} {et~al.}(2020){Matr{\`a}}, {Dent}, {Wilner}, {Marino},
  {Wyatt}, {Marshall}, {Su}, {Chavez}, {Hales}, {Hughes}, {Greaves}, \&
  {Corder}}]{matra:2020}
{Matr{\`a}}, L., {Dent}, W. R.~F., {Wilner}, D.~J., {et~al.} 2020,
  \href{http://dx.doi.org/10.3847/1538-4357/aba0a4}{\color{magenta}\apj},
  \href{https://ui.adsabs.harvard.edu/abs/2020ApJ...898..146M}{898, 146}

\bibitem[{{Meech} {et~al.}(2005){Meech}, {Ageorges}, {A'Hearn}, {Arpigny},
  {Ates}, {Aycock}, {Bagnulo}, {Bailey}, {Barber}, {Barrera}, {Barrena},
  {Bauer}, {Belton}, {Bensch}, {Bhattacharya}, {Biver}, {Blake},
  {Bockel{\'e}e-Morvan}, {Boehnhardt}, {Bonev}, {Bonev}, {Buie}, {Burton},
  {Butner}, {Cabanac}, {Campbell}, {Campins}, {Capria}, {Carroll}, {Chaffee},
  {Charnley}, {Cleis}, {Coates}, {Cochran}, {Colom}, {Conrad}, {Coulson},
  {Crovisier}, {deBuizer}, {Dekany}, {de L{\'e}on}, {Dello Russo}, {Delsanti},
  {DiSanti}, {Drummond}, {Dundon}, {Etzel}, {Farnham}, {Feldman},
  {Fern{\'a}ndez}, {Filipovic}, {Fisher}, {Fitzsimmons}, {Fong}, {Fugate},
  {Fujiwara}, {Fujiyoshi}, {Furusho}, {Fuse}, {Gibb}, {Groussin}, {Gulkis},
  {Gurwell}, {Hadamcik}, {Hainaut}, {Harker}, {Harrington}, {Harwit},
  {Hasegawa}, {Hergenrother}, {Hirst}, {Hodapp}, {Honda}, {Howell},
  {Hutsem{\'e}kers}, {Iono}, {Ip}, {Jackson}, {Jehin}, {Jiang}, {Jones},
  {Jones}, {Kadono}, {Kamath}, {K{\"a}ufl}, {Kasuga}, {Kawakita}, {Kelley},
  {Kerber}, {Kidger}, {Kinoshita}, {Knight}, {Lara}, {Larson}, {Lederer},
  {Lee}, {Levasseur-Regourd}, {Li}, {Li}, {Licandro}, {Lin}, {Lisse},
  {LoCurto}, {Lovell}, {Lowry}, {Lyke}, {Lynch}, {Ma}, {Magee-Sauer},
  {Maheswar}, {Manfroid}, {Marco}, {Martin}, {Melnick}, {Miller}, {Miyata},
  {Moriarty-Schieven}, {Moskovitz}, {Mueller}, {Mumma}, {Muneer}, {Neufeld},
  {Ootsubo}, {Osip}, {Pandea}, {Pantin}, {Paterno-Mahler}, {Patten},
  {Penprase}, {Peck}, {Petitpas}, {Pinilla-Alonso}, {Pittichova}, {Pompei},
  {Prabhu}, {Qi}, {Rao}, {Rauer}, {Reitsema}, {Rodgers}, {Rodriguez}, {Ruane},
  {Ruch}, {Rujopakarn}, {Sahu}, {Sako}, {Sakon}, {Samarasinha}, {Sarkissian},
  {Saviane}, {Schirmer}, {Schultz}, {Schulz}, {Seitzer}, {Sekiguchi}, {Selman},
  {Serra-Ricart}, {Sharp}, {Snell}, {Snodgrass}, {Stallard}, {Stecklein},
  {Sterken}, {St{\"u}we}, {Sugita}, {Sumner}, {Suntzeff}, {Swaters},
  {Takakuwa}, {Takato}, {Thomas-Osip}, {Thompson}, {Tokunaga}, {Tozzi}, {Tran},
  {Troy}, {Trujillo}, {Van Cleve}, {Vasundhara}, {Vazquez}, {Vilas},
  {Villanueva}, {von Braun}, {Vora}, {Wainscoat}, {Walsh}, {Watanabe},
  {Weaver}, {Weaver}, {Weiler}, {Weissman}, {Welsh}, {Wilner}, {Wolk},
  {Womack}, {Wooden}, {Woodney}, {Woodward}, {Wu}, {Wu}, {Yamashita}, {Yang},
  {Yang}, {Yokogawa}, {Zook}, {Zauderer}, {Zhao}, {Zhou}, \&
  {Zucconi}}]{meech:2005}
{Meech}, K.~J., {Ageorges}, N., {A'Hearn}, M.~F., {et~al.} 2005,
  \href{http://dx.doi.org/10.1126/science.1118978}{\color{magenta}Science},
  \href{https://ui.adsabs.harvard.edu/abs/2005Sci...310..265M}{310, 265}

\bibitem[{{Mennesson} {et~al.}(2013){Mennesson}, {Absil}, {Lebreton},
  {Augereau}, {Serabyn}, {Colavita}, {Millan-Gabet}, {Liu}, {Hinz}, \&
  {Th{\'e}bault}}]{mennesson:2013}
{Mennesson}, B., {Absil}, O., {Lebreton}, J., {et~al.} 2013,
  \href{http://dx.doi.org/10.1088/0004-637X/763/2/119}{\color{magenta}\apj},
  \href{https://ui.adsabs.harvard.edu/abs/2013ApJ...763..119M}{763, 119}

\bibitem[{{Mennesson} {et~al.}(2014){Mennesson}, {Millan-Gabet}, {Serabyn},
  {Colavita}, {Absil}, {Bryden}, {Wyatt}, {Danchi}, {Defr{\`e}re}, {Dor{\'e}},
  {Hinz}, {Kuchner}, {Ragland}, {Scott}, {Stapelfeldt}, {Traub}, \&
  {Woillez}}]{mennesson:2014}
{Mennesson}, B., {Millan-Gabet}, R., {Serabyn}, E., {et~al.} 2014,
  \href{http://dx.doi.org/10.1088/0004-637X/797/2/119}{\color{magenta}\apj},
  \href{https://ui.adsabs.harvard.edu/abs/2014ApJ...797..119M}{797, 119}

\bibitem[{{M{\'e}rand} {et~al.}(2006){M{\'e}rand}, {Coud{\'e} du Foresto},
  {Kellerer}, {ten Brummelaar}, {Reess}, \& {Ziegler}}]{merand:2006}
{M{\'e}rand}, A., {Coud{\'e} du Foresto}, V., {Kellerer}, A., {et~al.} 2006, in
  Society of Photo-Optical Instrumentation Engineers (SPIE) Conference Series,
  Vol. 6268, Society of Photo-Optical Instrumentation Engineers (SPIE)
  Conference Series, ed. J.~D. {Monnier}, M.~{Sch{\"o}ller}, \& W.~C. {Danchi},
  \href{https://ui.adsabs.harvard.edu/abs/2006SPIE.6268E..1FM}{62681F}

\bibitem[{{Mie}(1908)}]{mie:1908}
{Mie}, G. 1908,
  \href{http://dx.doi.org/10.1002/andp.19083300302}{\color{magenta}Annalen der
  Physik}, \href{https://ui.adsabs.harvard.edu/abs/1908AnP...330..377M}{330,
  377}

\bibitem[{{Millan-Gabet} {et~al.}(2011){Millan-Gabet}, {Serabyn}, {Mennesson},
  {Traub}, {Barry}, {Danchi}, {Kuchner}, {Stark}, {Ragland}, {Hrynevych},
  {Woillez}, {Stapelfeldt}, {Bryden}, {Colavita}, \&
  {Booth}}]{millan_gabet:2011}
{Millan-Gabet}, R., {Serabyn}, E., {Mennesson}, B., {et~al.} 2011,
  \href{http://dx.doi.org/10.1088/0004-637X/734/1/67}{\color{magenta}\apj},
  \href{https://ui.adsabs.harvard.edu/abs/2011ApJ...734...67M}{734, 67}

\bibitem[{{Monnier} {et~al.}(2012){Monnier}, {Che}, {Zhao}, {Ekstr{\"o}m},
  {Maestro}, {Aufdenberg}, {Baron}, {Georgy}, {Kraus}, {McAlister}, {Pedretti},
  {Ridgway}, {Sturmann}, {Sturmann}, {ten Brummelaar}, {Thureau}, {Turner}, \&
  {Tuthill}}]{monnier:2012}
{Monnier}, J.~D., {Che}, X., {Zhao}, M., {et~al.} 2012,
  \href{http://dx.doi.org/10.1088/2041-8205/761/1/L3}{\color{magenta}\apjl},
  \href{https://ui.adsabs.harvard.edu/abs/2012ApJ...761L...3M}{761, L3}

\bibitem[{{Mukai} {et~al.}(1974){Mukai}, {Yamamoto}, {Hasegawa}, {Fujiwara}, \&
  {Koike}}]{mukai:1974}
{Mukai}, T., {Yamamoto}, T., {Hasegawa}, H., {Fujiwara}, A., \& {Koike}, C.
  1974, \pasj, \href{https://ui.adsabs.harvard.edu/abs/1974PASJ...26..445M}{26,
  445}

\bibitem[{{Pearce} {et~al.}(2022){Pearce}, {Kirchschlager}, {Rouill{\'e}},
  {Ertel}, {Bensberg}, {Krivov}, {Booth}, {Wolf}, \& {Augereau}}]{pearce:2022b}
{Pearce}, T.~D., {Kirchschlager}, F., {Rouill{\'e}}, G., {et~al.} 2022,
  \href{http://dx.doi.org/10.1093/mnras/stac2773}{\color{magenta}\mnras},
  \href{https://ui.adsabs.harvard.edu/abs/2022MNRAS.517.1436P}{517, 1436}

\bibitem[{{Pearce} {et~al.}(2020){Pearce}, {Krivov}, \& {Booth}}]{pearce:2020}
{Pearce}, T.~D., {Krivov}, A.~V., \& {Booth}, M. 2020,
  \href{http://dx.doi.org/10.1093/mnras/staa2514}{\color{magenta}\mnras},
  \href{https://ui.adsabs.harvard.edu/abs/2020MNRAS.498.2798P}{498, 2798}

\bibitem[{{Perez} \& {Granger}(2007)}]{perez:granger:2007}
{Perez}, F. \& {Granger}, B.~E. 2007,
  \href{http://dx.doi.org/10.1109/MCSE.2007.53}{\color{magenta}Computing in
  Science and Engineering},
  \href{https://ui.adsabs.harvard.edu/abs/2007CSE.....9c..21P}{9, 21}

\bibitem[{{Peterson} {et~al.}(2006{\natexlab{a}}){Peterson}, {Hummel}, {Pauls},
  {Armstrong}, {Benson}, {Gilbreath}, {Hindsley}, {Hutter}, {Johnston},
  {Mozurkewich}, \& {Schmitt}}]{peterson:2006a}
{Peterson}, D.~M., {Hummel}, C.~A., {Pauls}, T.~A., {et~al.}
  2006{\natexlab{a}},
  \href{http://dx.doi.org/10.1086/497981}{\color{magenta}\apj},
  \href{https://ui.adsabs.harvard.edu/abs/2006ApJ...636.1087P}{636, 1087}

\bibitem[{{Peterson} {et~al.}(2006{\natexlab{b}}){Peterson}, {Hummel}, {Pauls},
  {Armstrong}, {Benson}, {Gilbreath}, {Hindsley}, {Hutter}, {Johnston},
  {Mozurkewich}, \& {Schmitt}}]{peterson:2006b}
{Peterson}, D.~M., {Hummel}, C.~A., {Pauls}, T.~A., {et~al.}
  2006{\natexlab{b}},
  \href{http://dx.doi.org/10.1038/nature04661}{\color{magenta}\nat},
  \href{https://ui.adsabs.harvard.edu/abs/2006Natur.440..896P}{440, 896}

\bibitem[{{Pilbratt} {et~al.}(2010){Pilbratt}, {Riedinger}, {Passvogel},
  {Crone}, {Doyle}, {Gageur}, {Heras}, {Jewell}, {Metcalfe}, {Ott}, \&
  {Schmidt}}]{pilbratt:2010}
{Pilbratt}, G.~L., {Riedinger}, J.~R., {Passvogel}, T., {et~al.} 2010,
  \href{http://dx.doi.org/10.1051/0004-6361/201014759}{\color{magenta}\aap},
  \href{https://ui.adsabs.harvard.edu/abs/2010A&A...518L...1P}{518, L1}

\bibitem[{{Poynting}(1904)}]{poynting:1904}
{Poynting}, J.~H. 1904,
  \href{http://dx.doi.org/10.1098/rsta.1904.0012}{\color{magenta}Philosophical
  Transactions of the Royal Society of London Series A},
  \href{https://ui.adsabs.harvard.edu/abs/1904RSPTA.202..525P}{202, 525}

\bibitem[{{Raymond} \& {Bonsor}(2014)}]{raymond:bonsor:2014}
{Raymond}, S.~N. \& {Bonsor}, A. 2014,
  \href{http://dx.doi.org/10.1093/mnrasl/slu048}{\color{magenta}\mnras},
  \href{https://ui.adsabs.harvard.edu/abs/2014MNRAS.442L..18R}{442, L18}

\bibitem[{{Richichi} \& {Percheron}(2005)}]{richichi:percheron:2005}
{Richichi}, A. \& {Percheron}, I. 2005,
  \href{http://dx.doi.org/10.1051/0004-6361:20042257}{\color{magenta}\aap},
  \href{https://ui.adsabs.harvard.edu/abs/2005A&A...434.1201R}{434, 1201}

\bibitem[{{Rieke} {et~al.}(2016){Rieke}, {G{\'a}sp{\'a}r}, \&
  {Ballering}}]{rieke:2016}
{Rieke}, G.~H., {G{\'a}sp{\'a}r}, A., \& {Ballering}, N.~P. 2016,
  \href{http://dx.doi.org/10.3847/0004-637X/816/2/50}{\color{magenta}\apj},
  \href{https://ui.adsabs.harvard.edu/abs/2016ApJ...816...50R}{816, 50}

\bibitem[{{Rigley} \& {Wyatt}(2020)}]{rigley:2020}
{Rigley}, J.~K. \& {Wyatt}, M.~C. 2020,
  \href{http://dx.doi.org/10.1093/mnras/staa2029}{\color{magenta}\mnras},
  \href{https://ui.adsabs.harvard.edu/abs/2020MNRAS.497.1143R}{497, 1143}

\bibitem[{{Robertson}(1937)}]{robertson:1937}
{Robertson}, H.~P. 1937,
  \href{http://dx.doi.org/10.1093/mnras/97.6.423}{\color{magenta}\mnras},
  \href{https://ui.adsabs.harvard.edu/abs/1937MNRAS..97..423R}{97, 423}

\bibitem[{{Rouleau} \& {Martin}(1991)}]{rouleau:martin:1991}
{Rouleau}, F. \& {Martin}, P.~G. 1991,
  \href{http://dx.doi.org/10.1086/170382}{\color{magenta}\apj},
  \href{https://ui.adsabs.harvard.edu/abs/1991ApJ...377..526R}{377, 526}

\bibitem[{{Royer} {et~al.}(2007){Royer}, {Zorec}, \& {G{\'o}mez}}]{royer:2007}
{Royer}, F., {Zorec}, J., \& {G{\'o}mez}, A.~E. 2007,
  \href{http://dx.doi.org/10.1051/0004-6361:20065224}{\color{magenta}\aap},
  \href{https://ui.adsabs.harvard.edu/abs/2007A&A...463..671R}{463, 671}

\bibitem[{{Rusk}(1987)}]{rusk:1987}
{Rusk}, E.~T. 1987,
  \href{http://dx.doi.org/10.1086/165544}{\color{magenta}\apj},
  \href{https://ui.adsabs.harvard.edu/abs/1987ApJ...320..315R}{320, 315}

\bibitem[{{Sch{\"o}ller} {et~al.}(2003){Sch{\"o}ller}, {Gitton}, {Argomedo},
  {Ballester}, {Bauvir}, {Van Boekel}, {Cantzler}, {Correia}, {Cotton},
  {Delplancke}, {Derie}, {Duhoux}, {Erm}, {di Folco}, {Coud{\'e} du Foresto},
  {Gennai}, {Gilli}, {Giordano}, {Glindemann}, {Guisard}, {Gutierrez},
  {Housen}, {Huedepohl}, {Huxley}, {Jackisch}, {Jaffe}, {Kervella}, {van
  Kesteren}, {Kiekebusch}, {Koehler}, {Leveque}, {Longinotti}, {Menardi},
  {Morel}, {Noethe}, {Paresce}, {Percheron}, {Phan Duc}, {Pino}, {Rabeling},
  {Ramirez}, {Robbe-Dubois}, {Richichi}, {Rijo}, {Sabet}, {Sandrock},
  {Segransan}, {Spyromilio}, {Tamai}, {Tarenghi}, {Wallander}, {Wilhelm}, \&
  {Wittkowski}}]{schoeller:2003}
{Sch{\"o}ller}, M., {Gitton}, P.~B., {Argomedo}, J., {et~al.} 2003, in Society
  of Photo-Optical Instrumentation Engineers (SPIE) Conference Series, Vol.
  4838, Interferometry for Optical Astronomy II, ed. W.~A. {Traub},
  \href{https://ui.adsabs.harvard.edu/abs/2003SPIE.4838..870S}{870--880}

\bibitem[{{Sch{\"o}ller} \& {Glindemann}(2003)}]{schoeller:glindemann:2003}
{Sch{\"o}ller}, M. \& {Glindemann}, A. 2003, in ESA Special Publication, Vol.
  539, Earths: DARWIN/TPF and the Search for Extrasolar Terrestrial Planets,
  ed. M.~{Fridlund}, T.~{Henning}, \& H.~{Lacoste},
  \href{https://ui.adsabs.harvard.edu/abs/2003ESASP.539..109S}{109--120}

\bibitem[{{Seager} {et~al.}(2000){Seager}, {Whitney}, \&
  {Sasselov}}]{seager:2000}
{Seager}, S., {Whitney}, B.~A., \& {Sasselov}, D.~D. 2000,
  \href{http://dx.doi.org/10.1086/309292}{\color{magenta}\apj},
  \href{https://ui.adsabs.harvard.edu/abs/2000ApJ...540..504S}{540, 504}

\bibitem[{{Sekanina} \& {Miller}(1973)}]{sekanina:miller:1973}
{Sekanina}, Z. \& {Miller}, F.~D. 1973,
  \href{http://dx.doi.org/10.1126/science.179.4073.565}{\color{magenta}Science},
  \href{https://ui.adsabs.harvard.edu/abs/1973Sci...179..565S}{179, 565}

\bibitem[{{Selina} {et~al.}(2018){Selina}, {Murphy}, {McKinnon}, {Beasley},
  {Butler}, {Carilli}, {Clark}, {Erickson}, {Grammer}, {Jackson}, {Kent},
  {Mason}, {Morgan}, {Ojeda}, {Shillue}, {Sturgis}, \& {Urbain}}]{selina:2018}
{Selina}, R.~J., {Murphy}, E.~J., {McKinnon}, M., {et~al.} 2018, in Society of
  Photo-Optical Instrumentation Engineers (SPIE) Conference Series, Vol. 10700,
  Ground-based and Airborne Telescopes VII, ed. H.~K. {Marshall} \&
  J.~{Spyromilio},
  \href{https://ui.adsabs.harvard.edu/abs/2018SPIE10700E..1OS}{107001O}

\bibitem[{{Serabyn} {et~al.}(2012){Serabyn}, {Mennesson}, {Colavita},
  {Koresko}, \& {Kuchner}}]{serabyn:2012}
{Serabyn}, E., {Mennesson}, B., {Colavita}, M.~M., {Koresko}, C., \& {Kuchner},
  M.~J. 2012,
  \href{http://dx.doi.org/10.1088/0004-637X/748/1/55}{\color{magenta}\apj},
  \href{https://ui.adsabs.harvard.edu/abs/2012ApJ...748...55S}{748, 55}

\bibitem[{{Sezestre} {et~al.}(2019){Sezestre}, {Augereau}, \&
  {Th{\'e}bault}}]{sezestre:2019}
{Sezestre}, {\'E}., {Augereau}, J.~C., \& {Th{\'e}bault}, P. 2019,
  \href{http://dx.doi.org/10.1051/0004-6361/201935250}{\color{magenta}\aap},
  \href{https://ui.adsabs.harvard.edu/abs/2019A&A...626A...2S}{626, A2}

\bibitem[{{Sibthorpe} {et~al.}(2018){Sibthorpe}, {Kennedy}, {Wyatt},
  {Lestrade}, {Greaves}, {Matthews}, \& {Duch{\^e}ne}}]{sibthorpe:2018}
{Sibthorpe}, B., {Kennedy}, G.~M., {Wyatt}, M.~C., {et~al.} 2018,
  \href{http://dx.doi.org/10.1093/mnras/stx3188}{\color{magenta}\mnras},
  \href{https://ui.adsabs.harvard.edu/abs/2018MNRAS.475.3046S}{475, 3046}

\bibitem[{{Sibthorpe} {et~al.}(2010){Sibthorpe}, {Vandenbussche}, {Greaves},
  {Pantin}, {Olofsson}, {Acke}, {Barlow}, {Blommaert}, {Bouwman}, {Brandeker},
  {Cohen}, {De Meester}, {Dent}, {di Francesco}, {Dominik}, {Fridlund}, {Gear},
  {Glauser}, {Gomez}, {Hargrave}, {Harvey}, {Henning}, {Heras}, {Hogerheijde},
  {Holland}, {Ivison}, {Leeks}, {Lim}, {Liseau}, {Matthews}, {Naylor},
  {Pilbratt}, {Polehampton}, {Regibo}, {Royer}, {Sicilia-Aguilar}, {Swinyard},
  {Waelkens}, {Walker}, \& {Wesson}}]{sibthorpe:2010}
{Sibthorpe}, B., {Vandenbussche}, B., {Greaves}, J.~S., {et~al.} 2010,
  \href{http://dx.doi.org/10.1051/0004-6361/201014574}{\color{magenta}\aap},
  \href{https://ui.adsabs.harvard.edu/abs/2010A&A...518L.130S}{518, L130}

\bibitem[{{Spitzer Science Center (SSC)} \& {Infrared Science Archive
  (IRSA)}(2021)}]{spitzer_science_center:IRSA:2021}
{Spitzer Science Center (SSC)} \& {Infrared Science Archive (IRSA)}. 2021,
  \href{https://ui.adsabs.harvard.edu/abs/2021yCat.2368....0S}{VizieR Online
  Data Catalog, II/368}

\bibitem[{{Stam}(2008)}]{stam:2008}
{Stam}, D.~M. 2008,
  \href{http://dx.doi.org/10.1051/0004-6361:20078358}{\color{magenta}\aap},
  \href{https://ui.adsabs.harvard.edu/abs/2008A&A...482..989S}{482, 989}

\bibitem[{{Stam} {et~al.}(2004){Stam}, {Hovenier}, \& {Waters}}]{stam:2004}
{Stam}, D.~M., {Hovenier}, J.~W., \& {Waters}, L.~B.~F.~M. 2004,
  \href{http://dx.doi.org/10.1051/0004-6361:20041578}{\color{magenta}\aap},
  \href{https://ui.adsabs.harvard.edu/abs/2004A&A...428..663S}{428, 663}

\bibitem[{{Stamm} {et~al.}(2019){Stamm}, {Czechowski}, {Mann}, {Baumann}, \&
  {Myrvang}}]{stamm:2019}
{Stamm}, J., {Czechowski}, A., {Mann}, I., {Baumann}, C., \& {Myrvang}, M.
  2019,
  \href{http://dx.doi.org/10.1051/0004-6361/201834727}{\color{magenta}\aap},
  \href{https://ui.adsabs.harvard.edu/abs/2019A&A...626A.107S}{626, A107}

\bibitem[{{Stolker} {et~al.}(2017){Stolker}, {Min}, {Stam}, {Molli{\`e}re},
  {Dominik}, \& {Waters}}]{stolker:2017}
{Stolker}, T., {Min}, M., {Stam}, D.~M., {et~al.} 2017,
  \href{http://dx.doi.org/10.1051/0004-6361/201730780}{\color{magenta}\aap},
  \href{https://ui.adsabs.harvard.edu/abs/2017A&A...607A..42S}{607, A42}

\bibitem[{{Su} {et~al.}(2016){Su}, {Rieke}, {Defr{\'e}re}, {Wang}, {Lai},
  {Wilner}, {van Lieshout}, \& {Lee}}]{su:2016}
{Su}, K. Y.~L., {Rieke}, G.~H., {Defr{\'e}re}, D., {et~al.} 2016,
  \href{http://dx.doi.org/10.3847/0004-637X/818/1/45}{\color{magenta}\apj},
  \href{https://ui.adsabs.harvard.edu/abs/2016ApJ...818...45S}{818, 45}

\bibitem[{{Su} {et~al.}(2013){Su}, {Rieke}, {Malhotra}, {Stapelfeldt},
  {Hughes}, {Bonsor}, {Wilner}, {Balog}, {Watson}, {Werner}, \&
  {Misselt}}]{su:2013}
{Su}, K. Y.~L., {Rieke}, G.~H., {Malhotra}, R., {et~al.} 2013,
  \href{http://dx.doi.org/10.1088/0004-637X/763/2/118}{\color{magenta}\apj},
  \href{https://ui.adsabs.harvard.edu/abs/2013ApJ...763..118S}{763, 118}

\bibitem[{{Su} {et~al.}(2006){Su}, {Rieke}, {Stansberry}, {Bryden},
  {Stapelfeldt}, {Trilling}, {Muzerolle}, {Beichman}, {Moro-Martin}, {Hines},
  \& {Werner}}]{su:2006}
{Su}, K.~Y.~L., {Rieke}, G.~H., {Stansberry}, J.~A., {et~al.} 2006,
  \href{http://dx.doi.org/10.1086/508649}{\color{magenta}\apj},
  \href{https://ui.adsabs.harvard.edu/abs/2006ApJ...653..675S}{653, 675}

\bibitem[{{ten Brummelaar} {et~al.}(2005){ten Brummelaar}, {McAlister},
  {Ridgway}, {Bagnuolo}, {Turner}, {Sturmann}, {Sturmann}, {Berger}, {Ogden},
  {Cadman}, {Hartkopf}, {Hopper}, \& {Shure}}]{ten_brummelaar:2005}
{ten Brummelaar}, T.~A., {McAlister}, H.~A., {Ridgway}, S.~T., {et~al.} 2005,
  \href{http://dx.doi.org/10.1086/430729}{\color{magenta}\apj},
  \href{https://ui.adsabs.harvard.edu/abs/2005ApJ...628..453T}{628, 453}

\bibitem[{{Thureau} {et~al.}(2014){Thureau}, {Greaves}, {Matthews}, {Kennedy},
  {Phillips}, {Booth}, {Duch{\^e}ne}, {Horner}, {Rodriguez}, {Sibthorpe}, \&
  {Wyatt}}]{thureau:2014}
{Thureau}, N.~D., {Greaves}, J.~S., {Matthews}, B.~C., {et~al.} 2014,
  \href{http://dx.doi.org/10.1093/mnras/stu1864}{\color{magenta}\mnras},
  \href{https://ui.adsabs.harvard.edu/abs/2014MNRAS.445.2558T}{445, 2558}

\bibitem[{{Torres} {et~al.}(2010){Torres}, {Andersen}, \&
  {Gim{\'e}nez}}]{torres:2010}
{Torres}, G., {Andersen}, J., \& {Gim{\'e}nez}, A. 2010,
  \href{http://dx.doi.org/10.1007/s00159-009-0025-1}{\color{magenta}\aapr},
  \href{https://ui.adsabs.harvard.edu/abs/2010A&ARv..18...67T}{18, 67}

\bibitem[{{van Belle} {et~al.}(2006){van Belle}, {Ciardi}, {ten Brummelaar},
  {McAlister}, {Ridgway}, {Berger}, {Goldfinger}, {Sturmann}, {Sturmann},
  {Turner}, {Boden}, {Thompson}, \& {Coyne}}]{van_belle:2006}
{van Belle}, G.~T., {Ciardi}, D.~R., {ten Brummelaar}, T., {et~al.} 2006,
  \href{http://dx.doi.org/10.1086/498334}{\color{magenta}\apj},
  \href{https://ui.adsabs.harvard.edu/abs/2006ApJ...637..494V}{637, 494}

\bibitem[{{van Belle} {et~al.}(2001){van Belle}, {Ciardi}, {Thompson},
  {Akeson}, \& {Lada}}]{van_belle:2001}
{van Belle}, G.~T., {Ciardi}, D.~R., {Thompson}, R.~R., {Akeson}, R.~L., \&
  {Lada}, E.~A. 2001,
  \href{http://dx.doi.org/10.1086/322340}{\color{magenta}\apj},
  \href{https://ui.adsabs.harvard.edu/abs/2001ApJ...559.1155V}{559, 1155}

\bibitem[{{van Belle} \& {von Braun}(2009)}]{van_belle:von_braun:2009}
{van Belle}, G.~T. \& {von Braun}, K. 2009,
  \href{http://dx.doi.org/10.1088/0004-637X/694/2/1085}{\color{magenta}\apj},
  \href{https://ui.adsabs.harvard.edu/abs/2009ApJ...694.1085V}{694, 1085}

\bibitem[{{van Leeuwen}(2007)}]{van_leeuwen:2007}
{van Leeuwen}, F. 2007,
  \href{http://dx.doi.org/10.1051/0004-6361:20078357}{\color{magenta}\aap},
  \href{https://ui.adsabs.harvard.edu/abs/2007A&A...474..653V}{474, 653}

\bibitem[{{van Lieshout} {et~al.}(2014{\natexlab{a}}){van Lieshout}, {Dominik},
  {Kama}, \& {Min}}]{van_lieshout:2014a}
{van Lieshout}, R., {Dominik}, C., {Kama}, M., \& {Min}, M. 2014{\natexlab{a}},
  \href{http://dx.doi.org/10.1051/0004-6361/201322090}{\color{magenta}\aap},
  \href{https://ui.adsabs.harvard.edu/abs/2014A&A...571A..51V}{571, A51}

\bibitem[{{van Lieshout} {et~al.}(2014{\natexlab{b}}){van Lieshout}, {Min}, \&
  {Dominik}}]{van_lieshout:2014b}
{van Lieshout}, R., {Min}, M., \& {Dominik}, C. 2014{\natexlab{b}},
  \href{http://dx.doi.org/10.1051/0004-6361/201424876}{\color{magenta}\aap},
  \href{https://ui.adsabs.harvard.edu/abs/2014A&A...572A..76V}{572, A76}

\bibitem[{{White} {et~al.}(2017){White}, {Boley}, {Dent}, {Ford}, \&
  {Corder}}]{white:2017}
{White}, J.~A., {Boley}, A.~C., {Dent}, W.~R.~F., {Ford}, E.~B., \& {Corder},
  S. 2017,
  \href{http://dx.doi.org/10.1093/mnras/stw3303}{\color{magenta}\mnras},
  \href{https://ui.adsabs.harvard.edu/abs/2017MNRAS.466.4201W}{466, 4201}

\bibitem[{{White} {et~al.}(2013){White}, {Huber}, {Maestro}, {Bedding},
  {Ireland}, {Baron}, {Boyajian}, {Che}, {Monnier}, {Pope}, {Roettenbacher},
  {Stello}, {Tuthill}, {Farrington}, {Goldfinger}, {McAlister}, {Schaefer},
  {Sturmann}, {Sturmann}, {ten Brummelaar}, \& {Turner}}]{white:2013}
{White}, T.~R., {Huber}, D., {Maestro}, V., {et~al.} 2013,
  \href{http://dx.doi.org/10.1093/mnras/stt802}{\color{magenta}\mnras},
  \href{https://ui.adsabs.harvard.edu/abs/2013MNRAS.433.1262W}{433, 1262}

\bibitem[{{Wiktorowicz} {et~al.}(2015){Wiktorowicz}, {Nofi}, {Jontof-Hutter},
  {Kopparla}, {Laughlin}, {Hermis}, {Yung}, \& {Swain}}]{wiktorowicz:2015}
{Wiktorowicz}, S.~J., {Nofi}, L.~A., {Jontof-Hutter}, D., {et~al.} 2015,
  \href{http://dx.doi.org/10.1088/0004-637X/813/1/48}{\color{magenta}\apj},
  \href{https://ui.adsabs.harvard.edu/abs/2015ApJ...813...48W}{813, 48}

\bibitem[{{Williams} \& {Gaidos}(2008)}]{williams:gaidos:2008}
{Williams}, D.~M. \& {Gaidos}, E. 2008,
  \href{http://dx.doi.org/10.1016/j.icarus.2008.01.002}{\color{magenta}\icarus},
  \href{https://ui.adsabs.harvard.edu/abs/2008Icar..195..927W}{195, 927}

\bibitem[{{Wolf} \& {Voshchinnikov}(2004)}]{wolf:voshchinnikov:2004}
{Wolf}, S. \& {Voshchinnikov}, N.~V. 2004,
  \href{http://dx.doi.org/10.1016/j.cpc.2004.06.070}{\color{magenta}Computer
  Physics Communications},
  \href{https://ui.adsabs.harvard.edu/abs/2004CoPhC.162..113W}{162, 113}

\bibitem[{{Wyatt} {et~al.}(2007){Wyatt}, {Smith}, {Greaves}, {Beichman},
  {Bryden}, \& {Lisse}}]{wyatt:2007a}
{Wyatt}, M.~C., {Smith}, R., {Greaves}, J.~S., {et~al.} 2007,
  \href{http://dx.doi.org/10.1086/510999}{\color{magenta}\apj},
  \href{https://ui.adsabs.harvard.edu/abs/2007ApJ...658..569W}{658, 569}

\bibitem[{{Wyatt} \& {Whipple}(1950)}]{wyatt:whipple:1950}
{Wyatt}, S.~P. \& {Whipple}, F.~L. 1950,
  \href{http://dx.doi.org/10.1086/145244}{\color{magenta}\apj},
  \href{https://ui.adsabs.harvard.edu/abs/1950ApJ...111..134W}{111, 134}

\bibitem[{{Zhao} {et~al.}(2009){Zhao}, {Monnier}, {Pedretti}, {Thureau},
  {M{\'e}rand}, {ten Brummelaar}, {McAlister}, {Ridgway}, {Turner}, {Sturmann},
  {Sturmann}, {Goldfinger}, \& {Farrington}}]{zhao:2009}
{Zhao}, M., {Monnier}, J.~D., {Pedretti}, E., {et~al.} 2009,
  \href{http://dx.doi.org/10.1088/0004-637X/701/1/209}{\color{magenta}\apj},
  \href{https://ui.adsabs.harvard.edu/abs/2009ApJ...701..209Z}{701, 209}

\bibitem[{{Zugger} {et~al.}(2010){Zugger}, {Kasting}, {Williams}, {Kane}, \&
  {Philbrick}}]{zugger:2010}
{Zugger}, M.~E., {Kasting}, J.~F., {Williams}, D.~M., {Kane}, T.~J., \&
  {Philbrick}, C.~R. 2010,
  \href{http://dx.doi.org/10.1088/0004-637X/723/2/1168}{\color{magenta}\apj},
  \href{https://ui.adsabs.harvard.edu/abs/2010ApJ...723.1168Z}{723, 1168}

\bibitem[{{Zugger} {et~al.}(2011{\natexlab{a}}){Zugger}, {Kasting}, {Williams},
  {Kane}, \& {Philbrick}}]{zugger:2011a}
{Zugger}, M.~E., {Kasting}, J.~F., {Williams}, D.~M., {Kane}, T.~J., \&
  {Philbrick}, C.~R. 2011{\natexlab{a}},
  \href{http://dx.doi.org/10.1088/0004-637X/739/1/55}{\color{magenta}\apj},
  \href{https://ui.adsabs.harvard.edu/abs/2011ApJ...739...55Z}{739, 55}

\bibitem[{{Zugger} {et~al.}(2011{\natexlab{b}}){Zugger}, {Kasting}, {Williams},
  {Kane}, \& {Philbrick}}]{zugger:2011b}
{Zugger}, M.~E., {Kasting}, J.~F., {Williams}, D.~M., {Kane}, T.~J., \&
  {Philbrick}, C.~R. 2011{\natexlab{b}},
  \href{http://dx.doi.org/10.1088/0004-637X/739/1/12}{\color{magenta}\apj},
  \href{https://ui.adsabs.harvard.edu/abs/2011ApJ...739...12Z}{739, 12}

\end{thebibliography}


\begin{appendix}

\section{Photometric measurements at wavelengths \texorpdfstring{$\bm{\lambda} \geq 24\,\upmu\mathrm{m}$}{p}}

The photometric measurements from the literature at wavelengths $\lambda \geq
\SI{24}{\upmu\m}$ are presented in Table~\ref{table_obs_longer_wavelengths}.
\begin{sidewaystable}
 \setlength{\tabcolsep}{3.88pt} 
 \centering
 \caption{Photometric values at specified observing wavelengths from the
          literature.}
 \begin{tabular}{c|cccccccccccccc}
  \toprule
  HD number & \multicolumn{14}{c}{$F^\nu$ /\SI{}{Jy}} \\
            & \SI{24}{\upmu\m} & \SI{25}{\upmu\m} & \SI{60}{\upmu\m} & \SI{70}{\upmu\m} & \SI{100}{\upmu\m} & \SI{160}{\upmu\m} & \SI{170}{\upmu\m} & \SI{250}{\upmu\m} & \SI{350}{\upmu\m} & \SI{450}{\upmu\m} & \SI{500}{\upmu\m} & \SI{850}{\upmu\m} & \SI{870}{\upmu\m} & \SI{1300}{\upmu\m} \\
  \midrule
  \phantom{1}10700 & -- & 1.319$\,$\tablefootmark{a} & 0.433$\,$\tablefootmark{b} & 0.303$\,$\tablefootmark{c} & 0.360$\,$\tablefootmark{d} & 0.111$\,$\tablefootmark{c} & 0.125$\,$\tablefootmark{b} & 0.035$\,$\tablefootmark{c} & $<$ 0.028$\,$\tablefootmark{c} & 0.0252$\,$\tablefootmark{e} & $<$ 0.020$\,$\tablefootmark{c} & 0.0044$\,$\tablefootmark{e} & -- & 0.00069$\,$\tablefootmark{$\star$, f} \\
  \phantom{1}22484 & 0.535$\,$\tablefootmark{g} & 0.511$\,$\tablefootmark{a} & 0.141$\,$\tablefootmark{b} & 0.120$\,$\tablefootmark{h} & 0.076$\,$\tablefootmark{a} & 0.026$\,$\tablefootmark{a} & 0.007$\,$\tablefootmark{b} & -- & -- & -- & -- & -- & -- & -- \\
  \phantom{1}56537 & 0.326$\,$\tablefootmark{g} & 0.586$\,$\tablefootmark{i} & 0.177$\,$\tablefootmark{i} & 0.03522$\,$\tablefootmark{g} & 0.01308$\,$\tablefootmark{j} & 0.00292$\,$\tablefootmark{j} & -- & -- & -- & -- & -- & -- & -- & -- \\
  102647 & 1.59968$\,$\tablefootmark{i} & 1.563$\,$\tablefootmark{a} & 0.784$\,$\tablefootmark{b} & 0.67647$\,$\tablefootmark{i} & 0.47554$\,$\tablefootmark{j} & 0.20415$\,$\tablefootmark{j} & -- & 0.051$\,$\tablefootmark{k} & 0.018$\,$\tablefootmark{k} & 0.075$\,$\tablefootmark{e} & 0.0028$\,$\tablefootmark{k} & 0.0045$\,$\tablefootmark{e} & -- & -- \\
  172167 & 8.900$\,$\tablefootmark{i} & 8.156$\,$\tablefootmark{a} & 6.530$\,$\tablefootmark{b} & 10.12$\,$\tablefootmark{l} & 5.844$\,$\tablefootmark{j} & 4.430$\,$\tablefootmark{j} & 2.621$\,$\tablefootmark{b} & 1.68$\,$\tablefootmark{l} & 0.61$\,$\tablefootmark{l} & 0.229$\,$\tablefootmark{e} & 0.21$\,$\tablefootmark{l} & 0.0344$\,$\tablefootmark{e} & -- & 0.002495$\,$\tablefootmark{$\star$, m} \\
  177724 & 0.49795$\,$\tablefootmark{n} & 0.7360$\,$\tablefootmark{o} & 0.5587$\,$\tablefootmark{o} & -- & 6.914$\,$\tablefootmark{o} & -- & -- & -- & -- & -- & -- & -- & -- & -- \\
  187642 & -- & 5.462$\,$\tablefootmark{a} & 1.010$\,$\tablefootmark{b} & -- & 0.32902$\,$\tablefootmark{j} & 0.14826$\,$\tablefootmark{j} & -- & -- & -- & -- & -- & -- & -- & -- \\
  203280 & -- & 1.241$\,$\tablefootmark{a} & 0.253$\,$\tablefootmark{b} & -- & 12.56$\,$\tablefootmark{o} & -- & -- & -- & -- & -- & -- & -- & -- & -- \\
  216956 & 3.850$\,$\tablefootmark{i} & 3.417$\,$\tablefootmark{a} & 6.930$\,$\tablefootmark{b} & 0.54$\,$\tablefootmark{$\star$, p} & 10.970$\,$\tablefootmark{j} & 0.124$\,$\tablefootmark{$\star$, p} & -- & 0.054$\,$\tablefootmark{$\star$, p} & 0.022$\,$\tablefootmark{$\star$, p} & 0.475$\,$\tablefootmark{e} & 0.01$\,$\tablefootmark{$\star$, p} & 0.0912$\,$\tablefootmark{e} & 0.001789$\,$\tablefootmark{q} & 0.0009$\,$\tablefootmark{$\star$, r} \\
  \bottomrule
 \end{tabular}
 \tablefoot{\tablefoottext{$\star$} \mbox{Flux} density of the central point source.}
 \tablebib{
           \tablefoottext{a}{\citet{laureijs:2002};}
           \tablefoottext{b}{\citet{habing:2001};}
           \tablefoottext{c}{\citet{lawler:2014};}
           \tablefoottext{d}{\citet{greaves:2004b};}
           \tablefoottext{e}{\citet{holland:2017};}
           \tablefoottext{f}{\citet{macgregor:2016};}
           \tablefoottext{g}{\citet{gaspar:2013};}
           \tablefoottext{h}{\citet{sibthorpe:2018};}
           \tablefoottext{i}{\citet{su:2006};}
           \tablefoottext{j}{\citet{thureau:2014};}
           \tablefoottext{k}{\citet{churcher:2011};}
           \tablefoottext{l}{\citet{sibthorpe:2010};}
           \tablefoottext{m}{\citet[][derived as mean of the \SI{12}{\m} data in
                                      their Table~1]{matra:2020};}
           \tablefoottext{n}{\citet{spitzer_science_center:IRSA:2021};}
           \tablefoottext{o}{\citet{hindsley:harrington:1994};}
           \tablefoottext{p}{\citet{acke:2012};}
           \tablefoottext{q}{\citet{su:2016};}
           \tablefoottext{r}{\citet{white:2017}.}
           }
 \label{table_obs_longer_wavelengths}
\end{sidewaystable}


\section{Constraints on hot exozodiacal dust parameters}

\begin{figure*}
 \resizebox{\hsize}{!}{
           \includegraphics{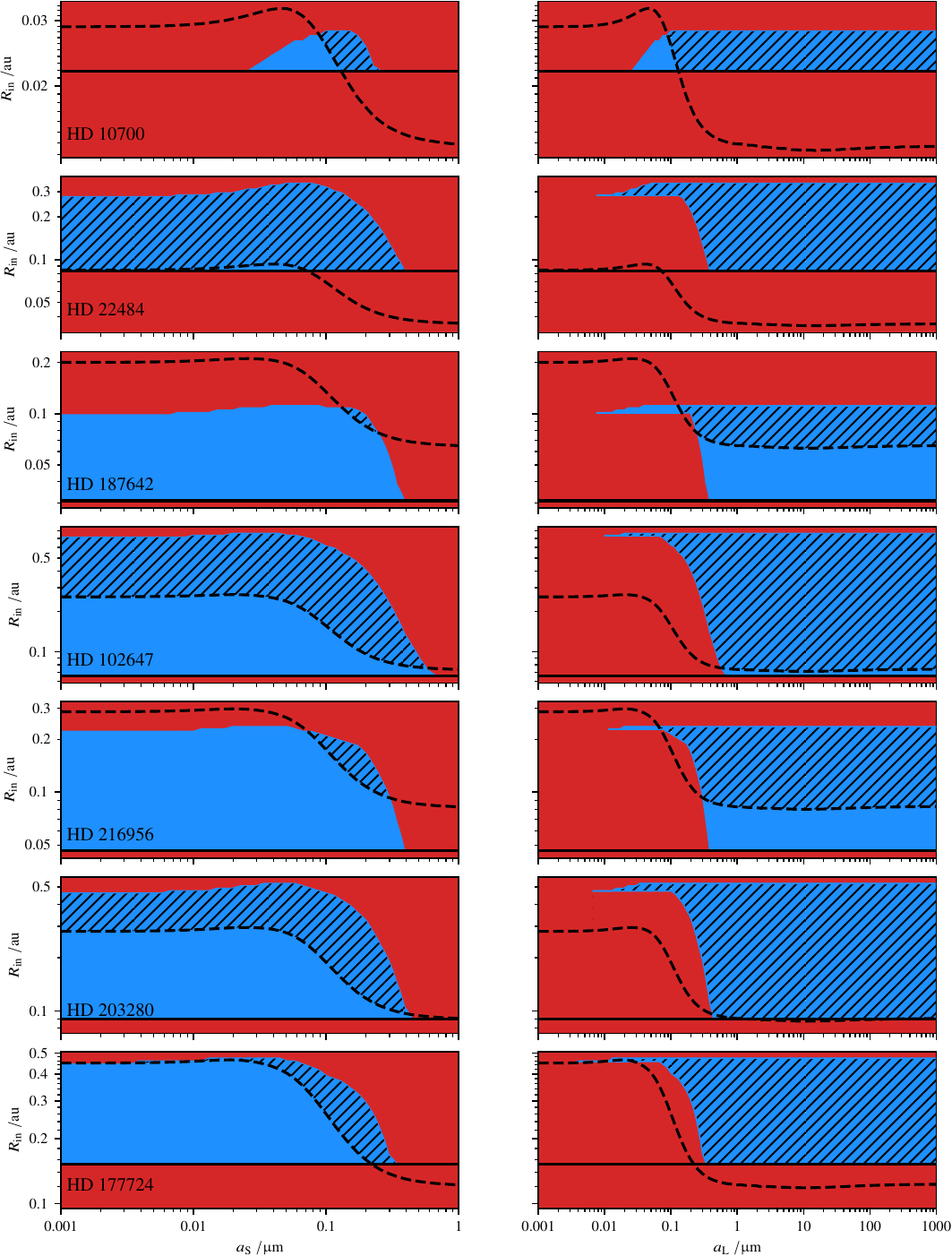}}
 \caption{
 Same as Fig.~\ref{fig_valid_Rin_and_a_HD172167}, but for all other investigated
 targets.}
 \label{fig_valid_Rin_and_a_all_stars}
\end{figure*}

The values of all constraints shown in Fig.~\ref{fig_valid_param_ranges}
are listed in Table~\ref{table_appdx_all_constraints}. Figure~\ref{fig_valid_Rin_and_a_all_stars} shows the same as
Fig.~\ref{fig_valid_Rin_and_a_HD172167}, but for all other investigated systems.
\begin{sidewaystable*}
 \centering
 \caption{Full set of constraints on hot exozodiacal dust parameters, as shown in
          Fig.~\ref{fig_valid_param_ranges}.}
 \begin{tabular}{c|c|c|cccccccc}
  \toprule
  \multicolumn{3}{c|}{} & HD~10700 & HD~22484 & HD~187642 & HD~102647 & HD~216956 & HD~203280 & HD~177724 & HD~172167 \\

  \midrule
  \multirow{6}{*}{$R\sbs{in}$ /\SI{}{au}} & \multirow{2}{*}{\FpNIRunsca, \FmMIRunsca} & max. & --   & 0.17 & --   & 0.49 & --   & --   & --   & -- \\
                                          &                                           & min. & --   & --   & --   & 0.08 & --   & --   & --   & -- \\
                                          & \multirow{2}{*}{\FNIRunsca, \FMIRunsca}   & max. & --   & 0.34 & --   & 0.76 & 0.22 & --   & 0.47 & 0.29 \\
                                          &                                           & min. & --   & --   & --   & 0.08 & 0.10 & --   & --   & 0.15 \\
                                          & \multirow{2}{*}{\FmNIRunsca, \FpMIRunsca} & max. & 0.03 & 0.50 & 0.11 & 1.11 & 0.58 & 0.52 & 1.29 & 0.68 \\
                                          &                                           & min. & --   & --   & 0.08 & 0.07 & 0.09 & 0.10 & --   & 0.14 \\

  \midrule
  \multirow{6}{*}{$M\ / \SI{}{M_\oplus}$} & \multirow{2}{*}{\FpNIRunsca, \FmMIRunsca} & max. & --                              & $1.5 \times 10^{-7\phantom{0}}$ & --                              & $4.0 \times 10^{-7\phantom{0}}$ & --                              & --                              & --           & -- \\
                                          &                                           & min. & --                              & \num{4.7e-10}                   & --                              & \num{3.5e-10}                   & --                              & --                              & --           & -- \\
                                          & \multirow{2}{*}{\FNIRunsca, \FMIRunsca}   & max. & --                              & $8.6 \times 10^{-7\phantom{0}}$ & --                              & $1.0 \times 10^{-6\phantom{0}}$ & $6.1 \times 10^{-8\phantom{0}}$ & --                              & \num{3.8e-7} & $1.2 \times 10^{-7\phantom{0}}$ \\
                                          &                                           & min. & --                              & \num{7.3e-10}                   & --                              & \num{3.9e-10}                   & \num{2.2e-10}                   & --                              & \num{1.0e-9} & \num{5.5e-10} \\
                                          & \multirow{2}{*}{\FmNIRunsca, \FpMIRunsca} & max. & $3.0 \times 10^{-9\phantom{1}}$ & $2.3 \times 10^{-6\phantom{0}}$ & $4.1 \times 10^{-8\phantom{0}}$ & $2.4 \times 10^{-6\phantom{0}}$ & $3.8 \times 10^{-7\phantom{0}}$ & $4.1 \times 10^{-7\phantom{0}}$ & \num{3.4e-6} & $4.2 \times 10^{-7\phantom{0}}$ \\
                                          &                                           & min. & \num{2.5e-11}                   & \num{9.9e-10}                   & \num{5.7e-10}                   & \num{4.7e-10}                   & \num{2.9e-10}                   & \num{4.0e-10}                   & \num{1.6e-9} & \num{5.6e-10} \\

  \midrule
  \multirow{6}{*}{$a\sbs{L}$ /\SI{}{\upmu\m}} & \multirow{2}{*}{\FpNIRunsca, \FmMIRunsca} & max. & --   & --    & --   & --    & --   & --    & --    & -- \\
                                              &                                           & min. & --   & 0.005 & --   & 0.004 & --   & --    & --    & -- \\
                                              & \multirow{2}{*}{\FNIRunsca, \FMIRunsca}   & max. & --   & --    & --   & --    & --   & --    & --    & -- \\
                                              &                                           & min. & --   & 0.008 & --   & 0.01  & 0.09 & --    & 0.004 & 0.11 \\
                                              & \multirow{2}{*}{\FmNIRunsca, \FpMIRunsca} & max. & --   & --    & --   & --    & --   & --    & --    & -- \\
                                              &                                           & min. & 0.10 & 0.01  & 0.17 & 0.009 & 0.01 & 0.007 & 0.01  & 0.01 \\
  \midrule
  \multirow{6}{*}{$a\sbs{S}$ /\SI{}{\upmu\m}} & \multirow{2}{*}{\FpNIRunsca, \FmMIRunsca} & max. & --   & 0.26 & --   & 0.39 & --   & --   & --   & -- \\
                                              &                                           & min. & --   & --   & --   & --   & --   & --   & --   & -- \\
                                              & \multirow{2}{*}{\FNIRunsca, \FMIRunsca}   & max. & --   & 0.34 & --   & 0.51 & 0.30 & --   & 0.30 & 0.26 \\
                                              &                                           & min. & --   & --   & --   & --   & 0.08 & --   & --   & 0.10 \\
                                              & \multirow{2}{*}{\FmNIRunsca, \FpMIRunsca} & max. & 0.23 & 0.44 & 0.23 & 0.87 & 0.39 & 0.39 & 0.44 & 0.34 \\
                                              &                                           & min. & 0.09 & --   & 0.15 & --   & --   & --   & --   & -- \\

  \bottomrule
 \end{tabular}
 \label{table_appdx_all_constraints}
\end{sidewaystable*}

\end{appendix}

\end{document}